\documentclass[aps,prx,superscriptaddress,twocolumn,amsmath,longbibliography]{revtex4-2}

\usepackage[utf8]{inputenc}
\usepackage{longtable}% Tables exceeding column/page width
\usepackage{amsmath}%
\usepackage{amsfonts}%
\usepackage{amssymb}%
\usepackage{graphicx}
\usepackage[colorlinks=true,linkcolor=blue,urlcolor=black,hyperfootnotes=true,citecolor=blue]{hyperref}
\usepackage{ulem} % Use then \sout{} to cross a word out
\usepackage[product-units=power, separate-uncertainty=true, multi-part-units=single]{siunitx} %SI units
\usepackage[usenames,dvipsnames]{xcolor}
\usepackage{threeparttable}
\usepackage{multirow}

\usepackage{kantlipsum,setspace}
\newcommand{\fig}{Fig.~\ref}
\newcommand{\figs}{Figs.~\ref}
\newcommand{\dIdU}{$\dd I / \dd U$}
\newcommand{\dd}{\text{d}}
\newcommand{\CuBiSe}{${\mathrm{Cu}}_x{\mathrm{Bi}}_{2}{\mathrm{Se}}_{3}$}  
\newcommand{\BiSe}{${\mathrm{Bi}}_{2}{\mathrm{Se}}_{3}$}
\newcommand{\scanp}{Scan parameters:}
\newcommand{\stabp}{Stabilization parameters:}

\newcommand{\etal}{\textit{et al.}}

\def\bk{{\bold{k}}}

\newcommand{\REV}[1]{\textcolor{black}{#1}}
\newcommand{\JB}[1]{\textcolor{black}{#1}}
\newcommand{\JHB}[1]{\textcolor{black}{#1}}
\newcommand{\YA}[1]{\textcolor{black}{#1}}

\usepackage{array}
\newcommand{\PreserveBackslash}[1]{\let\temp=\\#1\let\\=\temp}
\newcolumntype{C}[1]{>{\PreserveBackslash\centering}p{#1}}
\newcolumntype{R}[1]{>{\PreserveBackslash\raggedleft}p{#1}}
\newcolumntype{L}[1]{>{\PreserveBackslash\raggedright}p{#1}}

\begin{document}

\title{Rotation of gap nodes in the topological superconductor \textbf{$\mathrm{\mathbf{Cu}}_{\mathbf{x}}\mathrm{\mathbf{(PbSe)}}_{\mathbf{5}}(\mathrm{\mathbf{Bi}}_{\mathbf{2}}\mathrm{\mathbf{Se}}_{\mathbf{3}})_{\mathbf{6}}$}}

\author{Mahasweta Bagchi} \affiliation{Physics Institute II, University of Cologne, D-50937 K\"oln, Germany}

\author{Jens Brede} \email{brede@ph2.uni-koeln.de} \affiliation{Physics Institute II, University of Cologne, D-50937 K\"oln, Germany}

\author{Aline Ramires} \email{aline.ramires@psi.ch} \affiliation{Paul Scherrer Institute, CH-5232 Villigen PSI, Switzerland}

\author{Yoichi Ando }   \email{ando@ph2.uni-koeln.de}    \affiliation{Physics Institute II, University of Cologne, D-50937 K\"oln, Germany}

\date{\today}

\begin{abstract}
Among the family of odd-parity topological superconductors derived from \BiSe{}, $\mathrm{Cu}_{x}(\mathrm{PbSe})_{5}(\mathrm{Bi}_{2}\mathrm{Se}_{3})_{6}$ (CPSBS) has been elucidated to have gap nodes. Although the nodal gap structure has been established by specific-heat and thermal-conductivity measurements, there has been no direct observation of the superconducting gap of CPSBS using scanning tunnelling spectroscopy (STS). Here we report the first STS experiments on CPSBS down to 0.35~K, which found that the vortices generated by out-of-plane magnetic fields have an elliptical shape, reflecting the anisotropic gap structure. The orientation of the gap minima is found to be aligned with the bulk direction when the surface lattice image shows twofold symmetry, but, surprisingly, it is rotated by 30$^{\circ}$ when twofold symmetry is absent. In addition, the superconducting gap spectra in zero magnetic field suggest that the gap nodes are most likely lifted. We argue that only an emergent symmetry at the surface, allowing for a linear superposition of gap functions with different symmetries in the bulk, can lead to the rotation of the gap nodes. The absence of inversion symmetry at the surface additionally lifts the nodes. This result establishes the subtle but crucial role of crystalline symmetry in topological superconductivity.
\end{abstract}

\maketitle
\newpage

\section{Introduction}

Topological superconductivity is a current hot topic in condensed matter physics due to its close relevance to Majorana fermions \cite{Sato2017}. However, not many materials have been conclusively identified as topological superconductors. The family of bulk superconductors derived from Bi$_2$Se$_3$ presents a rare case, in which odd-parity topological superconductivity has been well established \cite{Sato2017, Yonezawa2019}. In this class of materials, despite the three-fold rotational symmetry of the lattice, bulk superconducting (SC) properties consistently show peculiar twofold symmetry \cite{Matano2016, Yonezawa2017, Liu2015, Pan2016, Shen2017, Asaba2017, Andersen2018} that points to the realization of an odd-parity gap function with $E_u$ symmetry. Although this gap function is unconventional and strongly anisotropic, the superconductivity is nonetheless protected from disorder due to the generalized Anderson's theorem, thanks to the additional orbital degrees of freedom and layered structure \cite{Scheurer2015, Andersen2020, Zinkl2022}. 

Interestingly, the $E_u$-symmetric gap function under $D_{3d}$ symmetry, \JHB{conventionally called $\Delta_4$ \cite{Fu2010,Fu2014},} is generally a linear superposition of two basis functions, conventionally called $\Delta_{4x}$ and $\Delta_{4y}$, which have nodes along a mirror plane of the crystal lattice or normal to it, respectively. The coefficients of the superposition form the nematic director $\bf{n}$ \cite{Fu2014}. In the presence of a principal rotation axis with threefold symmetry, there are three degenerate superpositions of these basis functions, corresponding to three distinct nematic directors. The selection of one of these superpositions endows a nematic character to the SC state. Note that only the mirror-symmetry-protected nodes are expected to be robust under the $D_{3d}$ symmetry, while others can be lifted by perturbations such as a warping term in the normal state electronic structure \cite{Fu2014}. Elucidating the factors that dictate the nematic axis is important not only for understanding the topological superconductivity in the Bi$_2$Se$_3$-based compounds but also for finding ways to manipulate the SC gap  \cite{Kostylev2020}.  

Experimentally, the orientation of the gap minima differs among experiments even for the same compound \cite{Yonezawa2019}. For example, in Cu$_x$Bi$_2$Se$_3$, the direction of the gap minima has been reported to be 90$^{\circ}$ rotated between bulk \cite{Yonezawa2017} and surface \cite{Tao2018} measurements. In this regard, there is a complexity arising from the threefold rotational symmetry of the Bi$_2$Se$_3$ lattice, which allows for three equivalent rotational domains \cite{Yonezawa2017,Yonezawa2019}; when contributions from two or more domains are superposed, the apparent symmetry may look like, e.g., $\Delta_{4x}$ even when the true symmetry is $\Delta_{4y}$ \cite{Yonezawa2017,Kawai2020}. 

Fortunately, this complexity is absent in the superconductor $\mathrm{Cu}_{x}(\mathrm{PbSe})_{5}(\mathrm{Bi}_{2}\mathrm{Se}_{3})_{6}$ (hereafter called CPSBS) \cite{Sasaki2014, Andersen2018, Andersen2020} which has a monoclinic crystal structure and a topologically nontrivial twofold symmetric gap function \cite{Fu2010, Ando2015, Sato2017}. Specifically, Andersen \etal{} \cite{Andersen2018} showed that the gap function in CPSBS has nodes located on the unique crystallographic mirror plane, giving rise to nodal superconductivity in the bulk, which was evinced by specific-heat \cite{Andersen2018} and thermal-conductivity \cite{Andersen2020} measurements in the mK-regime. In this work, we use scanning tunneling microscopy (STM) and spectroscopy (STS) to directly access the SC gap on the surface of CPSBS. On a relatively clean surface, we found that vortices generated under out-of-plane magnetic fields are elongated in a twofold-symmetric manner, and the elongation occurs in the direction perpendicular to the bulk gap nodes. This is contrary to the naive expectation that the coherence length should be longer along the direction of the gap nodes \cite{Tao2018}, but it is actually consistent with recent theoretical calculations which showed that the vortex anisotropy in a $p$-wave superconductor should rotate by 90$^{\circ}$ as a result of impurity scattering \cite{Sera2020}. On a more disordered surface where the twofold lattice symmetry is smeared, we found that the vortex anisotropy axis is rotated by 30$^{\circ}$, pointing to the rotation of the gap nodes/minima on the surface. Our symmetry analysis shows that rotation of the gap nodes is indeed possible in the presence of an emergent symmetry at the surface and, furthermore, the inversion symmetry breaking at the surface would lead to lifting of the nodes. The latter conclusion is consistent with the observed gap spectra. Our work hence offers a framework to understand the intricate relation between crystal symmetry and the gap function in a topological superconductor.

\begin{figure*}[ht]
	\includegraphics[width=\textwidth]{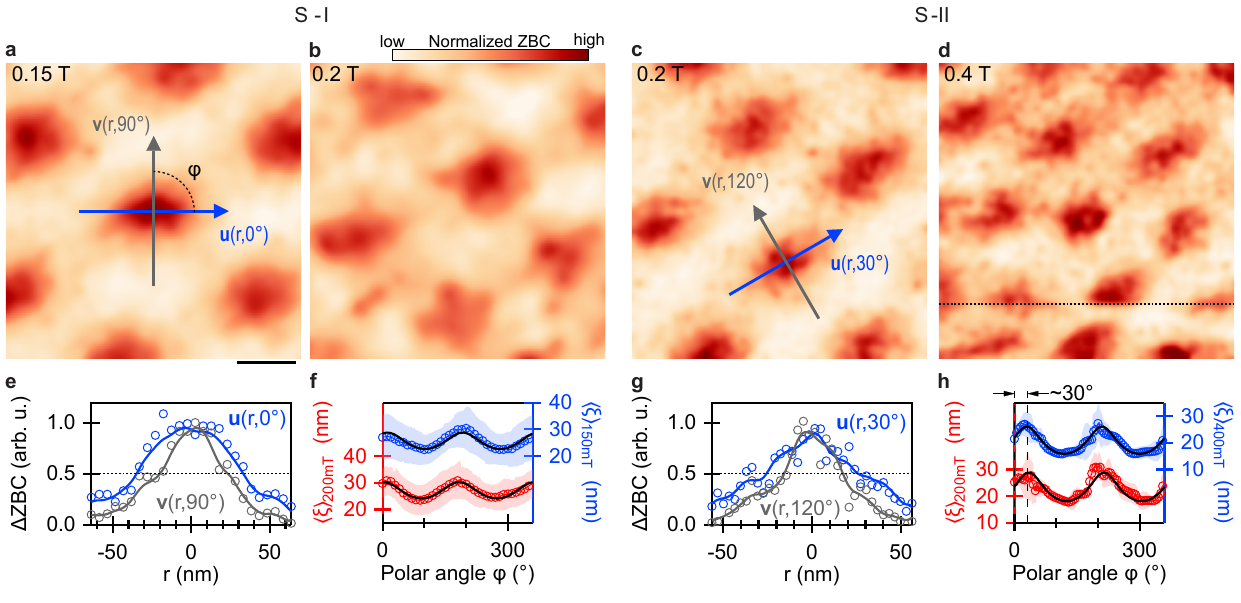}
	\caption{{\bf Elongated vortices in CPSBS.} (a,b) Normalized zero-bias conductance (ZBC) maps taken on sample S-I (a,b) in the out-of-plane magnetic field of 0.15 and 0.2~T. Stabilization parameters: $U=5$~mV, $I=200$~pA and $U_{\text{mod}}= 100$~$\mu$V$_{\text{p}}$. (c,d) Similar maps for sample S-II in 0.2 T and 0.4 T. Stabilization parameters: $U=1$~mV, $I=50$~pA and $U_{\text{mod}}= 100$~$\mu$V$_{\text{p}}$. Gaussian smoothing was applied to all maps to filter spatial variations of the ZBC smaller than the vortex size (unsmoothed data is shown in \cite{SM}). The discontinuity [dotted line in (d)] is due to the vortex lattice motion during the measurement.
 (e,g) Line profiles of the change in normalized ZBC ($\Delta$ZBC) taken across a vortex centre along the vectors shown in (a) and (c); solid lines and open circles are smoothed and raw data, respectively. 
 (f,h) Angular dependence of the average vortex radius $\langle \xi \rangle$. Open circles denote the average of 53 (56) vortices measured for sample S-I at 0.15~T (0.2~T); for sample S-II, the average was from 6 (7) vortices at 0.2~T (0.4~T). The shaded area indicates the standard deviation and the solid black line is a fit to the ellipse equation. The polar angle was measured from the horizontal axis, which is parallel to the monoclinic $b$-axis.
	}
	\label{fig:vortex}
\end{figure*}

\section{Results}

\subsection{Elongated vortices}

%\begin{figure}[]
	%\includegraphics[width=0.95\textwidth]{fig1_new.png}
	%\caption{{\bf Elongated vortices in CPSBS.} (a,b) Normalized zero-bias conductance (ZBC) maps taken in the out-of-plane magnetic field of \textit{B} = 0.2~T (a) and 0.4~T (b) with stabilization parameters $U=1$~mV, $I=50$~pA and $U_{\text{mod}}= 100$~$\mu$V$_{\text{p}}$. The discontinuity in (b) is likely due to the vortex lattice motion during the measurement.
 %(c) Line profiles of the ZBC taken along the long and short axes shown in (a). Solid lines are the fits to Eq.~(1), yielding the anisotropic $\xi_{\rm GL}$. (d) Angular dependence of $\xi_{\mathrm{GL}}$ on the surface plane extracted from a vortex in 0.2 and 0.4 T encircled in (a) and (b); the angle was measured from the horizontal axis, which is parallel to the monoclinic $b$-axis. 
	%}
	%\label{fig:vortex}
%\end{figure}

%[$\text{nomalized ZBC}=\left(\text{ZBC}-\text{ZBC}(r=50~\text{nm})\right)/\left(\text{ZBC}(r=0~\text{nm})-\text{ZBC}(r=50~\text{nm})\right)$] 

We examine the SC gap structure on the surface of CPSBS by applying an out-of-plane magnetic field and imaging the vortex lattice of CPSBS in the mixed state. Even though a major portion of the cleaved surface of CPSBS does not show superconductivity, we were able to observe superconductivity on the sample surface at roughly 17\% of the total scan area (see \cite{SM} for details), which is slightly better than the case of Cu$_x$Bi$_2$Se$_3$ \cite{Tao2018}.
Figure~\ref{fig:vortex}(a,b) and (c,d) shows the spatially resolved normalized zero-bias conductance (ZBC) on the superconducting surface of CPSBS for two different samples, S-I and S-II, and for out-of-plane magnetic fields between 0.15 and 0.4~T (additional data for other fields are shown in \cite{SM}). A vortex lattice is clearly resolved, and the vortex density increases with increasing field. Importantly, all the observed vortices are deformed --- the vortices in sample S-I are roughly elongated along the horizontal axis, while those in sample S-II are rotated by about 30$^{\circ}$ in comparison.
%The horizontal and vertical frames of these figures are taken to be along the crystallographic $a$ and $b$ axes, which are identified from the topographic image discussed later. It is noteworthy that the long axis of the elongated vortices are roughly 30$^{\circ}$ rotated from the $a$-axis, although there are fluctuations in the rotation angle.

Naively, one would expect that elongated vortices reflect an anisotropy in the Ginzburg-Landau (GL) in-plane coherence length, which results from the SC gap anisotropy in \textit{k}-space, with the longest (shortest) coherence length associated with directions for which the gap value is the smallest (largest) \cite{Tao2018}. However, it was recently pointed by theory that the local density of states (LDOS) around vortices imaged by STM experiments can acquire different geometries depending on the strength of impurity scattering \cite{Sera2020}. In particular, it was shown that for a $p$-wave superconductor, the LDOS is elongated along the the direction perpendicular to the gap nodes, in contrast to the naive GL prediction for clean systems. We will discuss the relation between the vortex elongation and the gap nodes/minima in Sec. II-C.

%However, it was recently pointed out that one needs to take into account the impurity scattering in the vortex core which contributes to the local density of states (LDOS) imaged in an STM experiment, and the impurity scattering in $p$-wave SC tends to make the LDOS spread out more in the direction perpendicular to the gap nodes \cite{Sera2020}. We will discuss the relation between the vortex elongation and the gap nodes/minima in Sec. II-C.

It is prudent to mention that a vortex-shape anisotropy can also be caused by a Fermi-velocity anisotropy \cite{Odobesko2020, Kim2021}. However, the ARPES measurements on superconducting CPSBS \cite{Nakayama2015} found no such anisotropy within the experimental error of $\sim$2\%. Even if there were some unexpected Fermi-velocity anisotropy in the region of the vortex lattice, we do not expect it can lead to an anisotropy of the vortex shape in the present case, because CPSBS is in the dirty limit \cite{SM} and a vortex-shape anisotropy cannot result purely from a Fermi-velocity anisotropy in the dirty limit \cite{Odobesko2020}. Also, a vortex can appear elongated if the magnetic field is not perpendicular to the sample surface \cite{Galvis2018}; to dismiss this possibility, we intentionally tilted the applied magnetic field by about $10^\circ$ both along the short and long vortex axes (see Fig.~S9 in \cite{SM}) and observed no significant change in the anisotropy. Hence, it can be concluded that the elongation of the vortex stems from the anisotropic gap.

%\begin{equation} 
	%%\label{eq:vortex_prof}
%\xi(\varphi) = \frac{\xi_{a}\xi_{b}}{\sqrt{(\xi_{b}\cos{\varphi})^2+(\xi_{a}\sin{\varphi})^2}}. \\
%\end{equation}

To quantify the vortex elongation, we first determine the vortex lattice (see \cite{SM} for details) and then take line-cuts of the ZBC at each lattice site corresponding to a vortex centre. Examples of such line-cuts are shown in \fig{fig:vortex} (e) and (g) for directions crossing the vortex shown in (a) and (c), respectively. We follow Sera \textit{et al.}~\cite{Sera2020} to define the vortex-core radius $\xi$ as the half width at half maximum, and determined it as a function of the polar angle $\varphi$. We obtained this $\xi(\varphi)$ for 53 (56) vortices in sample S-I at 0.15~T (0.2~T) and for 6 (7) vortices in sample S-II at 0.2~T (0.4~T). Note that $\xi$ is not necessarily equal to the GL coherence length due to the scattering-induced LDOS \cite{Sera2020}. The $\xi(\varphi)$ data are averaged over all vortices in each data set to yield $\langle \xi(\varphi) \rangle$, which is plotted for sample S-I in \fig{fig:vortex} (f) and for sample S-II in (h). The observed $\varphi$-dependence is reasonably well fit by the ellipse equation, and the fitting gives the ratio between the major and minor axes, $\gamma \equiv \langle\xi_{\mathrm{major}}\rangle/\langle\xi_{\mathrm{minor}}\rangle$, of $\sim$1.25 ($\sim$1.55) for sample S-I (S-II). The fitting also gives the rotation angle of $\sim$30$^\circ$ for the vortices in sample S-II.

 %We obtain $\xi_{\mathrm{GL}}$ by fitting the line-cut data to the formula typically used for STM analysis of vortex profiles \cite{Eskildsen2002,Bergeal2006},
%\begin{align} 
	%\label{eq:vortex_prof}
%\sigma(r,0) = \sigma_{0} + (1-\sigma_{0})\left(1-\mathrm{tanh}\left[r/(\sqrt{2}\xi_{\mathrm{GL}}) \right] \right),
%\end{align}
%where $\sigma_{0}$ is the normalized ZBC away from a vortex centre and $r$ is the distance to the vortex core center. The result of such fittings for the two axes are shown in \fig{fig:vortex}(c). The angular dependence of $\xi_{\mathrm{GL}}$ obtained from similar fits is shown in \fig{fig:vortex}(d) for both 0.2 and 0.4 T. These data give the anisotropy ratio $\gamma_{\mathrm{GL}}=\xi_{\mathrm{long}}/\xi_{\mathrm{short}}\approx$ 1.7$\pm$0.3 (1.6$\pm$0.3) for 0.2~T (0.4~T). Tao \etal{}~\cite{Tao2018} observed similar values of $\gamma_{\mathrm{GL}}$ on the surface of \CuBiSe{}. 

%We also calculated the angular dependence of $\xi_{\mathrm{GL}}$ for all the vortices within the field of view.

\subsection{Superconducting gap spectra}

\begin{figure*}[ht]
	\includegraphics[scale=0.95]{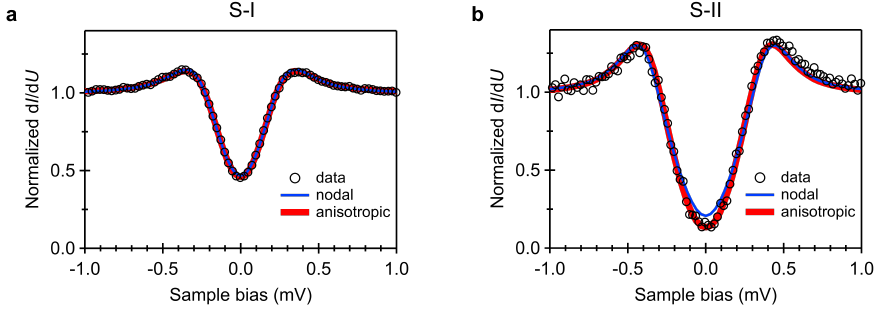}
	\caption{{\bf Superconducting gap spectra.} Representative high-resolution gap spectra taken in zero field, at the base temperature of 0.35~K and in the same area as the vortex maps. The spectrum in (a) is a result of averaging 10000 individual spectra covering an area of $(500~\text{nm})^2$, while the data in (b) is after averaging 5000 individual spectra taken at the same spot to enhance the resolution. The fits of the data to Eq.~(\ref{eq:dynes2}) assuming two different gap functions [nodal (blue) and node-lifted anisotropic (red) types] are overlayed on the data. In these fits, the effective temperature was fixed at 0.7 K. 
	The set of parameters ($\Delta_0$, $\Delta_1$, $\Gamma$) obtained from the fits in (a) for sample S-I are (0, 0.26, 0.03) and (0.07, 0.18, 0.05) for the nodal and node-lifted scenarios, respectively, and those for sample S-II in (b) are (0, 0.39, 0) and (0.11, 0.26, 0.01) for the two scenarios (all in mV unit).
	Any error of the fitting routine is well below the experimental uncertainty due to the spatial inhomogeneity of the total gap magnitude, which was about 0.05~\text{meV} in each area \cite{SM}. 
	\stabp{} $U=5 $~mV, $I=500$~pA for sample S-I (a) and $U=3$~mV, $I=50$~pA for sample S-II (b). 
	}
	\label{fig:sc_gapb}
\end{figure*}

As discussed above, our observation of elongated vortices points to an anisotropic SC gap. To investigate the anisotropy in the SC gap function, we analyzed the representative \dIdU-spectra (\fig{fig:sc_gapb}) measured in zero field at the lowest fridge temperature of 0.35 K  for samples S-I and S-II. We fit the spectra by using a generalization of Dynes formula \cite{Dynes1978} for the momentum-resolved superconducting DOS, 
\begin{align}
	\label{eq:dynes2}
	N_{\mathbf{k}}(E)= \left| Re\left[ (E-i\Gamma)/\sqrt{(E-i\Gamma)^{2}-\Delta^{2}_{\bold{k}}}\, \right] \right|,	
\end{align} 
where we assume a circular Fermi surface. $\Gamma$ is an effective broadening parameter due to pair-breaking scattering and $\Delta_{\bold{k}}$ is the SC gap that can have a $\bold{k}$-dependence. We fit a twofold symmetric gap with $\Delta_{\bold{k}}=\Delta_{0}+ \Delta_{1}|\cos\theta_\bold{k}|$: when the gap is nodal, $\Delta_{0}$ = 0. 
The tunnelling conductance \dIdU{} is given by
\begin{align}
\frac{dI}{dU} \propto \int N_{\bold{k}}(E)f'(E+eU)d\bold{k}dE,
\end{align}
where $f(E)$ is the Fermi-Dirac distribution function at the effective temperature $T_{\mathrm{eff}}$. The effective temperature of our STM experiments at the fridge temperature of 0.35 K was independently determined by a measurement of pure Nb \cite{Bagchi2022} to be 0.7 K, so we fixed $T_{\mathrm{eff}}$ = 0.7 K and used $\Gamma$, $\Delta_0$ and $\Delta_1$ as fitting parameters.

In \fig{fig:sc_gapb}, we show fits to two different types of gap function: (i) nodal gap ($\Delta_{0}$ = 0) and (ii) twofold symmetric gap with lifted nodes ($\Delta_{0} \neq 0$). 
As is described in detail in \cite{SM}, the size of the SC gap on the CPSBS surface varies with location. Thanks to a relatively large local gap, the data in \fig{fig:sc_gapb}(b) from sample S-II have a low ZBC which helps to infer if the nodes are lifted: The nodal fit of the spectrum yields a ZBC that is higher than the data even for $\Gamma$ = 0, which is clearly unreasonable given the dirty-limit nature of CPSBS \cite{SM}.  On the other hand, a reasonable fit is obtained for the anisotropic gap with lifted nodes, yielding $\Delta_{0}$ = 0.11~meV and $\Delta_{1}$ = 0.26~meV with $\Gamma$ = 10 $\mu$eV.  Hence, the data in \fig{fig:sc_gapb}(b) strongly suggest that the nodes are lifted at the surface. 
\JB{Here we should note that the conclusion of lifted nodes is based on the fits to the commonly-assumed sinusoidal gap function, and it may not be valid if the anisotropic gap has a non-sinusoidal function with unusually steep node.} 
The large anisotropy ratio ($\Delta_{0}$+$\Delta_{1}$)/$\Delta_{0} \simeq$ 3.4 implies a pronounced minima in the gap function, which is consistent with the finite quasiparticle scattering even in the absence of vortices.
Due to a smaller local gap resulting in a large ZBC, the data in \fig{fig:sc_gapb}(a) from sample S-I do not allow us to distinguish between nodal and the node-lifted scenarios; nevertheless, the fit with the anisotropic gap function yields a reasonable result with an anisotropy ratio of $\sim$3.6.
In fact, the spectra obtained at all the SC regions are consistent with the anisotropic gap function (see Fig.~S11 in \cite{SM}), even though the value of $\Delta_{0}$ and $\Delta_1$ varies significantly. This large variation appears to reflect the fact the the superconductivity in CPSBS (and in all other Bi$_2$Se$_3$-based superconductors) is weakened or disappear at a larger part of the surface, whose origin is a topic of on-going research: For example, in a recent paper \cite{Bagchi2022} it was proposed that a strong electric field due to \JB{intrinsic} surface band bending may break Cooper pairs near the surface. In the case of CPSBS, the strength of surface band bending would vary depending on the density of Cu dopants found on the surface.
%In any case, the large variation of $\Delta_{\bold{k}}$ at different regions of the surface makes the comparison between $T_c$ and $\Delta_{\bold{k}}$  meaningless.

\JB{To strengthen the conclusion of the anisotropic gap, we performed additional experiments on sample S-III to measure the dependence of the SC gap spectra on the direction of the in-plane magnetic field, $\varphi$ (see \cite{SM} for details). The $\varphi$-dependence of the spectra, in particular the conductance at zero bias, is clearly twofold symmetric (see Fig. S12 in  \cite{SM}). A similar phenomenon was reported for \CuBiSe{} \cite{Tao2018} and was taken as additional evidence for an anisotropic gap.
These observations are also in good agreement with the theoretical prediction by Nagai~\cite{Nagai2014}, who showed that the angular dependence of the zero-energy density of states has deep minima when the in-plane magnetic field is aligned with the direction of the nodes in the $\Delta_4$ gap realized in \CuBiSe{}.}

\subsection{Orientation of the gap minima}

\begin{figure*}[hbt]
	\includegraphics[width=0.95\textwidth]{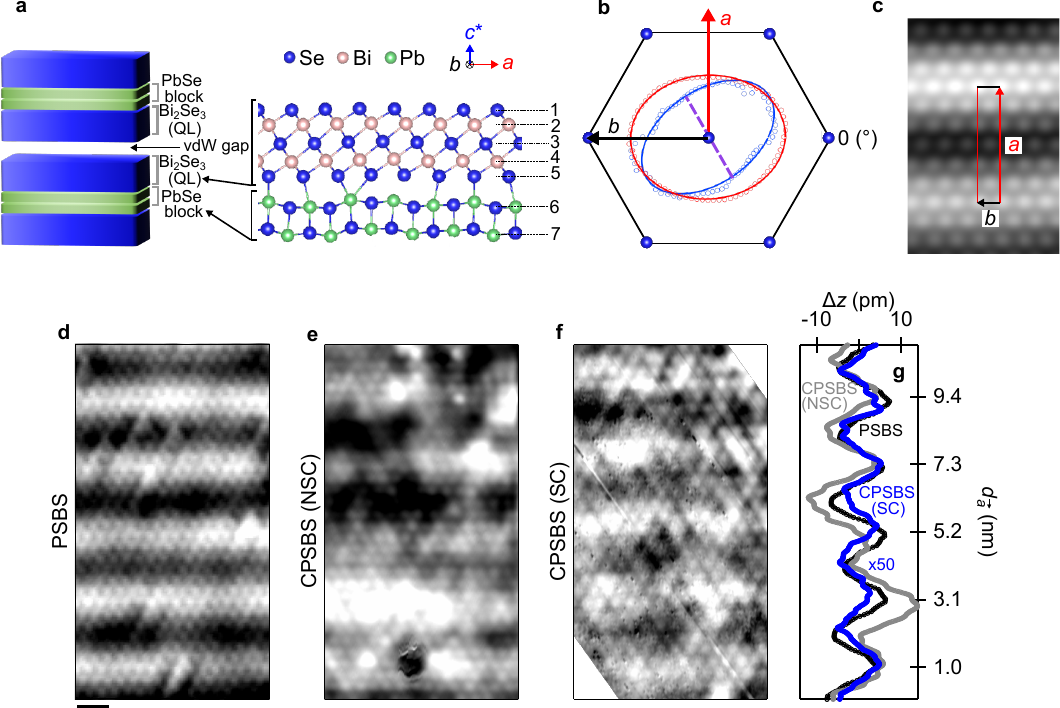}
	\caption{{\bf Crystal structure and the gap anisotropy.} (a) Schematics of the crystal structure of PSBS. (b) Average profile (open circles) and the fit to the ellipse equation (solid line) of vortices at 0.2 T for sample S-I (red) and S-II (blue) [the same data as in \figs{fig:vortex}(f) and (h)] with respect to the top Se lattice and the monoclinic \textit{a} and \textit{b} axes; the mirror plane is parallel to $a$. (c) Fourier-filtered image of (d) overlayed with the monoclinic unit vectors; the stripe periodicity of $\sim$2.1~nm agrees with the $a$ unit length. (d) Atomically resolved topmost Se layer on the cleaved surface of PSBS showing 1D stripes running across the whole surface. The stripe pattern is also observed on the non-superconducting areas of the CPSBS surface in sample S-II (e) and the superconducting area of sample S-I (f). Scale bar corresponds to 1~nm. (g) Averaged STM height profile along the vertical direction in (d) to (f). \scanp{} $U=900$~mV, $I=20$~nA for (d); $U=30$~mV, $I=500$~pA for (e); $U=900$~mV, $I=200$~pA for (f).
		}
\label{fig:topograph}
\end{figure*}

We now turn to the topographic images of CPSBS and those of pristine $(\mathrm{PbSe})_{5}(\mathrm{Bi}_{2}\mathrm{Se}_{3})_{6}$ (called PSBS) to identify the orientation of the gap minima.  Figure \ref{fig:topograph}(a) shows the schematics of the crystal structure of PSBS/CPSBS. Upon cleaving the PSBS crystal, one usually obtains a surface that is terminated by a single quintuple layer (QL) of the \BiSe{} unit on top of the PbSe layer \cite{Nakayama2019}. Figure \ref{fig:topograph}(d) shows a typical topograph on such a surface, where a clear one-dimensional (1D) stripe pattern with atomically resolved top Se layer is observed. Although the lattice in CPSBS is more disordered due to Cu intercalation, a similar stripe pattern is observed in atomic-resolution images on a SC area of CPSBS of sample S-I [\fig{fig:topograph}(f)] as well as on a non-superconducting (NSC) area of sample S-II [\fig{fig:topograph}(e)]). The stripe corrugation is less pronounced in the SC area of sample S-I [Fig.~\ref{fig:vortex}(g)].

The 1D stripe can be understood as a commensuration effect (similar to a moir\'e pattern) arising from the stacking of the square PbSe lattice and the hexagonal Bi$_2$Se$_3$ lattice. One can see in \fig{fig:topograph}(a) that the crystal structure repeats every six Se atoms in the layer 5 and every five Pb atoms in layer 6 along the $a$-axis, and this repetition defines the unit length along the $a$-axis. As shown in \fig{fig:topograph}(c), the stripe periodicity agrees with the length of the monoclinic $a$-axis, which clearly indicates that the stripes come from this commensuration of Se and Pb sublattices. This in turn allows for a unique determination of the in-plane monoclinic lattice vectors on the hexagonal top Se layer, i.e., the stripes are running along the $b$-axis. While we were not able to resolve the crystal lattice in the SC area of sample S-II where the vortex lattice shown in Fig.~\ref{fig:vortex}(c,d) was recorded (a similar case was reported by Tao \etal{} \cite{Tao2018} for \CuBiSe), we know from X-ray diffraction analysis that in our CPSBS sample the orientation of the monoclinic axes is macroscopically identical. This is further verified in STM since we only observed one fixed orientation of the 1D-stripe with respect to our scan coordinates, globally. 

\begin{figure*}[ht]
	\includegraphics[width=0.6\textwidth]{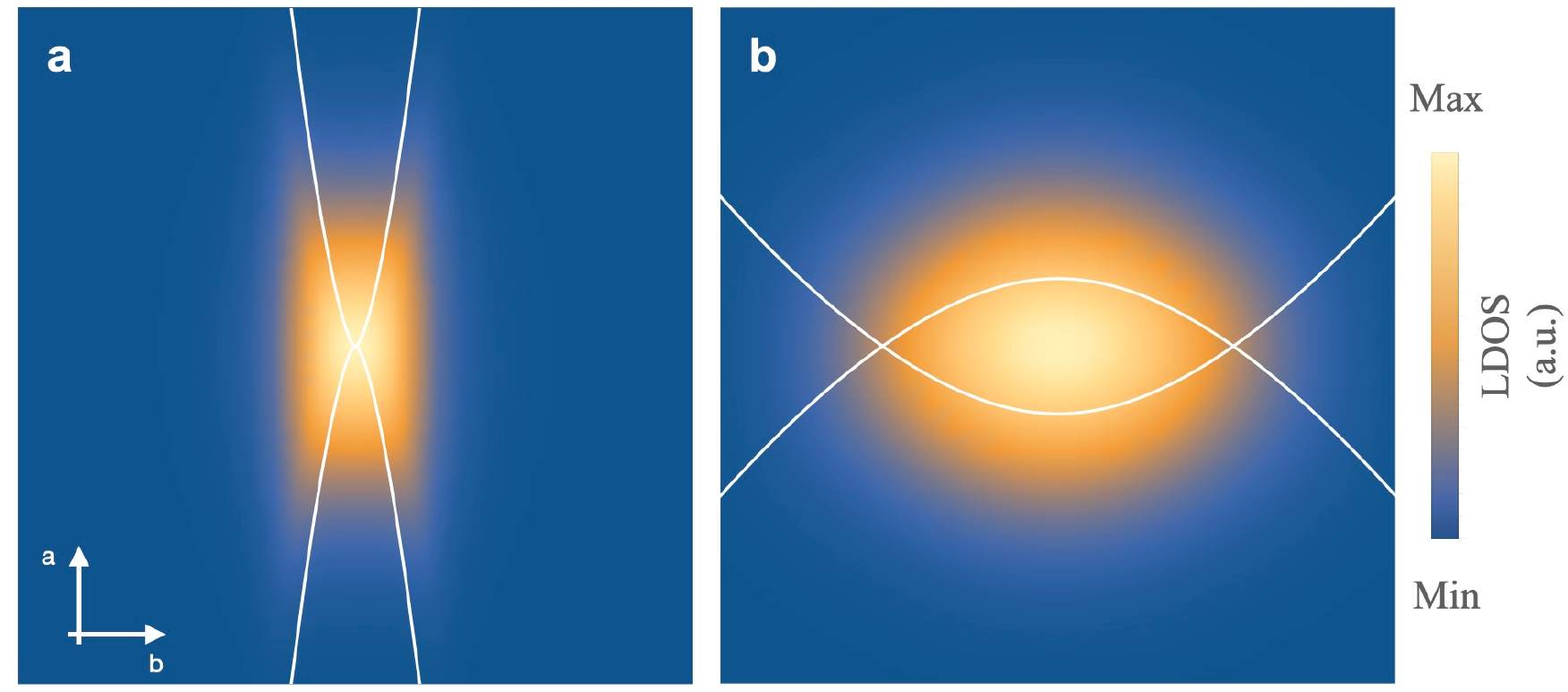}
	\caption{\JHB{ {\bf  LDOS from the quasiparticle trajectory picture.} The enveloping curves of the quasiparticle paths at which the LDOS diverges are shown in white lines. The density plots correspond to the LDOS obtained by Green's functions with poles determined by the enveloping curves smeared by $\delta/\Delta_\infty = 0.05$ and with an isotropic exponential decay characterized by a length scale equal to $\xi_0/400$. Here $\Delta_\infty$ is the gap magnitude in the bulk, $\epsilon$ is the bias energy scale, and $\xi_0$ the coherence length. The field of view corresponds to 0.2$\xi_0$ along the a- and b-directions. (a) Clean limit, with $\epsilon/\Delta_\infty =10^{-4}$. (b) Dirty limit, with $\epsilon/\Delta_\infty = 0.02$. 
%The plots follow the analytical form given by Nagai {\it et al.} \cite{Nagai2006} [Eq. (3.45) of their paper] for a nodal superconductor with twofold symmetry, $\lambda(\theta) = \cos(\theta)$. 
}}
		\label{fig:quasiclassical}
	\end{figure*}

In \fig{fig:topograph}(b), we replot the anisotropic $\xi$ with respect to the hexagonal lattice and the monoclinic axes for both samples S-I and S-II. Focusing first on the data for S-I, it is to be remarked that the vortex core is elongated roughly along the monoclinic $b$-axis, whereas the gap nodes in the bulk lie along the $a$-axis \cite{Andersen2018}. This fact indicates that the vortex-shape anisotropy is not dictated by the GL coherence length anisotropy but by the maxima of the LDOS in the presence of impurity scattering. \JHB{A recent theoretical work has reported that the maxima in the LDOS of a p-wave twofold symmetric superconductor can be rotated by 90$^\circ$ in the presence of impurity scattering \cite{Sera2020}.  This result can be understood in terms of the quasiparticle trajectory picture obtained within the quasiclassical theory of superconductivity \cite{Eilenberger1968, Kita2015, Kopnin2001}. The LDOS at point $\bold{r}$ and energy $\epsilon$ for an isotropic two-dimensional Fermi surface can be generally evaluated as \cite{Nagai2006}}
\begin{eqnarray}
\text{LDOS}(\bold{r}, \epsilon) = - \nu(0) \int \frac{d\Omega_\bold{k}}{4 \pi} \text{Re}[\text{tr}(\hat{g}^R(\bold{r},\tilde{\bold{k}}, \epsilon))],
\end{eqnarray}
\JHB{where $\nu(0)$ is density of states at the Fermi energy and $\hat{g}^R(\bold{r}, \bold{\tilde{k}}, \epsilon)$ is the retarded Green's function obtained from the quasiclassical Eilenberger equation written as matrices in spin space, and $\tilde{\bold{k}}=\bold{k}/k_F$ is the momentum normalized by the Fermi momentum $k_F$. An analytic form for $\hat{g}^R(\bold{r}, \bold{\tilde{k}}, \epsilon)$ around a vortex core can be obtained by a convenient parametrization of the Green's functions that cast the Eilenberger equation in terms of matrix Riccati equations. At low energies and near the vortex core, one can perform a Kramer-Pesch approximation, expanding the Green's function in $\epsilon/\Delta_\infty$ and $y/\xi_0$, where $\Delta_\infty$ corresponds to the bulk gap, $y$ is the quasiparticle trajectory impact parameter, $\xi_0$ is the superconducting coherence length $\xi_0 = v_F/(\pi\Delta_\infty)$, and $v_F$ the Fermi velocity. Within this expansion, the Ricatti equations form a set of inhomogeneous linear differential equations with closed form solutions, from which it is possible to identify the condition for the divergence of the LDOS for an arbitrary anisotropic superconducting gap. This condition is given by a parametric equation in the in-plane angle $\theta$ around the vortex, which defines an enveloping function for quasiparticle paths given by}

\begin{eqnarray}
\begin{pmatrix}
\frac{x}{\xi_0}
\\
\frac{y}{\xi_0}
\end{pmatrix}
=
\frac{\epsilon}{\Delta_\infty \lambda^2(\theta)}
\begin{pmatrix}
\frac{2}{\lambda(\theta)}\frac{\partial\lambda(\theta)}{\partial\theta} \cos\theta - \sin\theta
\\
\frac{2}{\lambda(\theta)}\frac{\partial\lambda(\theta)}{\partial\theta} \sin\theta + \cos\theta
\end{pmatrix},
\end{eqnarray}
\JHB{where $\lambda(\bold{k}) = \sqrt{\frac{1}{2} \text{tr} [\hat{\Delta}^\dagger(\bold{k}) \hat{\Delta}(\bold{k})]}$, with $ \hat{\Delta}(\bold{k})$ the superconducting order parameter matrix written in spin space. For a superconductor having a twofold-symmetric gap, the simplest form of gap anisotropy is captured by $\lambda(\theta) = \lambda_0 |\cos\theta|$, with the corresponding envelope function and LDOS  shown in Fig. \ref{fig:quasiclassical}. The LDOS is obtained from the poles determined by the enveloping function smeared by a small factor $\delta/\Delta_\infty$ associated with the presence of phonons or impurity scattering. Note that in the dirty limit the anisotropy of the LDOS is rotated by $90^o$ w.r.t. the anisotropy in the clean limit, the latter dictated by the coherence length anisotropy. For details, see \cite{Nagai2006}. A similar approach has been useful for understanding the rotation of vortices as a function of applied bias \cite{Ichioka1996, Kaneko2012}.}

It is worthwhile to note that $p$-wave superconductivity is commonly known to be fragile against impurity scattering; however, in CPSBS the generalized Anderson theorem \cite{Andersen2020} protects the unconventional pairing even in the dirty limit. Note also that this impurity effect on the vortex shape was not considered in the previous work on \CuBiSe ~\cite{Tao2018}, which concluded that the gap minima at the surface are 90$^{\circ}$ rotated compared to the bulk.

%which occurs in the direction perpendicular to the gap nodes in a $p$-wave superconductor \cite{Sera2020}. Note that in usual cases, $p$-wave superconductivity is killed by impurity scattering, but in CPSBS the generalized Anderson theorem \cite{Andersen2020} protects the unconventional pairing even in the dirty limit. Note also that this impurity effect on the vortex shape was not considered in the previous work on \CuBiSe ~\cite{Tao2018}, which concluded that the gap minima at the surface are 90$^{\circ}$ rotated compared to the bulk.
% A similar elongation of vortex cores perpendicular to the direction of line nodes was observed previously in the superconductor YNi$_2$B$_2$C~\cite{Kaneko2012}. 

Interestingly, the elongation axis of the vortices in sample S-II shown in \fig{fig:vortex}(c,d) is 30$^{\circ}$ rotated from that in sample S-I. Following the conclusion that the vortex elongation occurs in the direction of gap maxima in CPSBS, the data in \fig{fig:vortex}(c,d) suggest that on the SC surface of sample S-II, the gap minima are rotated from the monoclinic $a$ axis by 30$^{\circ}$, which is perpendicular to one of the mirror planes of $D_{3d}$ symmetry and corresponds to the $\Delta_{4y}$ gap. 
\JB{In correspondence with this result, the in-plane magnetic-field-direction dependence of the gap spectra observed in sample S-III, which was discussed in Sec. II-B, also shows that the gap minima is rotated from the $a$ axis by $\sim$20$^{\circ}$ (see \cite{SM} for details). As shown in Fig. S13 in \cite{SM}, the SC areas of both samples S-II and S-III are so disordered that the stripe pattern indicating the twofold symmetry of the lattice is no longer observed. This implies that the threefold symmetry of the Bi$_2$Se$_3$ QLs is effectively restored due to disorder in both samples.} 
In the next section, we argue that the rotation of the gap minima observed on the SC surface of samples S-II \JB{and S-III} can be understood as a consequence of this emergent symmetry.

%It is prudent to note, that only few regions on the surface show superconductivity, and we have to assume that the observed rotation of the gap minima in these regions is connected to the superconductivity of the bulk. However, as shown in \cite{SM}, we measure roughly the same orientation of the gap minima in both samples S-II and S-III for superconducting regions which differ in LDOS, superconducting gap magnitude, and topographic features but share the absence of twofold symmetry at the surface. Therefore, our observations are consistent with electric field effects being responsible for the general observability of superconductivity at the surface while the rotation of the gap minima is linked to

\section{Theoretical analysis}

\begin{figure*}[htb]
	\includegraphics[width=\textwidth]{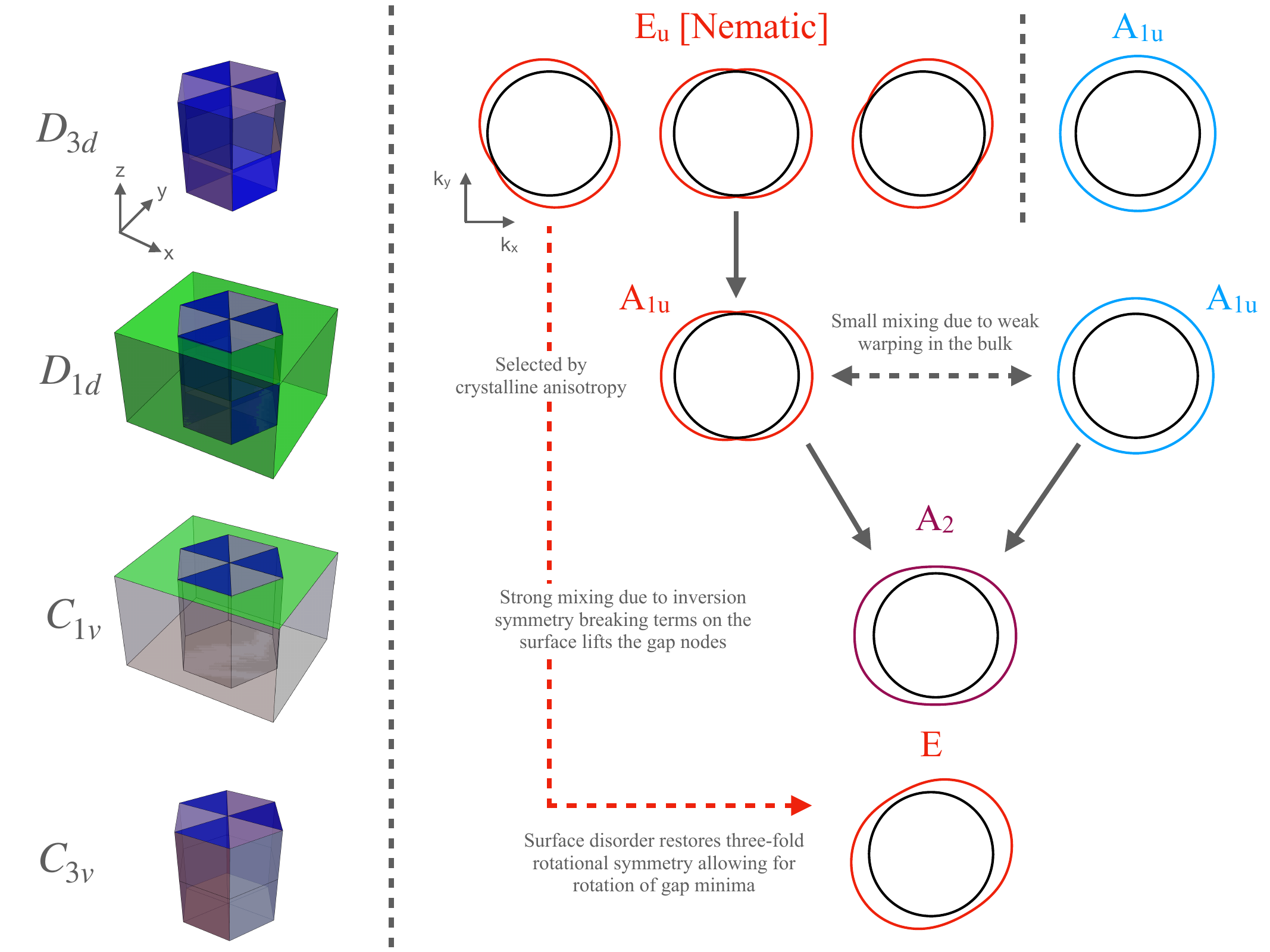}
	\caption{{\bf Symmetry analysis of the superconducting order parameter.} 
Left: Representative objects following the same point group symmetries as Bi$_2$Se$_3$ ($D_{3d})$, CPSBS ($D_{1d}$), and the surface of CPSBS ($C_{1v}$ or $C_{3v}$, without or with disorder, respectively). The colour of the objects matches the corresponding Bi$_2$Se$_3$ and PbSe blocks in Figure \ref{fig:topograph}(a). Right: SC gaps (coloured lines) at the Fermi surfaces (black circles) for the symmetry channels relevant for the discussion in the text. More details on the parameters used to generate the figures for each symmetry scenario are given in the Supplemental Material \cite{SM}. 
		\label{fig:theory}}
	\end{figure*}

We now present a symmetry analysis which provides a consistent picture for the above observations, \JB{under the assumption that the superconductivity observed on the surface inherits the unconventional pairing from the bulk. We emphasize that, although the nature of the superconductivity on the surface is apparently different from that of the bulk, it must be unconventional itself, because vortices cannot be anisotropic in the dirty limit of a conventional superconductor.}

We start the discussion from the perspective of the Bi$_2$Se$_3$ QLs with $D_{3d}$ point group symmetry. The minimal model for the normal state electronic structure that properly captures the topological properties of the bands is given in terms of two effective orbitals with opposite parity formed by a symmetric, labelled as $1$, or antisymmetric, $2$, combination of $p_z$ orbitals within the QLs \cite{Zhang2009, Liu2010}. In the orbital-spin basis $\Phi^\dagger_\bk  = (c_{1\uparrow}^\dagger, c_{1\downarrow}^\dagger, c_{2\uparrow}^\dagger, c_{2\downarrow}^\dagger)_\bk$, the normal-state Hamiltonian can be parametrized as:
\begin{eqnarray}\label{Eq:H0}
\hat{H}_0(\bk)= \sum_{a,b}  h_{ab}(\bk) \hat{\tau}_a\otimes\hat{\sigma}_b,
\end{eqnarray}
where $\hat{\tau}_{a=1,2,3}$ are Pauli matrices encoding the orbital degrees of freedom (DOF), $\hat{\sigma}_{b=1,2,3}$ are Pauli matrices encoding the spin DOF, and $\hat{\tau}_0$ and $\hat{\sigma}_0$ are two-dimensional identity matrices in orbital and spin space, respectively. In the presence of time-reversal [acting as $\hat{\Theta} = K \hat{\tau}_0 \otimes (i\hat{\sigma}_2)$, where $K$ stands for complex conjugation] and inversion (implemented as $\hat{P}=\hat{\tau}_3\otimes \hat{\sigma}_0$) symmetries, the only allowed terms in the Hamiltonian have indices $(a,b) = \{(0,0), (2,0),(3,0),(1,1),(1,2),(1,3)\}$. The properties of the $ \hat{\tau}_a\otimes\hat{\sigma}_b$ matrices under the point group operations allow us to associate each of these terms to a given irreducible representation (irrep) of $D_{3d}$, therefore constraining the momentum dependence of the form factors $h_{ab}(\bk)$ by symmetry. More details on the description of the normal state are given in \cite{SM}. The important features to keep in mind are the following: $(0,0)$ and $(3,0)$ correspond to intra-orbital hopping, $(2,0)$ corresponds to inter-orbital hopping, $(1,a)$, with $a=\{1,2,3\}$ correspond to spin-orbit coupling terms. In particular, $(1,3)$ is associated with trigonal warping \REV{and is very small within the parameter range of validity of this effective model \cite{Liu2010}.}

Following the parametrization of the normal state, the order parameters can be generally written as:
\begin{eqnarray}\label{Eq:Delta}
\hat{\Delta}(\bk) =  \sum_{a,b} d_{ab} (\bk)  \hat{\tau}_a \otimes \hat{\sigma}_b (i\hat{\sigma}_2) .
\end{eqnarray}
Focusing on local pairing mechanisms, the allowed momentum-independent gap matrices are associated with antisymmetric matrices $ \hat{\tau}_a \otimes \hat{\sigma}_b (i\hat{\sigma}_2)$. These can be classified according to the irreps of the of $D_{3d}$ point group, as displayed in the second column of Table \ref{Tab:SC} \cite{Fu2010}.

\begin{table}[t]
\begin{center}
    \begin{tabular}{| C{1cm} | C{1cm} | C{1cm} | C{1cm} | C{1cm} | C{1cm} |}
    \hline
   [a,b] & $D_{3d}$ &   $D_{1d}$ &   $C_{1v}$ & $C_{3v}$  \\ \hline
  [0,0] &   \multirow{2}{*}{$A_{1g}$} &  \multirow{2}{*}{$A_{1g}$} & \multirow{2}{*}{$A_{1}$}  & \multirow{2}{*}{$A_{1}$}  \\ \cline{1-1}
     	[3,0]& & & &   \\ \hline
	[2,3]& 	{$A_{1u}$}  &  $A_{1u}$  &  $A_2$ & $A_2$   \\ \hline
	[1,0] &	{$A_{2u}$}  &  $A_{2u}$   &  $A_1$  & $A_1$ \\ \hline
	[2,1] &	 \multirow{2}{*}{$E_{u}$}& $A_{2u}$  & $A_1$  & \multirow{2}{*}{$E$}
				  \\ \cline{1-1} \cline{3-4}
		[2,2]&	& $A_{1u}$  & $A_2$ &   \\ \hline
    \end{tabular}
               \end{center}
        \caption{\label{Tab:SC} \textbf{Classification of superconducting order parameters. } Here we focus on momentum-independent SC order parameters in the microscopic basis for materials in the family of Bi$_2$Se$_3$. The $[a,b]$ indexes in the first column correspond to the parametrization of the SC gap function according to Eq. \ref{Eq:Delta}. The second to fifth columns give the irreducible representation associated with each order parameter for the cases of $D_{3d}$, $D_{1d}$, $C_{1v}$, and $C_{3v}$ point group symmetry, respectively.}
\end{table}

The QLs of Bi$_2$Se$_3$ have $D_{3d}$ symmetry. In CPSBS, the presence of the PbSe layers reduces the point group symmetry from $D_{3d}$ to $D_{1d}$, and the irreps are mapped according to the third column of Table \ref{Tab:SC}. The SC order parameter in the bulk of CPSBS is believed to be of the form $[2,2]$, given its twofold symmetry and the presence of nodes along the mirror plane \cite{Andersen2018, Andersen2020}. This is an odd-parity order parameter which is inter-orbital and spin-triplet in nature. Note that, in the case of $D_{1d}$ symmetry, this order parameter belongs to $A_{1u}$ irrep. This is a one-dimensional irrep and the notion of nematicity does not apply as the threefold symmetry is explicitly broken by the lattice. Note, though, that the order parameter with indices $[2,3]$ belongs to the same $A_{1u}$ symmetry channel in $D_{1d}$. This means that an order parameter in $A_{1u}$ is generally a linear superposition of $[2,2]$ and $[2,3]$. Even if pairing is primarily driven by interactions promoting the order parameter $[2,2]$, the combination of spin-orbit coupling terms $(1,2)$ and $(1,3)$ in the normal state Hamiltonian could lead to the development of SC correlations with $[2,3]$ character. \REV{Nevertheless, as the trigonal warping term $(1, 3)$ is small in the Bi$_2$Se$_3$-family of compounds \cite{Liu2010}, the corresponding mixing should also be small in the bulk of CPSBS,} leading to a lifting of nodes that might be too small to be observed experimentally. This information is schematically conveyed in the second row of Figure \ref{fig:theory}.

At the surface of CPSBS inversion symmetry is broken and the point group is reduced to $C_{1v}$. The irreps associated to the order parameters with momentum-independent gap matrices are mapped according to the fourth column of Table \ref{Tab:SC}. Now the order parameter $[2,2]$ belongs to irrep $A_2$, and any order parameter in this symmetry channel should again be  a linear superposition of $[2,2]$ and $[2,3]$. Note that due to inversion symmetry breaking at the surface, the normal-state Hamiltonian includes all $(a,b)$ coefficients, allowing for multiple pairs of terms in the normal state to promote the mixing of the $[2,2]$ and $[2,3]$ order parameters (see detailed discussion in \cite{SM}). 
\REV{These surface terms in the normal-state Hamiltonian contribute further to the lifting of gap nodes, as illustrated in the third row of Figure \ref{fig:theory}, so that the node lifting at the surface would be stronger than that in the bulk. Here it should be emphasized that,} as any gap in $C_{1v}$ should still be symmetric or antisymmetric under reflections along the $k_yk_z$-plane, the gap cannot be rotated under these symmetry considerations. For a rotation of gap nodes or gap minima to take place, a mixing of order parameters in different symmetry channels of $C_{1v}$ would be required.

A possible origin of such a mixing is the strong disorder at the surface. \REV{The rotation of the gap minima was detected only in locations at which no stripe pattern could be observed by STM (see Fig. S13 in \cite{SM}),} suggesting that in these disordered areas the effect of the PbSe layers is weakened and the threefold rotational symmetry present in the Bi$_2$Se$_3$ layers is effectively restored. Under these considerations, the point group symmetry at the \REV{disordered} surface can be identified as $C_{3v}$. Interestingly, this point group has a two-dimensional irrep labelled as $E$, which would allow for mixing of gaps $[2,2]$ and $[2,1]$ that are associated with different irreps in $C_{1v}$. Under the $C_{3v}$ symmetry, by changing the mixing of these two order parameters in the absence of warping, we find that the gap minima can be tuned to any position along the circular Fermi surface. 
\REV{In particular, a ratio $d_{21}/d_{22} =$ 1 generates minima at $30^{\circ}$ from the bulk nodes. While the stability of this particular ratio requires an analysis of the energetics of the system, which is beyond the scope of this paper, the rotation by 30$^\circ$ corresponds to having the nodes along the $a$-axis of the hexagonal notation \cite{Andersen2018} for $D_{3d}$ symmetry. It is plausible that this ratio is in fact stable, as this is the direction of the nodes realized in \CuBiSe{}.}

The last row of Figure \ref{fig:theory} schematically shows a gap that could be generated at the surface under these considerations. \REV{Here, it should be remarked that the two-component nature of the order parameter in the quintuple layers of Bi$_2$Se$_3$-based materials with $D_{3d}$ symmetry is a necessary condition to explain the rotation of the gap at the surface of CPSBS. Therefore, the tunnelling spectra and the vortex anisotropy observed in our experiment provides one more piece of evidence for the intrinsically nematic nature of the SC gap in Bi$_2$Se$_3$-based materials.}

\section{Discussions and Conclusion}

We start our discussions by revisiting some of the experimental findings on \CuBiSe{} samples. Some of the discrepancies between the bulk \cite{Yonezawa2017} and the surface \cite{Tao2018} regarding the direction of the gap minima may be resolved by considering the effect of impurity scattering discussed in Sec. II-C. However, in the literature, even in the bulk measurements on \CuBiSe{}, there are reports which differ in the position of the gap minima with respect to the underlying lattice by 90$^\circ$ \cite{Yonezawa2017, Matano2016}. Recent Knight-shift measurements \cite{Kawai2020} ruled out a multidomain effect in the bulk of the sample which was previously proposed \cite{Yonezawa2017} as the possible reason for the different orientations of the gap. It was proposed in \cite{Kawai2020} that the specific local environment in the sample caused by lattice distortion or strain from dopant intercalation and/or quenching (which is necessary for obtaining superconductivity) may be responsible for determining the nematic axis. A high-resolution x-ray diffraction (XRD) experiment on Sr$_x$Bi$_2$Se$_3$ reported a tiny ($\sim$0.02\%) in-plane lattice distortion \cite{Kuntsevich2018}, while a multimodal synchrotron XRD experiment with a slightly lower resolution did not find any distortion \cite{Smylie2022}. Therefore, the situation in doped-\BiSe{} superconductors is complicated and it is still unclear what dictates the orientation of the gap minima in them. 

In contrast, the orientation of the gap minima in the bulk of CPSBS is robust due to the reduced symmetry of the crystal lattice which has only one mirror plane, and the rotation of the gap minima observed here is a pronounced manifestation of the decisive role of crystalline symmetry in determining the anisotropic axis of the SC gap. The rotation of the gap minima also signifies the intrinsically nematic nature of the SC order parameter in Bi$_2$Se$_3$-based materials, which suggests that the pairing mechanism must be the same for all superconductors in this family of materials.

It is useful to mention that according to the theoretical calculations reported in \cite{Sera2020}, the effect of impurity scattering is different  for $p$-wave and anisotropic $s$-wave order parameters even when the angular dependence of the gap magnitude $|\Delta_{\bold{k}}|$ is the same; in the latter case, there is no sign change in the anisotropic SC gap and the vortex shape becomes isotropic in the presence of strong scattering. Therefore, the observation of elongated vortices in the dirty limit gives additional evidence for the topological odd-parity gap function.

\YA{Given that the topological odd-parity superconductivity is realized in this family of superconductors, an important question is the observability of gapless Majorana surface states, the existence of which is guaranteed by topology. In the present case, the Majorana surface states are expected to comprise dispersive 2D gapless modes, and time-reversal symmetry dictates that they have a helical spin-momentum locking \cite{Sato2017}. The experimental difficulty in observing such Majorana surface states in this family of materials comes from the quasi-2D nature of the Fermi surface \cite{Sato2017} and the destruction of superconductivity at a major portion of the surface \cite{Levy2013, Tao2018, Bagchi2022}. The former implies that the Majorana surface states do not appear on the top surface, and the latter implies that the area where the superconductivity reaches the surface is surrounded by nonsuperconducting areas, causing the possible Majorana states (which may appear at the boundary) to merge with the surrounding metallic states, such that little spectral feature can be observed via STM.}

\YA{In this connection, it is worthwhile to mention that a peculiar zero-bias conductance peak (ZBCP) has been observed in point-contact spectroscopy on Cu$_x$Bi$_2$Se$_3$ \cite{Sasaki2011, Kirzhner2012, Peng2013} and on Nb$_x$Bi$_2$Se$_3$ \cite{Kurter2019}. While there is an argument \cite{Peng2013} that the ZBCP is simply due to Andreev reflections expected from the Blonder-Tinkham-Klapwijk (BTK) theory \cite{Blonder1982} for a high transparency contact, it was pointed out that the conductance dip which is consistently observed at the superconducting gap energy is incompatible with the BTK theory and points to $p$-wave pairing \cite{Kurter2019}. It is useful to note that those point-contact spectroscopy experiments had no problem in detecting superconductivity at the surface, which suggests that the mesoscopic metal in contact with the surface plays a role in locally restoring the superconductivity beneath the contact (via, e.g., electron transfer). One may further speculate that such a mesoscopic metal accesses the boundary of the locally-restored superconductivity. Hence, well-controlled point-contact spectroscopy experiments might be a viable way to address the Majorana surface states expected for this family of superconductors.}

The present result \YA{of our STM experiments} on CPSBS, taken together with the complications in the orientation of the gap minima in \CuBiSe{}, clearly shows that the odd-parity gap function is highly sensitive to crystal symmetry in the topological superconductors derived from \BiSe{}. The symmetry-based analysis of the possible superpositions of different gap functions presented here gives a useful framework to understand odd-parity topological superconductors.

\section{Acknowledgment:}
This project has received funding from the European Research Council (ERC) under the European Union's Horizon 2020 research and innovation programme (grant agreement No 741121) and was also funded by the DFG under CRC 1238 - 277146847 (Subprojects A04 and B06) as well as under Germany's Excellence Strategy - Cluster of Excellence Matter and Light for Quantum Computing (ML4Q) EXC 2004/1 - 390534769. AR is supported by the Swiss National Science Foundation through the Ambizione Grant No. 186043.\\
{\bf Data and materials availability}: Raw data used in the generation of main and supplementary figures are available in Zenodo with the identifier 
10.5281/zenodo.10684776.
%Data and materials availability: Raw data as well as all measurement, data-analysis, and simulation code used in the generation of main and supplementary figures are available in Zenodo with the identifier 10.1234/zenodo.xyz.
\section{Appendix}

{\bf Material:}
We grew $(\mathrm{PbSe})_{5}(\mathrm{Bi}_{2}\mathrm{Se}_{3})_{6}$ single crystals using a modified Bridgeman method as described previously \cite{Sasaki2014,Andersen2018}. Cu was electrochemically intercalated using the recipe of Kriener \etal{} \cite{Kriener2011} with a nominal $x$ value of 1.36. The SC shielding fraction of the resulting CPSBS sample was measured using a Quantum Design superconducting quantum interference device (SQUID) magnetometer (see Fig. S1 in \cite{SM}) and was 59\% for sample S-I and 61\% for sample S-II.

{\bf STM experiments:}
STM experiments were carried out under UHV conditions with a commercial system (Unisoku USM1300) operating at 0.35~K. STM images were recorded in the constant-current mode at the set current $I$ and sample bias voltage $U$. 
\dIdU{} curves and \dIdU{} maps were obtained either using a lock-in amplifier by modulating $U_{\mathrm{bias}}$ and demodulating $I$, or by recording a series of $I$-$U$ curves followed by numerical differentiation. The \dIdU{} maps displayed in \fig{fig:vortex} were smoothed by using a standard Gaussian filter with the smallest ($3\times3$ points) kernel, corresponding to $15 \times 15$~nm$^2$ (a), $20 \times 20$~nm$^2$ (b) and $9.35 \times 9.35$~nm$^2$ (c,d), with up to three iterations. The \dIdU{} spectra displayed in \fig{fig:sc_gapb} were obtained by taking the numerical
derivative of raw $I$-$U$ data and subsequently applying a simple low-pass filter (binominal 21 passes).
All STM data were analyzed using Igor Pro 9.
We used in-house electrochemically etched W tips, first prepared on the Cu(111) crystal. Tip forming is done until a clean signature of the surface state is observed in spectroscopy. Prior to STM measurements, the crystals were cleaved under UHV conditions as described in \cite{Bagchi2022}.
%\cite{Eskildsen2002, Bergeal2006, Renner1991, Nakayama2012}


%{\bf Author contributions:} Y.A. conceived the experiment and A.R. constructed the theory. M.B., supported by J.B. and Y.A., grew the crystals and performed the measurements and data analysis. M.B., A.R. and Y.A. wrote the manuscript with inputs from J.B.

%{\bf Competing Interests:} The authors declare that they have no competing financial interests.
 
%{\bf Correspondence:} Correspondence and requests for materials should be addressed to Y.A. (ando@ph2.uni-koeln.de).

%{\bf Data availability:} The data that support the findings of this study are available at the online depository figshare with the identifier **** and Supplementary Information. Source data are are provided with this paper.
%Additional data are available from the corresponding authors upon reasonable request.


\begin{thebibliography}{50}%
\makeatletter
\providecommand \@ifxundefined [1]{%
 \@ifx{#1\undefined}
}%
\providecommand \@ifnum [1]{%
 \ifnum #1\expandafter \@firstoftwo
 \else \expandafter \@secondoftwo
 \fi
}%
\providecommand \@ifx [1]{%
 \ifx #1\expandafter \@firstoftwo
 \else \expandafter \@secondoftwo
 \fi
}%
\providecommand \natexlab [1]{#1}%
\providecommand \enquote  [1]{``#1''}%
\providecommand \bibnamefont  [1]{#1}%
\providecommand \bibfnamefont [1]{#1}%
\providecommand \citenamefont [1]{#1}%
\providecommand \href@noop [0]{\@secondoftwo}%
\providecommand \href [0]{\begingroup \@sanitize@url \@href}%
\providecommand \@href[1]{\@@startlink{#1}\@@href}%
\providecommand \@@href[1]{\endgroup#1\@@endlink}%
\providecommand \@sanitize@url [0]{\catcode `\\12\catcode `\$12\catcode
  `\&12\catcode `\#12\catcode `\^12\catcode `\_12\catcode `\%12\relax}%
\providecommand \@@startlink[1]{}%
\providecommand \@@endlink[0]{}%
\providecommand \url  [0]{\begingroup\@sanitize@url \@url }%
\providecommand \@url [1]{\endgroup\@href {#1}{\urlprefix }}%
\providecommand \urlprefix  [0]{URL }%
\providecommand \Eprint [0]{\href }%
\providecommand \doibase [0]{https://doi.org/}%
\providecommand \selectlanguage [0]{\@gobble}%
\providecommand \bibinfo  [0]{\@secondoftwo}%
\providecommand \bibfield  [0]{\@secondoftwo}%
\providecommand \translation [1]{[#1]}%
\providecommand \BibitemOpen [0]{}%
\providecommand \bibitemStop [0]{}%
\providecommand \bibitemNoStop [0]{.\EOS\space}%
\providecommand \EOS [0]{\spacefactor3000\relax}%
\providecommand \BibitemShut  [1]{\csname bibitem#1\endcsname}%
\let\auto@bib@innerbib\@empty
%</preamble>
\bibitem [{\citenamefont {Sato}\ and\ \citenamefont {Ando}(2017)}]{Sato2017}%
  \BibitemOpen
  \bibfield  {author} {\bibinfo {author} {\bibfnamefont {M.}~\bibnamefont
  {Sato}}\ and\ \bibinfo {author} {\bibfnamefont {Y.}~\bibnamefont {Ando}},\
  }\bibfield  {title} {\bibinfo {title} {Topological superconductors: a
  review},\ }\href {https://doi.org/10.1088/1361-6633/aa6ac7} {\bibfield
  {journal} {\bibinfo  {journal} {Rep. Prog. Phys.}\ }\textbf {\bibinfo
  {volume} {80}},\ \bibinfo {pages} {076501} (\bibinfo {year}
  {2017})}\BibitemShut {NoStop}%
\bibitem [{\citenamefont {Yonezawa}(2019)}]{Yonezawa2019}%
  \BibitemOpen
  \bibfield  {author} {\bibinfo {author} {\bibfnamefont {S.}~\bibnamefont
  {Yonezawa}},\ }\bibfield  {title} {\bibinfo {title} {Nematic
  superconductivity in doped {Bi$_2$Se$_3$} topological superconductors},\
  }\href {https://doi.org/10.3390/condmat4010002} {\bibfield  {journal}
  {\bibinfo  {journal} {Condens. Matter}\ }\textbf {\bibinfo {volume} {4}},\
  \bibinfo {pages} {2} (\bibinfo {year} {2019})}\BibitemShut {NoStop}%
\bibitem [{\citenamefont {Matano}\ \emph {et~al.}(2016)\citenamefont {Matano},
  \citenamefont {Kriener}, \citenamefont {Segawa}, \citenamefont {Ando},\ and\
  \citenamefont {Zheng}}]{Matano2016}%
  \BibitemOpen
  \bibfield  {author} {\bibinfo {author} {\bibfnamefont {K.}~\bibnamefont
  {Matano}}, \bibinfo {author} {\bibfnamefont {M.}~\bibnamefont {Kriener}},
  \bibinfo {author} {\bibfnamefont {K.}~\bibnamefont {Segawa}}, \bibinfo
  {author} {\bibfnamefont {Y.}~\bibnamefont {Ando}},\ and\ \bibinfo {author}
  {\bibfnamefont {G.-q.}\ \bibnamefont {Zheng}},\ }\bibfield  {title} {\bibinfo
  {title} {Spin-rotation symmetry breaking in the superconducting state of
  {Cu$_x$Bi$_2$Se$_3$}},\ }\href {https://doi.org/10.1038/nphys3781} {\bibfield
   {journal} {\bibinfo  {journal} {Nat. Phys.}\ }\textbf {\bibinfo {volume}
  {12}},\ \bibinfo {pages} {852} (\bibinfo {year} {2016})}\BibitemShut
  {NoStop}%
\bibitem [{\citenamefont {Yonezawa}\ \emph {et~al.}(2017)\citenamefont
  {Yonezawa}, \citenamefont {Tajiri}, \citenamefont {Nakata}, \citenamefont
  {Nagai}, \citenamefont {Wang}, \citenamefont {Segawa}, \citenamefont {Ando},\
  and\ \citenamefont {Maeno}}]{Yonezawa2017}%
  \BibitemOpen
  \bibfield  {author} {\bibinfo {author} {\bibfnamefont {S.}~\bibnamefont
  {Yonezawa}}, \bibinfo {author} {\bibfnamefont {K.}~\bibnamefont {Tajiri}},
  \bibinfo {author} {\bibfnamefont {S.}~\bibnamefont {Nakata}}, \bibinfo
  {author} {\bibfnamefont {Y.}~\bibnamefont {Nagai}}, \bibinfo {author}
  {\bibfnamefont {Z.}~\bibnamefont {Wang}}, \bibinfo {author} {\bibfnamefont
  {K.}~\bibnamefont {Segawa}}, \bibinfo {author} {\bibfnamefont
  {Y.}~\bibnamefont {Ando}},\ and\ \bibinfo {author} {\bibfnamefont
  {Y.}~\bibnamefont {Maeno}},\ }\bibfield  {title} {\bibinfo {title}
  {Thermodynamic evidence for nematic superconductivity in
  {Cu$_x$Bi$_2$Se$_3$}},\ }\href {https://doi.org/10.1038/nphys3907} {\bibfield
   {journal} {\bibinfo  {journal} {Nat. Phys.}\ }\textbf {\bibinfo {volume}
  {13}},\ \bibinfo {pages} {123} (\bibinfo {year} {2017})}\BibitemShut
  {NoStop}%
\bibitem [{\citenamefont {Liu}\ \emph {et~al.}(2015)\citenamefont {Liu},
  \citenamefont {Yao}, \citenamefont {Shao}, \citenamefont {Zuo}, \citenamefont
  {Pi}, \citenamefont {Tan}, \citenamefont {Zhang},\ and\ \citenamefont
  {Zhang}}]{Liu2015}%
  \BibitemOpen
  \bibfield  {author} {\bibinfo {author} {\bibfnamefont {Z.}~\bibnamefont
  {Liu}}, \bibinfo {author} {\bibfnamefont {X.}~\bibnamefont {Yao}}, \bibinfo
  {author} {\bibfnamefont {J.}~\bibnamefont {Shao}}, \bibinfo {author}
  {\bibfnamefont {M.}~\bibnamefont {Zuo}}, \bibinfo {author} {\bibfnamefont
  {L.}~\bibnamefont {Pi}}, \bibinfo {author} {\bibfnamefont {S.}~\bibnamefont
  {Tan}}, \bibinfo {author} {\bibfnamefont {C.}~\bibnamefont {Zhang}},\ and\
  \bibinfo {author} {\bibfnamefont {Y.}~\bibnamefont {Zhang}},\ }\bibfield
  {title} {\bibinfo {title} {Superconductivity with topological surface state
  in {Sr$_x$Bi$_2$Se$_3$}},\ }\href {https://doi.org/10.1021/jacs.5b06815}
  {\bibfield  {journal} {\bibinfo  {journal} {J. Am. Chem. Soc.}\ }\textbf
  {\bibinfo {volume} {137}},\ \bibinfo {pages} {10512} (\bibinfo {year}
  {2015})}\BibitemShut {NoStop}%
\bibitem [{\citenamefont {Pan}\ \emph {et~al.}(2016)\citenamefont {Pan},
  \citenamefont {Nikitin}, \citenamefont {Araizi}, \citenamefont {Huang},
  \citenamefont {Matsushita}, \citenamefont {Naka},\ and\ \citenamefont
  {de~Visser}}]{Pan2016}%
  \BibitemOpen
  \bibfield  {author} {\bibinfo {author} {\bibfnamefont {Y.}~\bibnamefont
  {Pan}}, \bibinfo {author} {\bibfnamefont {A.~M.}\ \bibnamefont {Nikitin}},
  \bibinfo {author} {\bibfnamefont {G.~K.}\ \bibnamefont {Araizi}}, \bibinfo
  {author} {\bibfnamefont {Y.~K.}\ \bibnamefont {Huang}}, \bibinfo {author}
  {\bibfnamefont {Y.}~\bibnamefont {Matsushita}}, \bibinfo {author}
  {\bibfnamefont {T.}~\bibnamefont {Naka}},\ and\ \bibinfo {author}
  {\bibfnamefont {A.}~\bibnamefont {de~Visser}},\ }\bibfield  {title} {\bibinfo
  {title} {Rotational symmetry breaking in the topological superconductor
  {Sr$_x$Bi$_2$Se$_3$} probed by upper-critical field experiments},\ }\href
  {https://doi.org/10.1038/srep28632} {\bibfield  {journal} {\bibinfo
  {journal} {Sci. Rep.}\ }\textbf {\bibinfo {volume} {6}},\ \bibinfo {pages}
  {28632} (\bibinfo {year} {2016})}\BibitemShut {NoStop}%
\bibitem [{\citenamefont {Shen}\ \emph {et~al.}(2017)\citenamefont {Shen},
  \citenamefont {He}, \citenamefont {Yuan}, \citenamefont {Huang},
  \citenamefont {Cho}, \citenamefont {Lee}, \citenamefont {Hor}, \citenamefont
  {Law},\ and\ \citenamefont {Lortz}}]{Shen2017}%
  \BibitemOpen
  \bibfield  {author} {\bibinfo {author} {\bibfnamefont {J.}~\bibnamefont
  {Shen}}, \bibinfo {author} {\bibfnamefont {W.-Y.}\ \bibnamefont {He}},
  \bibinfo {author} {\bibfnamefont {N.~F.~Q.}\ \bibnamefont {Yuan}}, \bibinfo
  {author} {\bibfnamefont {Z.}~\bibnamefont {Huang}}, \bibinfo {author}
  {\bibfnamefont {C.-w.}\ \bibnamefont {Cho}}, \bibinfo {author} {\bibfnamefont
  {S.~H.}\ \bibnamefont {Lee}}, \bibinfo {author} {\bibfnamefont {Y.~S.}\
  \bibnamefont {Hor}}, \bibinfo {author} {\bibfnamefont {K.~T.}\ \bibnamefont
  {Law}},\ and\ \bibinfo {author} {\bibfnamefont {R.}~\bibnamefont {Lortz}},\
  }\bibfield  {title} {\bibinfo {title} {Nematic topological superconducting
  phase in {N}b-doped {Bi$_2$Se$_3$}},\ }\href
  {https://doi.org/10.1038/s41535-017-0064-1} {\bibfield  {journal} {\bibinfo
  {journal} {npj Quant. Mater.}\ }\textbf {\bibinfo {volume} {2}},\ \bibinfo
  {pages} {59} (\bibinfo {year} {2017})}\BibitemShut {NoStop}%
\bibitem [{\citenamefont {Asaba}\ \emph {et~al.}(2017)\citenamefont {Asaba},
  \citenamefont {Lawson}, \citenamefont {Tinsman}, \citenamefont {Chen},
  \citenamefont {Corbae}, \citenamefont {Li}, \citenamefont {Qiu},
  \citenamefont {Hor}, \citenamefont {Fu},\ and\ \citenamefont
  {Li}}]{Asaba2017}%
  \BibitemOpen
  \bibfield  {author} {\bibinfo {author} {\bibfnamefont {T.}~\bibnamefont
  {Asaba}}, \bibinfo {author} {\bibfnamefont {B.~J.}\ \bibnamefont {Lawson}},
  \bibinfo {author} {\bibfnamefont {C.}~\bibnamefont {Tinsman}}, \bibinfo
  {author} {\bibfnamefont {L.}~\bibnamefont {Chen}}, \bibinfo {author}
  {\bibfnamefont {P.}~\bibnamefont {Corbae}}, \bibinfo {author} {\bibfnamefont
  {G.}~\bibnamefont {Li}}, \bibinfo {author} {\bibfnamefont {Y.}~\bibnamefont
  {Qiu}}, \bibinfo {author} {\bibfnamefont {Y.~S.}\ \bibnamefont {Hor}},
  \bibinfo {author} {\bibfnamefont {L.}~\bibnamefont {Fu}},\ and\ \bibinfo
  {author} {\bibfnamefont {L.}~\bibnamefont {Li}},\ }\bibfield  {title}
  {\bibinfo {title} {Rotational symmetry breaking in a trigonal superconductor
  {N}b-doped {${\mathrm{Bi}}_{2}{\mathrm{Se}}_{3}$}},\ }\href
  {https://doi.org/10.1103/PhysRevX.7.011009} {\bibfield  {journal} {\bibinfo
  {journal} {Phys. Rev. X}\ }\textbf {\bibinfo {volume} {7}},\ \bibinfo {pages}
  {011009} (\bibinfo {year} {2017})}\BibitemShut {NoStop}%
\bibitem [{\citenamefont {Andersen}\ \emph {et~al.}(2018)\citenamefont
  {Andersen}, \citenamefont {Wang}, \citenamefont {Lorenz},\ and\ \citenamefont
  {Ando}}]{Andersen2018}%
  \BibitemOpen
  \bibfield  {author} {\bibinfo {author} {\bibfnamefont {L.}~\bibnamefont
  {Andersen}}, \bibinfo {author} {\bibfnamefont {Z.}~\bibnamefont {Wang}},
  \bibinfo {author} {\bibfnamefont {T.}~\bibnamefont {Lorenz}},\ and\ \bibinfo
  {author} {\bibfnamefont {Y.}~\bibnamefont {Ando}},\ }\bibfield  {title}
  {\bibinfo {title} {Nematic superconductivity in
  {${\mathrm{Cu}}_{1.5}{(\mathrm{PbSe})}_{5}{({\mathrm{Bi}}_{2}{\mathrm{Se}}_{3})}_{6}$}},\
  }\href {https://doi.org/10.1103/PhysRevB.98.220512} {\bibfield  {journal}
  {\bibinfo  {journal} {Phys. Rev. B}\ }\textbf {\bibinfo {volume} {98}},\
  \bibinfo {pages} {220512} (\bibinfo {year} {2018})}\BibitemShut {NoStop}%
\bibitem [{\citenamefont {Scheurer}\ \emph {et~al.}(2015)\citenamefont
  {Scheurer}, \citenamefont {Hoyer},\ and\ \citenamefont
  {Schmalian}}]{Scheurer2015}%
  \BibitemOpen
  \bibfield  {author} {\bibinfo {author} {\bibfnamefont {M.~S.}\ \bibnamefont
  {Scheurer}}, \bibinfo {author} {\bibfnamefont {M.}~\bibnamefont {Hoyer}},\
  and\ \bibinfo {author} {\bibfnamefont {J.}~\bibnamefont {Schmalian}},\
  }\bibfield  {title} {\bibinfo {title} {Pair breaking in multiorbital
  superconductors: An application to oxide interfaces},\ }\href
  {https://doi.org/10.1103/PhysRevB.92.014518} {\bibfield  {journal} {\bibinfo
  {journal} {Phys. Rev. B}\ }\textbf {\bibinfo {volume} {92}},\ \bibinfo
  {pages} {014518} (\bibinfo {year} {2015})}\BibitemShut {NoStop}%
\bibitem [{\citenamefont {Andersen}\ \emph {et~al.}(2020)\citenamefont
  {Andersen}, \citenamefont {Ramires}, \citenamefont {Wang}, \citenamefont
  {Lorenz},\ and\ \citenamefont {Ando}}]{Andersen2020}%
  \BibitemOpen
  \bibfield  {author} {\bibinfo {author} {\bibfnamefont {L.}~\bibnamefont
  {Andersen}}, \bibinfo {author} {\bibfnamefont {A.}~\bibnamefont {Ramires}},
  \bibinfo {author} {\bibfnamefont {Z.}~\bibnamefont {Wang}}, \bibinfo {author}
  {\bibfnamefont {T.}~\bibnamefont {Lorenz}},\ and\ \bibinfo {author}
  {\bibfnamefont {Y.}~\bibnamefont {Ando}},\ }\bibfield  {title} {\bibinfo
  {title} {Generalized {A}nderson's theorem for superconductors derived from
  topological insulators},\ }\href {https://doi.org/10.1126/sciadv.aay6502}
  {\bibfield  {journal} {\bibinfo  {journal} {Sci. Adv.}\ }\textbf {\bibinfo
  {volume} {6}},\ \bibinfo {pages} {eaay6502} (\bibinfo {year}
  {2020})}\BibitemShut {NoStop}%
\bibitem [{\citenamefont {Zinkl}\ and\ \citenamefont
  {Ramires}(2022)}]{Zinkl2022}%
  \BibitemOpen
  \bibfield  {author} {\bibinfo {author} {\bibfnamefont {B.}~\bibnamefont
  {Zinkl}}\ and\ \bibinfo {author} {\bibfnamefont {A.}~\bibnamefont
  {Ramires}},\ }\bibfield  {title} {\bibinfo {title} {Sensitivity of
  superconducting states to the impurity location in layered materials},\
  }\href {https://doi.org/10.1103/PhysRevB.106.014515} {\bibfield  {journal}
  {\bibinfo  {journal} {Phys. Rev. B}\ }\textbf {\bibinfo {volume} {106}},\
  \bibinfo {pages} {014515} (\bibinfo {year} {2022})}\BibitemShut {NoStop}%
\bibitem [{\citenamefont {Fu}\ and\ \citenamefont {Berg}(2010)}]{Fu2010}%
  \BibitemOpen
  \bibfield  {author} {\bibinfo {author} {\bibfnamefont {L.}~\bibnamefont
  {Fu}}\ and\ \bibinfo {author} {\bibfnamefont {E.}~\bibnamefont {Berg}},\
  }\bibfield  {title} {\bibinfo {title} {Odd-parity topological
  superconductors: {T}heory and application to
  {${\mathrm{Cu}}_{x}{\mathrm{Bi}}_{2}{\mathrm{Se}}_{3}$}},\ }\href
  {https://doi.org/10.1103/PhysRevLett.105.097001} {\bibfield  {journal}
  {\bibinfo  {journal} {Phys. Rev. Lett.}\ }\textbf {\bibinfo {volume} {105}},\
  \bibinfo {pages} {097001} (\bibinfo {year} {2010})}\BibitemShut {NoStop}%
\bibitem [{\citenamefont {Fu}(2014)}]{Fu2014}%
  \BibitemOpen
  \bibfield  {author} {\bibinfo {author} {\bibfnamefont {L.}~\bibnamefont
  {Fu}},\ }\bibfield  {title} {\bibinfo {title} {Odd-parity topological
  superconductor with nematic order: {A}pplication to
  {${\mathrm{Cu}}_{x}{\mathrm{Bi}}_{2}{\mathrm{Se}}_{3}$}},\ }\href
  {https://doi.org/10.1103/PhysRevB.90.100509} {\bibfield  {journal} {\bibinfo
  {journal} {Phys. Rev. B}\ }\textbf {\bibinfo {volume} {90}},\ \bibinfo
  {pages} {100509} (\bibinfo {year} {2014})}\BibitemShut {NoStop}%
\bibitem [{\citenamefont {Kostylev}\ \emph {et~al.}(2020)\citenamefont
  {Kostylev}, \citenamefont {Yonezawa}, \citenamefont {Wang}, \citenamefont
  {Ando},\ and\ \citenamefont {Maeno}}]{Kostylev2020}%
  \BibitemOpen
  \bibfield  {author} {\bibinfo {author} {\bibfnamefont {I.}~\bibnamefont
  {Kostylev}}, \bibinfo {author} {\bibfnamefont {S.}~\bibnamefont {Yonezawa}},
  \bibinfo {author} {\bibfnamefont {Z.}~\bibnamefont {Wang}}, \bibinfo {author}
  {\bibfnamefont {Y.}~\bibnamefont {Ando}},\ and\ \bibinfo {author}
  {\bibfnamefont {Y.}~\bibnamefont {Maeno}},\ }\bibfield  {title} {\bibinfo
  {title} {Uniaxial-strain control of nematic superconductivity in
  {Sr$_x$Bi$_2$Se$_3$}},\ }\href {https://doi.org/10.1038/s41467-020-17913-y}
  {\bibfield  {journal} {\bibinfo  {journal} {Nat. Commun.}\ }\textbf {\bibinfo
  {volume} {11}},\ \bibinfo {pages} {4152} (\bibinfo {year}
  {2020})}\BibitemShut {NoStop}%
\bibitem [{\citenamefont {Tao}\ \emph {et~al.}(2018)\citenamefont {Tao},
  \citenamefont {Yan}, \citenamefont {Liu}, \citenamefont {Wang}, \citenamefont
  {Ando}, \citenamefont {Wang}, \citenamefont {Zhang},\ and\ \citenamefont
  {Feng}}]{Tao2018}%
  \BibitemOpen
  \bibfield  {author} {\bibinfo {author} {\bibfnamefont {R.}~\bibnamefont
  {Tao}}, \bibinfo {author} {\bibfnamefont {Y.-J.}\ \bibnamefont {Yan}},
  \bibinfo {author} {\bibfnamefont {X.}~\bibnamefont {Liu}}, \bibinfo {author}
  {\bibfnamefont {Z.-W.}\ \bibnamefont {Wang}}, \bibinfo {author}
  {\bibfnamefont {Y.}~\bibnamefont {Ando}}, \bibinfo {author} {\bibfnamefont
  {Q.-H.}\ \bibnamefont {Wang}}, \bibinfo {author} {\bibfnamefont
  {T.}~\bibnamefont {Zhang}},\ and\ \bibinfo {author} {\bibfnamefont {D.-L.}\
  \bibnamefont {Feng}},\ }\bibfield  {title} {\bibinfo {title} {Direct
  visualization of the nematic superconductivity in
  {${\mathrm{Cu}}_{x}{\mathrm{Bi}}_{2}{\mathrm{Se}}_{3}$}},\ }\href
  {https://doi.org/10.1103/PhysRevX.8.041024} {\bibfield  {journal} {\bibinfo
  {journal} {Phys. Rev. X}\ }\textbf {\bibinfo {volume} {8}},\ \bibinfo {pages}
  {041024} (\bibinfo {year} {2018})}\BibitemShut {NoStop}%
\bibitem [{\citenamefont {Kawai}\ \emph {et~al.}(2020)\citenamefont {Kawai},
  \citenamefont {Wang}, \citenamefont {Kandori}, \citenamefont {Honoki},
  \citenamefont {Matano}, \citenamefont {Kambe},\ and\ \citenamefont
  {Zheng}}]{Kawai2020}%
  \BibitemOpen
  \bibfield  {author} {\bibinfo {author} {\bibfnamefont {T.}~\bibnamefont
  {Kawai}}, \bibinfo {author} {\bibfnamefont {C.~G.}\ \bibnamefont {Wang}},
  \bibinfo {author} {\bibfnamefont {Y.}~\bibnamefont {Kandori}}, \bibinfo
  {author} {\bibfnamefont {Y.}~\bibnamefont {Honoki}}, \bibinfo {author}
  {\bibfnamefont {K.}~\bibnamefont {Matano}}, \bibinfo {author} {\bibfnamefont
  {T.}~\bibnamefont {Kambe}},\ and\ \bibinfo {author} {\bibfnamefont {G.-q.}\
  \bibnamefont {Zheng}},\ }\bibfield  {title} {\bibinfo {title} {Direction and
  symmetry transition of the vector order parameter in topological
  superconductors {Cu$_x$Bi$_2$Se$_3$}},\ }\href
  {https://doi.org/10.1038/s41467-019-14126-w} {\bibfield  {journal} {\bibinfo
  {journal} {Nat. Commun.}\ }\textbf {\bibinfo {volume} {11}},\ \bibinfo
  {pages} {235} (\bibinfo {year} {2020})}\BibitemShut {NoStop}%
\bibitem [{\citenamefont {Sasaki}\ \emph {et~al.}(2014)\citenamefont {Sasaki},
  \citenamefont {Segawa},\ and\ \citenamefont {Ando}}]{Sasaki2014}%
  \BibitemOpen
  \bibfield  {author} {\bibinfo {author} {\bibfnamefont {S.}~\bibnamefont
  {Sasaki}}, \bibinfo {author} {\bibfnamefont {K.}~\bibnamefont {Segawa}},\
  and\ \bibinfo {author} {\bibfnamefont {Y.}~\bibnamefont {Ando}},\ }\bibfield
  {title} {\bibinfo {title} {Superconductor derived from a topological
  insulator heterostructure},\ }\href
  {https://doi.org/10.1103/PhysRevB.90.220504} {\bibfield  {journal} {\bibinfo
  {journal} {Phys. Rev. B}\ }\textbf {\bibinfo {volume} {90}},\ \bibinfo
  {pages} {220504} (\bibinfo {year} {2014})}\BibitemShut {NoStop}%
\bibitem [{\citenamefont {Ando}\ and\ \citenamefont {Fu}(2015)}]{Ando2015}%
  \BibitemOpen
  \bibfield  {author} {\bibinfo {author} {\bibfnamefont {Y.}~\bibnamefont
  {Ando}}\ and\ \bibinfo {author} {\bibfnamefont {L.}~\bibnamefont {Fu}},\
  }\bibfield  {title} {\bibinfo {title} {Topological crystalline insulators and
  topological superconductors: {F}rom concepts to materials},\ }\href
  {https://doi.org/10.1146/annurev-conmatphys-031214-014501} {\bibfield
  {journal} {\bibinfo  {journal} {Annu. Rev. Condens. Matter Phys.}\ }\textbf
  {\bibinfo {volume} {6}},\ \bibinfo {pages} {361} (\bibinfo {year}
  {2015})}\BibitemShut {NoStop}%
\bibitem [{\citenamefont {Sera}\ \emph {et~al.}(2020)\citenamefont {Sera},
  \citenamefont {Ueda}, \citenamefont {Adachi},\ and\ \citenamefont
  {Ichioka}}]{Sera2020}%
  \BibitemOpen
  \bibfield  {author} {\bibinfo {author} {\bibfnamefont {Y.}~\bibnamefont
  {Sera}}, \bibinfo {author} {\bibfnamefont {T.}~\bibnamefont {Ueda}}, \bibinfo
  {author} {\bibfnamefont {H.}~\bibnamefont {Adachi}},\ and\ \bibinfo {author}
  {\bibfnamefont {M.}~\bibnamefont {Ichioka}},\ }\bibfield  {title} {\bibinfo
  {title} {{R}elation of superconducting pairing symmetry and non-magnetic
  impurity effects in vortex states},\ }\href
  {https://doi.org/10.3390/sym12010175} {\bibfield  {journal} {\bibinfo
  {journal} {Symmetry}\ }\textbf {\bibinfo {volume} {12}},\ \bibinfo {pages}
  {175} (\bibinfo {year} {2020})}\BibitemShut {NoStop}%
\bibitem [{SM()}]{SM}%
  \BibitemOpen
  \href@noop {} {}\bibinfo {note} {See Supplemental Material at (URL) for additional
  data and discussions, which additionally cites Refs. \cite{Eskildsen2002, Bergeal2006, Renner1991, Nakayama2012}.}\BibitemShut {Stop}%
\bibitem [{\citenamefont {Odobesko}\ \emph {et~al.}(2020)\citenamefont
  {Odobesko}, \citenamefont {Friedrich}, \citenamefont {Zhang}, \citenamefont
  {Haldar}, \citenamefont {Heinze}, \citenamefont {Trauzettel},\ and\
  \citenamefont {Bode}}]{Odobesko2020}%
  \BibitemOpen
  \bibfield  {author} {\bibinfo {author} {\bibfnamefont {A.}~\bibnamefont
  {Odobesko}}, \bibinfo {author} {\bibfnamefont {F.}~\bibnamefont {Friedrich}},
  \bibinfo {author} {\bibfnamefont {S.-B.}\ \bibnamefont {Zhang}}, \bibinfo
  {author} {\bibfnamefont {S.}~\bibnamefont {Haldar}}, \bibinfo {author}
  {\bibfnamefont {S.}~\bibnamefont {Heinze}}, \bibinfo {author} {\bibfnamefont
  {B.}~\bibnamefont {Trauzettel}},\ and\ \bibinfo {author} {\bibfnamefont
  {M.}~\bibnamefont {Bode}},\ }\bibfield  {title} {\bibinfo {title}
  {Anisotropic vortices on superconducting {N}b(110)},\ }\href
  {https://doi.org/10.1103/PhysRevB.102.174502} {\bibfield  {journal} {\bibinfo
   {journal} {Phys. Rev. B}\ }\textbf {\bibinfo {volume} {102}},\ \bibinfo
  {pages} {174502} (\bibinfo {year} {2020})}\BibitemShut {NoStop}%
\bibitem [{\citenamefont {Kim}\ \emph {et~al.}(2021)\citenamefont {Kim},
  \citenamefont {Nagai}, \citenamefont {Rózsa}, \citenamefont {Schreyer},\
  and\ \citenamefont {Wiesendanger}}]{Kim2021}%
  \BibitemOpen
  \bibfield  {author} {\bibinfo {author} {\bibfnamefont {H.}~\bibnamefont
  {Kim}}, \bibinfo {author} {\bibfnamefont {Y.}~\bibnamefont {Nagai}}, \bibinfo
  {author} {\bibfnamefont {L.}~\bibnamefont {Rózsa}}, \bibinfo {author}
  {\bibfnamefont {D.}~\bibnamefont {Schreyer}},\ and\ \bibinfo {author}
  {\bibfnamefont {R.}~\bibnamefont {Wiesendanger}},\ }\bibfield  {title}
  {\bibinfo {title} {Anisotropic non-split zero-energy vortex bound states in a
  conventional superconductor},\ }\href {https://doi.org/10.1063/5.0055839}
  {\bibfield  {journal} {\bibinfo  {journal} {Appl. Phys. Rev.}\ }\textbf
  {\bibinfo {volume} {8}},\ \bibinfo {pages} {031417} (\bibinfo {year}
  {2021})}\BibitemShut {NoStop}%
\bibitem [{\citenamefont {Nakayama}\ \emph {et~al.}(2015)\citenamefont
  {Nakayama}, \citenamefont {Kimizuka}, \citenamefont {Tanaka}, \citenamefont
  {Sato}, \citenamefont {Souma}, \citenamefont {Takahashi}, \citenamefont
  {Sasaki}, \citenamefont {Segawa},\ and\ \citenamefont {Ando}}]{Nakayama2015}%
  \BibitemOpen
  \bibfield  {author} {\bibinfo {author} {\bibfnamefont {K.}~\bibnamefont
  {Nakayama}}, \bibinfo {author} {\bibfnamefont {H.}~\bibnamefont {Kimizuka}},
  \bibinfo {author} {\bibfnamefont {Y.}~\bibnamefont {Tanaka}}, \bibinfo
  {author} {\bibfnamefont {T.}~\bibnamefont {Sato}}, \bibinfo {author}
  {\bibfnamefont {S.}~\bibnamefont {Souma}}, \bibinfo {author} {\bibfnamefont
  {T.}~\bibnamefont {Takahashi}}, \bibinfo {author} {\bibfnamefont
  {S.}~\bibnamefont {Sasaki}}, \bibinfo {author} {\bibfnamefont
  {K.}~\bibnamefont {Segawa}},\ and\ \bibinfo {author} {\bibfnamefont
  {Y.}~\bibnamefont {Ando}},\ }\bibfield  {title} {\bibinfo {title}
  {Observation of two-dimensional bulk electronic states in the superconducting
  topological insulator heterostructure
  {${\mathrm{Cu}}_{x}{(\mathrm{PbSe})}_{5}{({\mathrm{Bi}}_{2}{\mathrm{Se}}_{3})}_{6}$}:
  {I}mplications for unconventional superconductivity},\ }\href
  {https://doi.org/10.1103/PhysRevB.92.100508} {\bibfield  {journal} {\bibinfo
  {journal} {Phys. Rev. B}\ }\textbf {\bibinfo {volume} {92}},\ \bibinfo
  {pages} {100508} (\bibinfo {year} {2015})}\BibitemShut {NoStop}%
\bibitem [{\citenamefont {Galvis}\ \emph {et~al.}(2018)\citenamefont {Galvis},
  \citenamefont {Herrera}, \citenamefont {Berthod}, \citenamefont {Vieira},
  \citenamefont {Guillamón},\ and\ \citenamefont {Suderow}}]{Galvis2018}%
  \BibitemOpen
  \bibfield  {author} {\bibinfo {author} {\bibfnamefont {J.~A.}\ \bibnamefont
  {Galvis}}, \bibinfo {author} {\bibfnamefont {E.}~\bibnamefont {Herrera}},
  \bibinfo {author} {\bibfnamefont {C.}~\bibnamefont {Berthod}}, \bibinfo
  {author} {\bibfnamefont {S.}~\bibnamefont {Vieira}}, \bibinfo {author}
  {\bibfnamefont {I.}~\bibnamefont {Guillamón}},\ and\ \bibinfo {author}
  {\bibfnamefont {H.}~\bibnamefont {Suderow}},\ }\bibfield  {title} {\bibinfo
  {title} {Tilted vortex cores and superconducting gap anisotropy in
  {2H-NbSe$_2$}},\ }\href {https://doi.org/10.1038/s42005-018-0028-1}
  {\bibfield  {journal} {\bibinfo  {journal} {Commun. Phys.}\ }\textbf
  {\bibinfo {volume} {1}},\ \bibinfo {pages} {30} (\bibinfo {year}
  {2018})}\BibitemShut {NoStop}%
\bibitem [{\citenamefont {Dynes}\ \emph {et~al.}(1978)\citenamefont {Dynes},
  \citenamefont {Narayanamurti},\ and\ \citenamefont {Garno}}]{Dynes1978}%
  \BibitemOpen
  \bibfield  {author} {\bibinfo {author} {\bibfnamefont {R.~C.}\ \bibnamefont
  {Dynes}}, \bibinfo {author} {\bibfnamefont {V.}~\bibnamefont
  {Narayanamurti}},\ and\ \bibinfo {author} {\bibfnamefont {J.~P.}\
  \bibnamefont {Garno}},\ }\bibfield  {title} {\bibinfo {title} {Direct
  measurement of quasiparticle-lifetime broadening in a strong-coupled
  superconductor},\ }\href {https://doi.org/10.1103/PhysRevLett.41.1509}
  {\bibfield  {journal} {\bibinfo  {journal} {Phys. Rev. Lett.}\ }\textbf
  {\bibinfo {volume} {41}},\ \bibinfo {pages} {1509} (\bibinfo {year}
  {1978})}\BibitemShut {NoStop}%
\bibitem [{\citenamefont {Bagchi}\ \emph {et~al.}(2022)\citenamefont {Bagchi},
  \citenamefont {Brede},\ and\ \citenamefont {Ando}}]{Bagchi2022}%
  \BibitemOpen
  \bibfield  {author} {\bibinfo {author} {\bibfnamefont {M.}~\bibnamefont
  {Bagchi}}, \bibinfo {author} {\bibfnamefont {J.}~\bibnamefont {Brede}},\ and\
  \bibinfo {author} {\bibfnamefont {Y.}~\bibnamefont {Ando}},\ }\bibfield
  {title} {\bibinfo {title} {Observability of superconductivity in {S}r-doped
  {${\mathrm{Bi}}_{2}{\mathrm{Se}}_{3}$} at the surface using scanning
  tunneling microscope},\ }\href
  {https://doi.org/10.1103/PhysRevMaterials.6.034201} {\bibfield  {journal}
  {\bibinfo  {journal} {Phys. Rev. Mater.}\ }\textbf {\bibinfo {volume} {6}},\
  \bibinfo {pages} {034201} (\bibinfo {year} {2022})}\BibitemShut {NoStop}%
\bibitem [{\citenamefont {Nagai}(2014)}]{Nagai2014}%
  \BibitemOpen
  \bibfield  {author} {\bibinfo {author} {\bibfnamefont {Y.}~\bibnamefont
  {Nagai}},\ }\bibfield  {title} {\bibinfo {title} {{F}ield-angle-dependent
  low-energy excitations around a vortex in the superconducting topological
  insulator {Cu$_\mathrm{x}$Bi$_2$Se$_3$}},\ }\href
  {https://doi.org/10.7566/JPSJ.83.063705} {\bibfield  {journal} {\bibinfo
  {journal} {J. Phys. Soc. Jpn.}\ }\textbf {\bibinfo {volume} {83}},\ \bibinfo
  {pages} {063705} (\bibinfo {year} {2014})}\BibitemShut {NoStop}%
\bibitem [{\citenamefont {Nakayama}\ \emph {et~al.}(2019)\citenamefont
  {Nakayama}, \citenamefont {Souma}, \citenamefont {Trang}, \citenamefont
  {Takane}, \citenamefont {Chen}, \citenamefont {Avila}, \citenamefont
  {Takahashi}, \citenamefont {Sasaki}, \citenamefont {Segawa}, \citenamefont
  {Asensio}, \citenamefont {Ando},\ and\ \citenamefont {Sato}}]{Nakayama2019}%
  \BibitemOpen
  \bibfield  {author} {\bibinfo {author} {\bibfnamefont {K.}~\bibnamefont
  {Nakayama}}, \bibinfo {author} {\bibfnamefont {S.}~\bibnamefont {Souma}},
  \bibinfo {author} {\bibfnamefont {C.~X.}\ \bibnamefont {Trang}}, \bibinfo
  {author} {\bibfnamefont {D.}~\bibnamefont {Takane}}, \bibinfo {author}
  {\bibfnamefont {C.}~\bibnamefont {Chen}}, \bibinfo {author} {\bibfnamefont
  {J.}~\bibnamefont {Avila}}, \bibinfo {author} {\bibfnamefont
  {T.}~\bibnamefont {Takahashi}}, \bibinfo {author} {\bibfnamefont
  {S.}~\bibnamefont {Sasaki}}, \bibinfo {author} {\bibfnamefont
  {K.}~\bibnamefont {Segawa}}, \bibinfo {author} {\bibfnamefont {M.~C.}\
  \bibnamefont {Asensio}}, \bibinfo {author} {\bibfnamefont {Y.}~\bibnamefont
  {Ando}},\ and\ \bibinfo {author} {\bibfnamefont {T.}~\bibnamefont {Sato}},\
  }\bibfield  {title} {\bibinfo {title} {Nanomosaic of topological {D}irac
  states on the surface of {Pb$_5$Bi$_{24}$Se$_{41}$} observed by
  nano-{ARPES}},\ }\href {https://doi.org/10.1021/acs.nanolett.9b00875}
  {\bibfield  {journal} {\bibinfo  {journal} {Nano Lett.}\ }\textbf {\bibinfo
  {volume} {19}},\ \bibinfo {pages} {3737} (\bibinfo {year}
  {2019})}\BibitemShut {NoStop}%
\bibitem [{\citenamefont {Eilenberger}(1968)}]{Eilenberger1968}%
  \BibitemOpen
  \bibfield  {author} {\bibinfo {author} {\bibfnamefont {G.}~\bibnamefont
  {Eilenberger}},\ }\bibfield  {title} {\bibinfo {title} {Transformation of
  {G}orkov's equation for type ii superconductors into transport-like
  equations},\ }\href {https://doi.org/10.1007/BF01379803} {\bibfield
  {journal} {\bibinfo  {journal} {Zeitschrift für Physik A Hadrons and
  nuclei}\ }\textbf {\bibinfo {volume} {214}},\ \bibinfo {pages} {195}
  (\bibinfo {year} {1968})}\BibitemShut {NoStop}%
\bibitem [{\citenamefont {Kita}(2015)}]{Kita2015}%
  \BibitemOpen
  \bibfield  {author} {\bibinfo {author} {\bibfnamefont {T.}~\bibnamefont
  {Kita}},\ }\bibinfo {title} {{G}or'kov, {E}ilenberger, and
  {G}inzburg--{L}andau equations},\ in\ \href
  {https://doi.org/10.1007/978-4-431-55405-9_14} {\emph {\bibinfo {booktitle}
  {Statistical Mechanics of Superconductivity}}}\ (\bibinfo  {publisher}
  {Springer Japan},\ \bibinfo {address} {Tokyo},\ \bibinfo {year} {2015})\ pp.\
  \bibinfo {pages} {201--227}\BibitemShut {NoStop}%
\bibitem [{\citenamefont {Kopnin}(2001)}]{Kopnin2001}%
  \BibitemOpen
  \bibfield  {author} {\bibinfo {author} {\bibfnamefont {N.}~\bibnamefont
  {Kopnin}},\ }\href
  {https://doi.org/10.1093/acprof:oso/9780198507888.001.0001} {\emph {\bibinfo
  {title} {{Theory of Nonequilibrium Superconductivity}}}}\ (\bibinfo
  {publisher} {Oxford University Press},\ \bibinfo {year} {2001})\BibitemShut
  {NoStop}%
\bibitem [{\citenamefont {Nagai}\ \emph {et~al.}(2006)\citenamefont {Nagai},
  \citenamefont {Ueno}, \citenamefont {Kato},\ and\ \citenamefont
  {Hayashi}}]{Nagai2006}%
  \BibitemOpen
  \bibfield  {author} {\bibinfo {author} {\bibfnamefont {Y.}~\bibnamefont
  {Nagai}}, \bibinfo {author} {\bibfnamefont {Y.}~\bibnamefont {Ueno}},
  \bibinfo {author} {\bibfnamefont {Y.}~\bibnamefont {Kato}},\ and\ \bibinfo
  {author} {\bibfnamefont {N.}~\bibnamefont {Hayashi}},\ }\bibfield  {title}
  {\bibinfo {title} {Analytical formulation of the local density of states
  around a vortex core in unconventional superconductors},\ }\href@noop {}
  {\bibfield  {journal} {\bibinfo  {journal} {J. Phys. Soc. Jpn.}\ }\textbf
  {\bibinfo {volume} {75}},\ \bibinfo {pages} {104701} (\bibinfo {year}
  {2006})}\BibitemShut {NoStop}%
\bibitem [{\citenamefont {Ichioka}\ \emph {et~al.}(1996)\citenamefont
  {Ichioka}, \citenamefont {Hayashi}, \citenamefont {Enomoto},\ and\
  \citenamefont {Machida}}]{Ichioka1996}%
  \BibitemOpen
  \bibfield  {author} {\bibinfo {author} {\bibfnamefont {M.}~\bibnamefont
  {Ichioka}}, \bibinfo {author} {\bibfnamefont {N.}~\bibnamefont {Hayashi}},
  \bibinfo {author} {\bibfnamefont {N.}~\bibnamefont {Enomoto}},\ and\ \bibinfo
  {author} {\bibfnamefont {K.}~\bibnamefont {Machida}},\ }\bibfield  {title}
  {\bibinfo {title} {Vortex structure in d-wave superconductors},\ }\href
  {https://doi.org/10.1103/PhysRevB.53.15316} {\bibfield  {journal} {\bibinfo
  {journal} {Phys. Rev. B}\ }\textbf {\bibinfo {volume} {53}},\ \bibinfo
  {pages} {15316} (\bibinfo {year} {1996})}\BibitemShut {NoStop}%
\bibitem [{\citenamefont {Kaneko}\ \emph {et~al.}(2012)\citenamefont {Kaneko},
  \citenamefont {Matsuba}, \citenamefont {Hafiz}, \citenamefont {Yamasaki},
  \citenamefont {Kakizaki}, \citenamefont {Nishida}, \citenamefont {Takeya},
  \citenamefont {Hirata}, \citenamefont {Kawakami}, \citenamefont {Mizushima},\
  and\ \citenamefont {Machida}}]{Kaneko2012}%
  \BibitemOpen
  \bibfield  {author} {\bibinfo {author} {\bibfnamefont {S.-i.}\ \bibnamefont
  {Kaneko}}, \bibinfo {author} {\bibfnamefont {K.}~\bibnamefont {Matsuba}},
  \bibinfo {author} {\bibfnamefont {M.}~\bibnamefont {Hafiz}}, \bibinfo
  {author} {\bibfnamefont {K.}~\bibnamefont {Yamasaki}}, \bibinfo {author}
  {\bibfnamefont {E.}~\bibnamefont {Kakizaki}}, \bibinfo {author}
  {\bibfnamefont {N.}~\bibnamefont {Nishida}}, \bibinfo {author} {\bibfnamefont
  {H.}~\bibnamefont {Takeya}}, \bibinfo {author} {\bibfnamefont
  {K.}~\bibnamefont {Hirata}}, \bibinfo {author} {\bibfnamefont
  {T.}~\bibnamefont {Kawakami}}, \bibinfo {author} {\bibfnamefont
  {T.}~\bibnamefont {Mizushima}},\ and\ \bibinfo {author} {\bibfnamefont
  {K.}~\bibnamefont {Machida}},\ }\bibfield  {title} {\bibinfo {title} {Quantum
  limiting behaviors of a vortex core in an anisotropic gap superconductor},\
  }\href {https://doi.org/10.1143/JPSJ.81.063701} {\bibfield  {journal}
  {\bibinfo  {journal} {J. Phys. Soc. Jpn.}\ }\textbf {\bibinfo {volume}
  {81}},\ \bibinfo {pages} {063701} (\bibinfo {year} {2012})}\BibitemShut
  {NoStop}%
\bibitem [{\citenamefont {Zhang}\ \emph {et~al.}(2009)\citenamefont {Zhang},
  \citenamefont {Liu}, \citenamefont {Qi}, \citenamefont {Dai}, \citenamefont
  {Fang},\ and\ \citenamefont {Zhang}}]{Zhang2009}%
  \BibitemOpen
  \bibfield  {author} {\bibinfo {author} {\bibfnamefont {H.}~\bibnamefont
  {Zhang}}, \bibinfo {author} {\bibfnamefont {C.-X.}\ \bibnamefont {Liu}},
  \bibinfo {author} {\bibfnamefont {X.-L.}\ \bibnamefont {Qi}}, \bibinfo
  {author} {\bibfnamefont {X.}~\bibnamefont {Dai}}, \bibinfo {author}
  {\bibfnamefont {Z.}~\bibnamefont {Fang}},\ and\ \bibinfo {author}
  {\bibfnamefont {S.-C.}\ \bibnamefont {Zhang}},\ }\bibfield  {title} {\bibinfo
  {title} {{Topological insulators in {Bi$_2$Se$_3$}, {Bi$_2$Te$_3$} and
  {Sb$_2$Te$_3$} with a single Dirac cone on the surface}},\ }\href
  {https://doi.org/10.1038/nphys1270} {\bibfield  {journal} {\bibinfo
  {journal} {Nature Physics}\ }\textbf {\bibinfo {volume} {5}},\ \bibinfo
  {pages} {438} (\bibinfo {year} {2009})}\BibitemShut {NoStop}%
\bibitem [{\citenamefont {Liu}\ \emph {et~al.}(2010)\citenamefont {Liu},
  \citenamefont {Qi}, \citenamefont {Zhang}, \citenamefont {Dai}, \citenamefont
  {Fang},\ and\ \citenamefont {Zhang}}]{Liu2010}%
  \BibitemOpen
  \bibfield  {author} {\bibinfo {author} {\bibfnamefont {C.-X.}\ \bibnamefont
  {Liu}}, \bibinfo {author} {\bibfnamefont {X.-L.}\ \bibnamefont {Qi}},
  \bibinfo {author} {\bibfnamefont {H.}~\bibnamefont {Zhang}}, \bibinfo
  {author} {\bibfnamefont {X.}~\bibnamefont {Dai}}, \bibinfo {author}
  {\bibfnamefont {Z.}~\bibnamefont {Fang}},\ and\ \bibinfo {author}
  {\bibfnamefont {S.-C.}\ \bibnamefont {Zhang}},\ }\bibfield  {title} {\bibinfo
  {title} {{Model Hamiltonian for topological insulators}},\ }\href
  {https://doi.org/10.1103/PhysRevB.82.045122} {\bibfield  {journal} {\bibinfo
  {journal} {Phys. Rev. B}\ }\textbf {\bibinfo {volume} {82}},\ \bibinfo
  {pages} {45122} (\bibinfo {year} {2010})}\BibitemShut {NoStop}%
\bibitem [{\citenamefont {Kuntsevich}\ \emph {et~al.}(2018)\citenamefont
  {Kuntsevich}, \citenamefont {Bryzgalov}, \citenamefont {Prudkoglyad},
  \citenamefont {Martovitskii}, \citenamefont {Selivanov},\ and\ \citenamefont
  {Chizhevskii}}]{Kuntsevich2018}%
  \BibitemOpen
  \bibfield  {author} {\bibinfo {author} {\bibfnamefont {A.~Y.}\ \bibnamefont
  {Kuntsevich}}, \bibinfo {author} {\bibfnamefont {M.~A.}\ \bibnamefont
  {Bryzgalov}}, \bibinfo {author} {\bibfnamefont {V.~A.}\ \bibnamefont
  {Prudkoglyad}}, \bibinfo {author} {\bibfnamefont {V.~P.}\ \bibnamefont
  {Martovitskii}}, \bibinfo {author} {\bibfnamefont {Y.~G.}\ \bibnamefont
  {Selivanov}},\ and\ \bibinfo {author} {\bibfnamefont {E.~G.}\ \bibnamefont
  {Chizhevskii}},\ }\bibfield  {title} {\bibinfo {title} {Structural distortion
  behind the nematic superconductivity in {Sr$_x$Bi$_2$Se$_3$}},\ }\href
  {https://doi.org/10.1088/1367-2630/aae595} {\bibfield  {journal} {\bibinfo
  {journal} {New J. Phys.}\ }\textbf {\bibinfo {volume} {20}},\ \bibinfo
  {pages} {103022} (\bibinfo {year} {2018})}\BibitemShut {NoStop}%
\bibitem [{\citenamefont {Smylie}\ \emph {et~al.}(2022)\citenamefont {Smylie},
  \citenamefont {Islam}, \citenamefont {Gu}, \citenamefont {Schneeloch},
  \citenamefont {Zhong}, \citenamefont {Rosenkranz}, \citenamefont {Kwok},\
  and\ \citenamefont {Welp}}]{Smylie2022}%
  \BibitemOpen
  \bibfield  {author} {\bibinfo {author} {\bibfnamefont {M.~P.}\ \bibnamefont
  {Smylie}}, \bibinfo {author} {\bibfnamefont {Z.}~\bibnamefont {Islam}},
  \bibinfo {author} {\bibfnamefont {G.~D.}\ \bibnamefont {Gu}}, \bibinfo
  {author} {\bibfnamefont {J.}~\bibnamefont {Schneeloch}}, \bibinfo {author}
  {\bibfnamefont {R.~D.}\ \bibnamefont {Zhong}}, \bibinfo {author}
  {\bibfnamefont {S.}~\bibnamefont {Rosenkranz}}, \bibinfo {author}
  {\bibfnamefont {W.~K.}\ \bibnamefont {Kwok}},\ and\ \bibinfo {author}
  {\bibfnamefont {U.}~\bibnamefont {Welp}},\ }\bibfield  {title} {\bibinfo
  {title} {Multimodal synchrotron {X}-ray diffraction across the
  superconducting transition of {Sr$_{0.1}$Bi$_2$Se$_3$}},\ }\href
  {https://doi.org/10.48550/arXiv.2207.13221} {\bibfield  {journal} {\bibinfo
  {journal} {arXiv:}\ ,\ \bibinfo {pages} {2207.13221}} (\bibinfo {year}
  {2022})}\BibitemShut {NoStop}%
\bibitem [{\citenamefont {Levy}\ \emph {et~al.}(2013)\citenamefont {Levy},
  \citenamefont {Zhang}, \citenamefont {Ha}, \citenamefont {Sharifi},
  \citenamefont {Talin}, \citenamefont {Kuk},\ and\ \citenamefont
  {Stroscio}}]{Levy2013}%
  \BibitemOpen
  \bibfield  {author} {\bibinfo {author} {\bibfnamefont {N.}~\bibnamefont
  {Levy}}, \bibinfo {author} {\bibfnamefont {T.}~\bibnamefont {Zhang}},
  \bibinfo {author} {\bibfnamefont {J.}~\bibnamefont {Ha}}, \bibinfo {author}
  {\bibfnamefont {F.}~\bibnamefont {Sharifi}}, \bibinfo {author} {\bibfnamefont
  {A.~A.}\ \bibnamefont {Talin}}, \bibinfo {author} {\bibfnamefont
  {Y.}~\bibnamefont {Kuk}},\ and\ \bibinfo {author} {\bibfnamefont {J.~A.}\
  \bibnamefont {Stroscio}},\ }\bibfield  {title} {\bibinfo {title}
  {Experimental evidence for {$s$}-wave pairing symmetry in superconducting
  {${\mathrm{Cu}}_{x}{\mathrm{Bi}}_{2}{\mathrm{Se}}_{3}$} single crystals using
  a scanning tunneling microscope},\ }\href
  {https://doi.org/10.1103/PhysRevLett.110.117001} {\bibfield  {journal}
  {\bibinfo  {journal} {Phys. Rev. Lett.}\ }\textbf {\bibinfo {volume} {110}},\
  \bibinfo {pages} {117001} (\bibinfo {year} {2013})}\BibitemShut {NoStop}%
\bibitem [{\citenamefont {Sasaki}\ \emph {et~al.}(2011)\citenamefont {Sasaki},
  \citenamefont {Kriener}, \citenamefont {Segawa}, \citenamefont {Yada},
  \citenamefont {Tanaka}, \citenamefont {Sato},\ and\ \citenamefont
  {Ando}}]{Sasaki2011}%
  \BibitemOpen
  \bibfield  {author} {\bibinfo {author} {\bibfnamefont {S.}~\bibnamefont
  {Sasaki}}, \bibinfo {author} {\bibfnamefont {M.}~\bibnamefont {Kriener}},
  \bibinfo {author} {\bibfnamefont {K.}~\bibnamefont {Segawa}}, \bibinfo
  {author} {\bibfnamefont {K.}~\bibnamefont {Yada}}, \bibinfo {author}
  {\bibfnamefont {Y.}~\bibnamefont {Tanaka}}, \bibinfo {author} {\bibfnamefont
  {M.}~\bibnamefont {Sato}},\ and\ \bibinfo {author} {\bibfnamefont
  {Y.}~\bibnamefont {Ando}},\ }\bibfield  {title} {\bibinfo {title}
  {Topological superconductivity in {Cu$_x$Bi$_2$Se$_3$}},\ }\href
  {https://doi.org/10.1103/PhysRevLett.107.217001} {\bibfield  {journal}
  {\bibinfo  {journal} {Phys. Rev. Lett.}\ }\textbf {\bibinfo {volume} {107}},\
  \bibinfo {pages} {217001} (\bibinfo {year} {2011})}\BibitemShut {NoStop}%
\bibitem [{\citenamefont {Kirzhner}\ \emph {et~al.}(2012)\citenamefont
  {Kirzhner}, \citenamefont {Lahoud}, \citenamefont {Chaska}, \citenamefont
  {Salman},\ and\ \citenamefont {Kanigel}}]{Kirzhner2012}%
  \BibitemOpen
  \bibfield  {author} {\bibinfo {author} {\bibfnamefont {T.}~\bibnamefont
  {Kirzhner}}, \bibinfo {author} {\bibfnamefont {E.}~\bibnamefont {Lahoud}},
  \bibinfo {author} {\bibfnamefont {K.~B.}\ \bibnamefont {Chaska}}, \bibinfo
  {author} {\bibfnamefont {Z.}~\bibnamefont {Salman}},\ and\ \bibinfo {author}
  {\bibfnamefont {A.}~\bibnamefont {Kanigel}},\ }\bibfield  {title} {\bibinfo
  {title} {Point-contact spectroscopy of {Cu$_{0.2}$Bi$_2$Se$_3$} single
  crystals},\ }\href {https://doi.org/10.1103/PhysRevB.86.064517} {\bibfield
  {journal} {\bibinfo  {journal} {Phys. Rev. B}\ }\textbf {\bibinfo {volume}
  {86}},\ \bibinfo {pages} {064517} (\bibinfo {year} {2012})}\BibitemShut
  {NoStop}%
\bibitem [{\citenamefont {Peng}\ \emph {et~al.}(2013)\citenamefont {Peng},
  \citenamefont {De}, \citenamefont {Lv}, \citenamefont {Wei},\ and\
  \citenamefont {Chu}}]{Peng2013}%
  \BibitemOpen
  \bibfield  {author} {\bibinfo {author} {\bibfnamefont {H.}~\bibnamefont
  {Peng}}, \bibinfo {author} {\bibfnamefont {D.}~\bibnamefont {De}}, \bibinfo
  {author} {\bibfnamefont {B.}~\bibnamefont {Lv}}, \bibinfo {author}
  {\bibfnamefont {F.}~\bibnamefont {Wei}},\ and\ \bibinfo {author}
  {\bibfnamefont {C.-W.}\ \bibnamefont {Chu}},\ }\bibfield  {title} {\bibinfo
  {title} {Absence of zero-energy surface bound states in {Cu$_x$Bi$_2$Se$_3$}
  studied via {A}ndreev reflection spectroscopy},\ }\href
  {https://doi.org/10.1103/PhysRevB.88.024515} {\bibfield  {journal} {\bibinfo
  {journal} {Phys. Rev. B}\ }\textbf {\bibinfo {volume} {88}},\ \bibinfo
  {pages} {024515} (\bibinfo {year} {2013})}\BibitemShut {NoStop}%
\bibitem [{\citenamefont {Kurter}\ \emph {et~al.}(2019)\citenamefont {Kurter},
  \citenamefont {Finck}, \citenamefont {Huemiller}, \citenamefont {Medvedeva},
  \citenamefont {Weis}, \citenamefont {Atkinson}, \citenamefont {Qiu},
  \citenamefont {Shen}, \citenamefont {Lee}, \citenamefont {Vojta},
  \citenamefont {Ghaemi}, \citenamefont {Hor},\ and\ \citenamefont
  {Van~Harlingen}}]{Kurter2019}%
  \BibitemOpen
  \bibfield  {author} {\bibinfo {author} {\bibfnamefont {C.}~\bibnamefont
  {Kurter}}, \bibinfo {author} {\bibfnamefont {A.~D.~K.}\ \bibnamefont
  {Finck}}, \bibinfo {author} {\bibfnamefont {E.~D.}\ \bibnamefont
  {Huemiller}}, \bibinfo {author} {\bibfnamefont {J.}~\bibnamefont
  {Medvedeva}}, \bibinfo {author} {\bibfnamefont {A.}~\bibnamefont {Weis}},
  \bibinfo {author} {\bibfnamefont {J.~M.}\ \bibnamefont {Atkinson}}, \bibinfo
  {author} {\bibfnamefont {Y.}~\bibnamefont {Qiu}}, \bibinfo {author}
  {\bibfnamefont {L.}~\bibnamefont {Shen}}, \bibinfo {author} {\bibfnamefont
  {S.~H.}\ \bibnamefont {Lee}}, \bibinfo {author} {\bibfnamefont
  {T.}~\bibnamefont {Vojta}}, \bibinfo {author} {\bibfnamefont
  {P.}~\bibnamefont {Ghaemi}}, \bibinfo {author} {\bibfnamefont {Y.~S.}\
  \bibnamefont {Hor}},\ and\ \bibinfo {author} {\bibfnamefont {D.~J.}\
  \bibnamefont {Van~Harlingen}},\ }\bibfield  {title} {\bibinfo {title}
  {Conductance spectroscopy of exfoliated thin flakes of
  {Nb$_x$Bi$_2$Se$_3$}},\ }\href {https://doi.org/10.1021/acs.nanolett.8b02954}
  {\bibfield  {journal} {\bibinfo  {journal} {Nano Lett.}\ }\textbf {\bibinfo
  {volume} {19}},\ \bibinfo {pages} {38} (\bibinfo {year} {2019})}\BibitemShut
  {NoStop}%
\bibitem [{\citenamefont {Blonder}\ \emph {et~al.}(1982)\citenamefont
  {Blonder}, \citenamefont {Tinkham},\ and\ \citenamefont
  {Klapwijk}}]{Blonder1982}%
  \BibitemOpen
  \bibfield  {author} {\bibinfo {author} {\bibfnamefont {G.~E.}\ \bibnamefont
  {Blonder}}, \bibinfo {author} {\bibfnamefont {M.}~\bibnamefont {Tinkham}},\
  and\ \bibinfo {author} {\bibfnamefont {T.~M.}\ \bibnamefont {Klapwijk}},\
  }\bibfield  {title} {\bibinfo {title} {Transition from metallic to tunneling
  regimes in superconducting microconstrictions: Excess current, charge
  imbalance, and supercurrent conversion},\ }\href@noop {} {\bibfield
  {journal} {\bibinfo  {journal} {Phys. Rev. B}\ }\textbf {\bibinfo {volume}
  {25}},\ \bibinfo {pages} {4515} (\bibinfo {year} {1982})}\BibitemShut
  {NoStop}%
\bibitem [{\citenamefont {Kriener}\ \emph {et~al.}(2011)\citenamefont
  {Kriener}, \citenamefont {Segawa}, \citenamefont {Ren}, \citenamefont
  {Sasaki}, \citenamefont {Wada}, \citenamefont {Kuwabata},\ and\ \citenamefont
  {Ando}}]{Kriener2011}%
  \BibitemOpen
  \bibfield  {author} {\bibinfo {author} {\bibfnamefont {M.}~\bibnamefont
  {Kriener}}, \bibinfo {author} {\bibfnamefont {K.}~\bibnamefont {Segawa}},
  \bibinfo {author} {\bibfnamefont {Z.}~\bibnamefont {Ren}}, \bibinfo {author}
  {\bibfnamefont {S.}~\bibnamefont {Sasaki}}, \bibinfo {author} {\bibfnamefont
  {S.}~\bibnamefont {Wada}}, \bibinfo {author} {\bibfnamefont {S.}~\bibnamefont
  {Kuwabata}},\ and\ \bibinfo {author} {\bibfnamefont {Y.}~\bibnamefont
  {Ando}},\ }\bibfield  {title} {\bibinfo {title} {Electrochemical synthesis
  and superconducting phase diagram of {Cu$_x$Bi$_2$Se$_3$}},\ }\href
  {https://doi.org/10.1103/PhysRevB.84.054513} {\bibfield  {journal} {\bibinfo
  {journal} {Phys. Rev. B}\ }\textbf {\bibinfo {volume} {84}},\ \bibinfo
  {pages} {054513} (\bibinfo {year} {2011})}\BibitemShut {NoStop}%
\bibitem [{\citenamefont {Eskildsen}\ \emph {et~al.}(2002)\citenamefont
  {Eskildsen}, \citenamefont {Kugler}, \citenamefont {Tanaka}, \citenamefont
  {Jun}, \citenamefont {Kazakov}, \citenamefont {Karpinski},\ and\
  \citenamefont {Fischer}}]{Eskildsen2002}%
  \BibitemOpen
  \bibfield  {author} {\bibinfo {author} {\bibfnamefont {M.~R.}\ \bibnamefont
  {Eskildsen}}, \bibinfo {author} {\bibfnamefont {M.}~\bibnamefont {Kugler}},
  \bibinfo {author} {\bibfnamefont {S.}~\bibnamefont {Tanaka}}, \bibinfo
  {author} {\bibfnamefont {J.}~\bibnamefont {Jun}}, \bibinfo {author}
  {\bibfnamefont {S.~M.}\ \bibnamefont {Kazakov}}, \bibinfo {author}
  {\bibfnamefont {J.}~\bibnamefont {Karpinski}},\ and\ \bibinfo {author}
  {\bibfnamefont {{\O}.}~\bibnamefont {Fischer}},\ }\bibfield  {title}
  {\bibinfo {title} {Vortex imaging in the {$\ensuremath{\pi}$} band of
  {M}agnesium {D}iboride},\ }\href
  {https://doi.org/10.1103/PhysRevLett.89.187003} {\bibfield  {journal}
  {\bibinfo  {journal} {Phys. Rev. Lett.}\ }\textbf {\bibinfo {volume} {89}},\
  \bibinfo {pages} {187003} (\bibinfo {year} {2002})}\BibitemShut {NoStop}%
\bibitem [{\citenamefont {Bergeal}\ \emph {et~al.}(2006)\citenamefont
  {Bergeal}, \citenamefont {Dubost}, \citenamefont {Noat}, \citenamefont
  {Sacks}, \citenamefont {Roditchev}, \citenamefont {Emery}, \citenamefont
  {Hérold}, \citenamefont {Marêché}, \citenamefont {Lagrange},\ and\
  \citenamefont {Loupias}}]{Bergeal2006}%
  \BibitemOpen
  \bibfield  {author} {\bibinfo {author} {\bibfnamefont {N.}~\bibnamefont
  {Bergeal}}, \bibinfo {author} {\bibfnamefont {V.}~\bibnamefont {Dubost}},
  \bibinfo {author} {\bibfnamefont {Y.}~\bibnamefont {Noat}}, \bibinfo {author}
  {\bibfnamefont {W.}~\bibnamefont {Sacks}}, \bibinfo {author} {\bibfnamefont
  {D.}~\bibnamefont {Roditchev}}, \bibinfo {author} {\bibfnamefont
  {N.}~\bibnamefont {Emery}}, \bibinfo {author} {\bibfnamefont
  {C.}~\bibnamefont {Hérold}}, \bibinfo {author} {\bibfnamefont {J.-F.}\
  \bibnamefont {Marêché}}, \bibinfo {author} {\bibfnamefont {P.}~\bibnamefont
  {Lagrange}},\ and\ \bibinfo {author} {\bibfnamefont {G.}~\bibnamefont
  {Loupias}},\ }\bibfield  {title} {\bibinfo {title} {Scanning tunneling
  spectroscopy on the novel superconductor {${\mathrm{CaC}}_{6}$}},\ }\href
  {https://doi.org/10.1103/PhysRevLett.97.077003} {\bibfield  {journal}
  {\bibinfo  {journal} {Phys. Rev. Lett.}\ }\textbf {\bibinfo {volume} {97}},\
  \bibinfo {pages} {077003} (\bibinfo {year} {2006})}\BibitemShut {NoStop}%
\bibitem [{\citenamefont {Renner}\ \emph {et~al.}(1991)\citenamefont {Renner},
  \citenamefont {Kent}, \citenamefont {Niedermann}, \citenamefont {Fischer},\
  and\ \citenamefont {L\'evy}}]{Renner1991}%
  \BibitemOpen
  \bibfield  {author} {\bibinfo {author} {\bibfnamefont {C.}~\bibnamefont
  {Renner}}, \bibinfo {author} {\bibfnamefont {A.~D.}\ \bibnamefont {Kent}},
  \bibinfo {author} {\bibfnamefont {P.}~\bibnamefont {Niedermann}}, \bibinfo
  {author} {\bibfnamefont {O.}~\bibnamefont {Fischer}},\ and\ \bibinfo {author}
  {\bibfnamefont {F.}~\bibnamefont {L\'evy}},\ }\bibfield  {title} {\bibinfo
  {title} {Scanning tunneling spectroscopy of a vortex core from the clean to
  the dirty limit},\ }\href {https://doi.org/10.1103/PhysRevLett.67.1650}
  {\bibfield  {journal} {\bibinfo  {journal} {Phys. Rev. Lett.}\ }\textbf
  {\bibinfo {volume} {67}},\ \bibinfo {pages} {1650} (\bibinfo {year}
  {1991})}\BibitemShut {NoStop}%
\bibitem [{\citenamefont {Nakayama}\ \emph {et~al.}(2012)\citenamefont
  {Nakayama}, \citenamefont {Eto}, \citenamefont {Tanaka}, \citenamefont
  {Sato}, \citenamefont {Souma}, \citenamefont {Takahashi}, \citenamefont
  {Segawa},\ and\ \citenamefont {Ando}}]{Nakayama2012}%
  \BibitemOpen
  \bibfield  {author} {\bibinfo {author} {\bibfnamefont {K.}~\bibnamefont
  {Nakayama}}, \bibinfo {author} {\bibfnamefont {K.}~\bibnamefont {Eto}},
  \bibinfo {author} {\bibfnamefont {Y.}~\bibnamefont {Tanaka}}, \bibinfo
  {author} {\bibfnamefont {T.}~\bibnamefont {Sato}}, \bibinfo {author}
  {\bibfnamefont {S.}~\bibnamefont {Souma}}, \bibinfo {author} {\bibfnamefont
  {T.}~\bibnamefont {Takahashi}}, \bibinfo {author} {\bibfnamefont
  {K.}~\bibnamefont {Segawa}},\ and\ \bibinfo {author} {\bibfnamefont
  {Y.}~\bibnamefont {Ando}},\ }\bibfield  {title} {\bibinfo {title}
  {Manipulation of topological states and the bulk band gap using natural
  heterostructures of a topological insulator},\ }\href
  {https://doi.org/10.1103/PhysRevLett.109.236804} {\bibfield  {journal}
  {\bibinfo  {journal} {Phys. Rev. Lett.}\ }\textbf {\bibinfo {volume} {109}},\
  \bibinfo {pages} {236804} (\bibinfo {year} {2012})}\BibitemShut {NoStop}%
\end{thebibliography}
\end{document}

\title{Supplemental Material for \\
%``Rotation of gap nodes in the topological superconductor \textbf{$\mathrm{\mathbf{Cu}}_{\mathbf{x}}\mathrm{\mathbf{(PbSe)}}_{\mathbf{5}}(\mathrm{\mathbf{Bi}}_{\mathbf{2}}\mathrm{\mathbf{Se}}_{\mathbf{3}})_{\mathbf{6}}$}'' by Mahasweta Bagchi, Jens Brede, Aline Ramires, and Yoichi Ando}
\vspace{1.5mm}
``Rotation of gap nodes in the topological superconductor \textbf{$\mathrm{\mathbf{Cu}}_{\mathbf{x}}\mathrm{\mathbf{(PbSe)}}_{\mathbf{5}}(\mathrm{\mathbf{Bi}}_{\mathbf{2}}\mathrm{\mathbf{Se}}_{\mathbf{3}})_{\mathbf{6}}$}'' by Mahasweta Bagchi,$^1$ Jens Brede,$^1$ Aline Ramires,$^2$ and Yoichi Ando$^1$\\
\vspace{1mm}
{\it $^1$Physics Institute II, University of Cologne, D-50937 K\"oln, Germany}\\
{\it $^2$Paul Scherrer Institute, CH-5232 Villigen PSI, Switzerland}}

\maketitle

%\vspace{-2cm}

\section{Sample characterization}

\begin{figure*}[ht]
	\centering
	\includegraphics[scale=1]{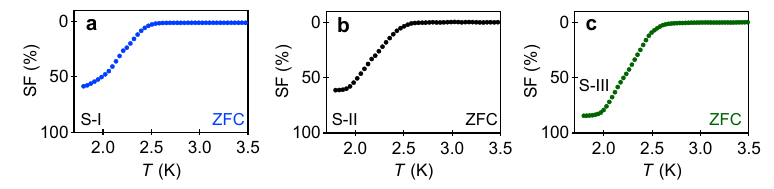}
	\caption{\JB{(a,b,c) Zero-field-cooled (ZFC) magnetization data showing shielding fractions of 59\% for Sample S-I (a), 61\% for Sample S-II (b) and 84\% for Sample S-III (c).}
	}
	\label{fig:squid_data}
\end{figure*}

Figure \ref{fig:squid_data} shows the shielding fractions (SF) of \JB{three} CPSBS samples that were synthesized from the same batch of PSBS crystals. The STM data obtained on sample S-I \JB{and on sample S-II is shown in the main text and the data on sample S-III is presented in this supplement.} \JB{Samples S-I to S-III show a bulk $T_{\rm{c}}$ between 2.5 and 2.6~K.}  

\vspace{-2mm}
\JB{\section{Supplemental data for sample S-I}}

\begin{figure*}[ht]
	\centering
	\includegraphics[width=0.9\textwidth]{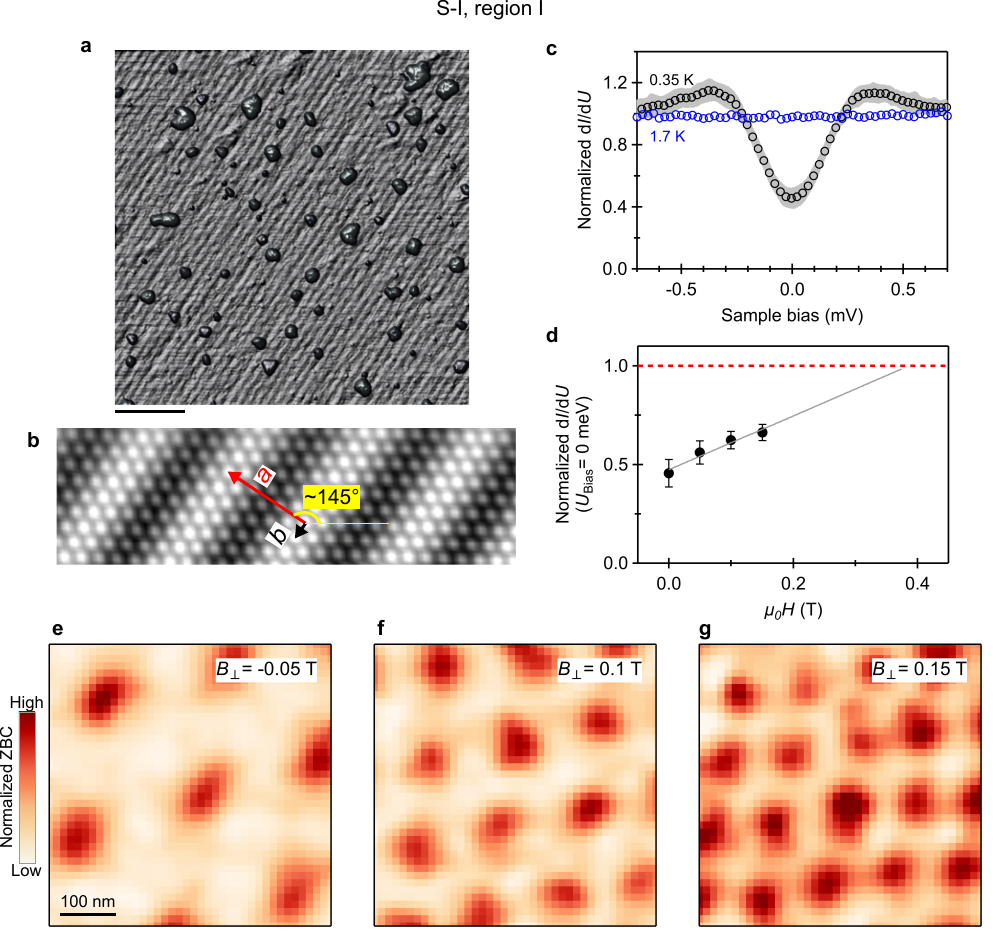}
	\caption{\JB{(a) Pseudo 3D representation of an STM topograph on the SC area of sample S-I. Scale bar corresponds to 20~nm. The structural 1D stripe is clearly observed in the region together with islands/clusters of Cu atoms. (b) Fourier-filtered atomic resolution image taken in the area as in (a) with the monoclinic \textit{a} and \textit{b} axes indicated. The \textit{a} axis is at an angle of about 145$^\circ$ with respect to the horizontal axis for \textit{as-acquired data}. (c) Normalized \dIdU{} spectra acquired in the region in (a) for a fridge temperatures of 0.35~K (gray open circles) and 1.7~K (blue open circles). The data points correspond to a spatial average of 100 STS over 500 by 500~nm$^{2}$. The light gray area is the standard deviation of the data, as described in the text. (d, e, f, g) Magnetic-field dependence of the normalized ZBC values (d) as obtained from the the zero field grid and grids taken in the presence of a magnetic field [presented in (e - f)]. Red dashed line corresponds to the normal-state conductance. Error bars in (d) correspond to the one sigma standard deviation of the ZBC value of the respective grid. Scan/stabilization parameters: (a) $U=900$~V, $I=100$~pA; (b) $U=900$~V, $I=200$~pA; (e, f, g) $U=5$~mV, $I=500$~pA (e, f) $I=100$~pA (g).}
	}
	\label{fig:sample3}
\end{figure*}

\JB{The superconducting regions of sample S-I clearly show the structural 1D stripe [Fig.~\ref{fig:sample3}(a)], allowing the determination of the monoclinic \textit{a} and \textit{b} axis. The monoclinic \textit{a} axis in this sample is about 145$^\circ$ with respect to the horizontal axis [Fig.~\ref{fig:sample3}(b)]. The spatially averaged superconducting gap measured in this region is shown in Fig.~\ref{fig:sample3}(c) (same data as Fig.~2(a) of the main text).  Moreover, we have experimentally ascertained that the $T_{\mathrm{c}}$ in this region is lower than 1.7~K.}

\begin{figure*}[ht]
	\centering
	\includegraphics[width=0.95\textwidth]{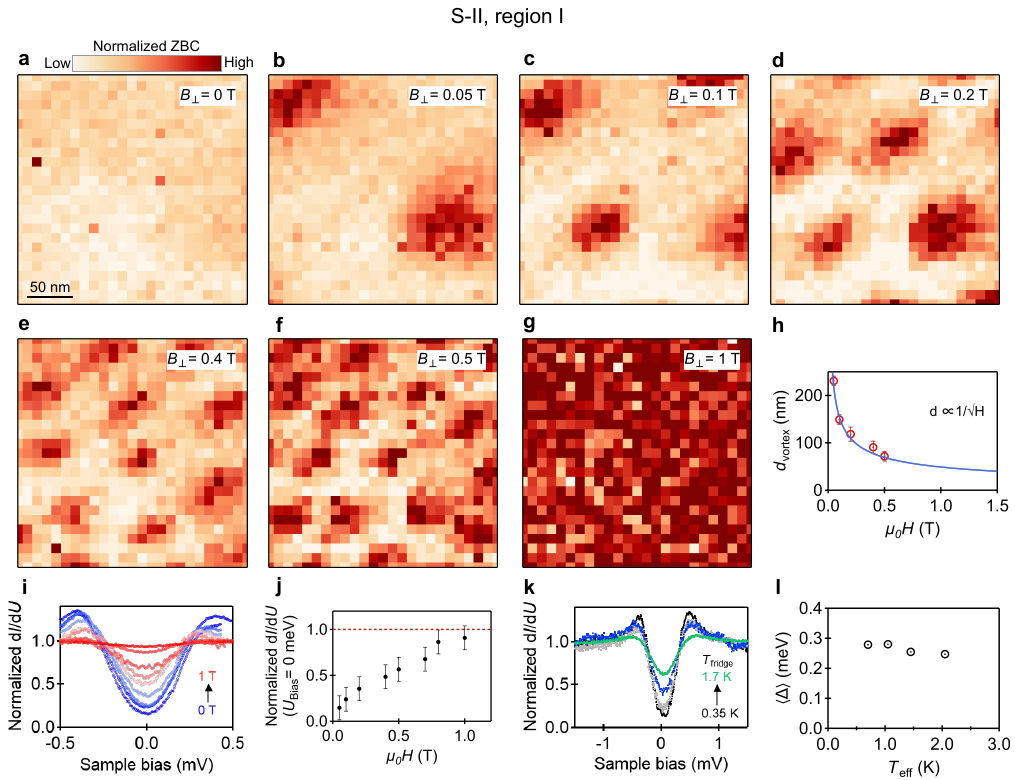}
	\caption{(a)-(g) Normalized ZBC maps showing the evolution of the vortex lattice in a field of view of 250~nm by 250~nm under various out-of-plane magnetic fields as indicated. The spectrum at each grid point was obtained in the \textit{I-V} mode. \stabp{} $U=3$~mV, $I=50$~pA (100 sweeps). A majority of the vortices are oriented along one direction in all cases. (h) The distance between vortices plotted as a function of the applied magnetic field. The blue curve shows the simulated profile assuming a perfect triangular lattice; deviation of the data points from the simulation curve indicates imperfect ordering. Error bars are given by the larger of the resolution of our STS grid ($\sim$10~nm) or the standard deviation of the nearest-neighbour distance of all lattice points in the field of view. (i) Normalized \dIdU{} spectra for various magnetic fields; the curves are obtained by averaging over the grids that lie outside the vortices; the plotted spectra correspond to 0, 0.05, 0.1, 0.2, 0.4, 0.5, 0.7, 0.8, and 1.0 T. (j) Magnetic-field dependence of the normalized ZBC values obtained from the data in (i); the red dashed line signifies the normal-state conductance. Errors bars correspond to a fixed value which is the standard deviation of the ZBC obtained from the zero field grid. (k) SC gap spectra measured in zero field at the fridge temperatures of 0.35, 0.7, 1.1 and 1.7~K; the black curve is also shown in Fig.~2 of the main text. (l) \JB{The average gap $\langle \Delta \rangle$ calculated from the $\Delta_0$ and $\Delta_1$ values obtained from the fits to the data in (k) plotted against the effective temperature. }
	}
	\label{fig:vortex_lattice}
\end{figure*}

\JB{When the external magnetic field is applied normal to the surface, a nearly hexagonal vortex lattice is clearly resolved [Fig.~\ref{fig:sample3}(e-g)]. We extrapolate the upper critical field at the surface to about 0.4~T from the magnetic-field dependence of the zero-bias conductance (ZBC) as shown in \fig{fig:vortex_lattice}(d).}

\JB{We note considerable variations among the shape of different vortices in \fig{fig:sample3}(e-g), but as discussed in the main text, on average, vortices are elongated roughly at $55^\circ$ from the horizontal (i.e. at $90^\circ$ with respect to the $a$-axis).} 

%\vspace{0.5cm}

\JB{\section{Supplemental data for sample S-II}}

We characterized the vortex area \JB{of sample S-II} by taking low-resolution STS-grids in various magnetic fields, as shown in \fig{fig:vortex_lattice}(a-g). The vortex lattice clearly evolves with the magnetic field until the superconductivity is fully suppressed at around $\sim$1~T. At low fields, a nearly triangular lattice is observed, in which the distance $d_{\mathrm{vortex}}$ between nearest-neighbors is given as $d_{\mathrm{vortex}} \approx 1.075\left({\Phi_{0}}/{B}\right)^{1/2}$. We plot the average distance between the vortices for varying magnetic fields together with the profile expected for a triangular lattice in \fig{fig:vortex_lattice}(h). We extract the superconducting (SC) gap at each field by averaging the spectra in-between the vortices for a small bias range around $E_{\mathrm{F}}$. These spectra change systematically as a function of the magnetic field [\fig{fig:vortex_lattice}(i)]. The magnetic-field dependence of the zero-bias conductance (ZBC) normalized to that at 0 T is shown in \fig{fig:vortex_lattice}(j). 

\JB{The temperature dependence of the SC gap spectra was measured up to the fridge temperature $T_{\rm fridge}$ of 1.7 K [\fig{fig:vortex_lattice}(k)]. Since the effective base temperature of 0.7~K calibrated for $T_{\rm{fridge}}$ = 0.35~K suggests that the noise in our STM system heats the electrons by $T_{\rm noise} \simeq$ 0.35 K, we fix $T_{\rm eff} = T_{\rm{fridge}} + 0.35$~K when we fit the spectra with the anisotropic gap function discussed in the main text. We can calculate the average gap $\langle \Delta \rangle =\Delta_0+(2/\pi)\Delta_1$ from the $\Delta_0$ and the $\Delta_1$ values obtained from the fits of the spectra, and the result is plotted in \fig{fig:vortex_lattice}(l) as a function of temperature.}

\pagebreak
\JB{\section{Fitting vortex profiles with the Ginzburg-Landau derived expression}}
\begin{figure}[ht]
	\centering
	\includegraphics[scale=0.9]{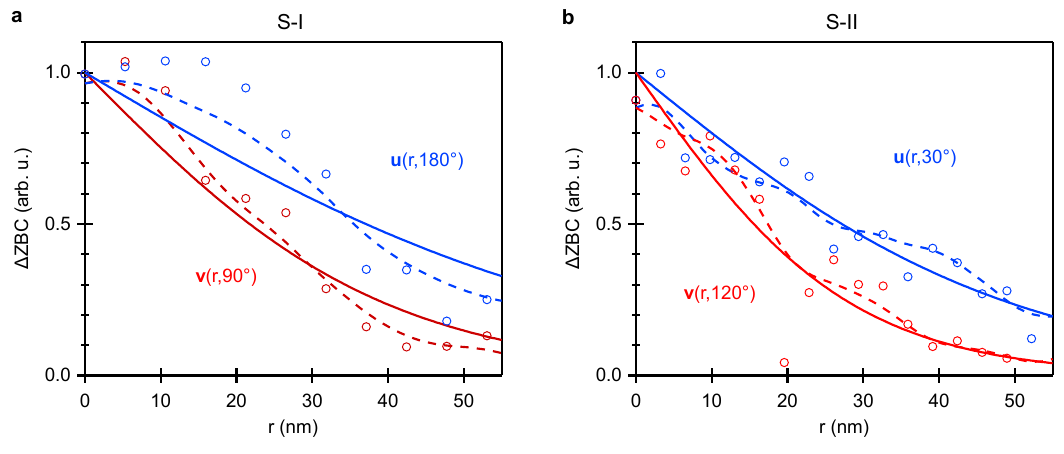}
	\caption{\JB{(a) Line profiles of the vortices taken from Fig.~1(e) of the main text. Open circles, dashed lines, and solid lines refer to raw data, smoothed curve, and GL fits to Eq.~(\ref{eq:vortex_prof}), respectively. \YA{The smoothing was performed for the analysis of the half width at half maximum mentioned in the text by using a standard Gaussian filter with the smallest (3$\times$3 pixels) kernel with up to three iterations.} Blue (red) data are taken along the major (minor) vortex axis; \YA{note that the images in Fig. 1(a,b) of the main text are rotated
clockwise by 55$^\circ$ to align the monoclinic $a$-axis with the vertical axis}. (b) The same as (a) but for line profiles shown in Fig.~1(g) of the main text. The GL fit gives $\xi_\mathrm{GL}=28\pm2$~nm ($47\pm6$~nm) for the short (long) axis for the vortex of sample S-I and $20\pm2$~nm ($35\pm2$~nm) for the vortex of sample S-II.}}
	\label{fig:ComparisonHWHMvsGL}
\end{figure}

\JB{Zero bias conductance (ZBC) profiles across vortex cores in STM experiments are commonly analyzed~\cite{Eskildsen2002,Bergeal2006} with the Ginzburg-Landau (GL) derived expression for the superconducting order parameter,
\begin{align} 
	\label{eq:vortex_prof}
\sigma(r,0) = \sigma_{0} + (1-\sigma_{0})\left(1-\mathrm{tanh}\left[r/(\sqrt{2}\xi_{\mathrm{GL}}) \right] \right),
\end{align}
where $\sigma_{0}$ is the normalized ZBC away from a vortex centre and $r$ is the distance to the vortex core centre.}
\JB{In \fig{fig:ComparisonHWHMvsGL} we plot line cuts of the vortex cores discussed in Fig.~1 of the main text together with the fit of the raw data to Eq.~(\ref{eq:vortex_prof}). 
\YA{The $r$-dependences of the ZBC in sample S-I does not really follow Eq.~(\ref{eq:vortex_prof}), but from this analysis}
we obtain $\xi_{\mathrm{GL}}$ of about 28~(47)~nm for the short (long) axis of the vortex in sample S-I [\fig{fig:ComparisonHWHMvsGL}~(a)] and about 20~(35)~nm in S-II [\fig{fig:ComparisonHWHMvsGL}~(b)].}
\JB{The values are in reasonable agreement with the estimate of $\xi_\mathrm{GL}$ for sample S-I (S-II) based on the measured upper critical field $H_{c2}\approx0.4$~T (1~T),  
$$\xi_{\mathrm{GL}}=\sqrt{\frac{\Phi_0}{2\pi \mu_0 H_{c2}}}\approx 40~\text{nm}~(18~\text{nm}),$$
with $\Phi_0$ the (superconducting) magnetic flux quantum and $\mu_0$ the vacuum permeability.}

\JB{\section{Extended data for Fig.~1 of the main text}}

\JB{Here, we give extended data for Fig.~1 of the main text consisting of the unsmoothed ZBC maps, the lattice vectors of the vortex lattice, and the comparison of the average vortex profile based on the analysis of smoothed data as discussed in the main text with the results obtained by fitting the raw data to Eq.~(\ref{eq:vortex_prof}).}

\begin{figure}[ht]
	\centering
	\includegraphics[scale=0.85]{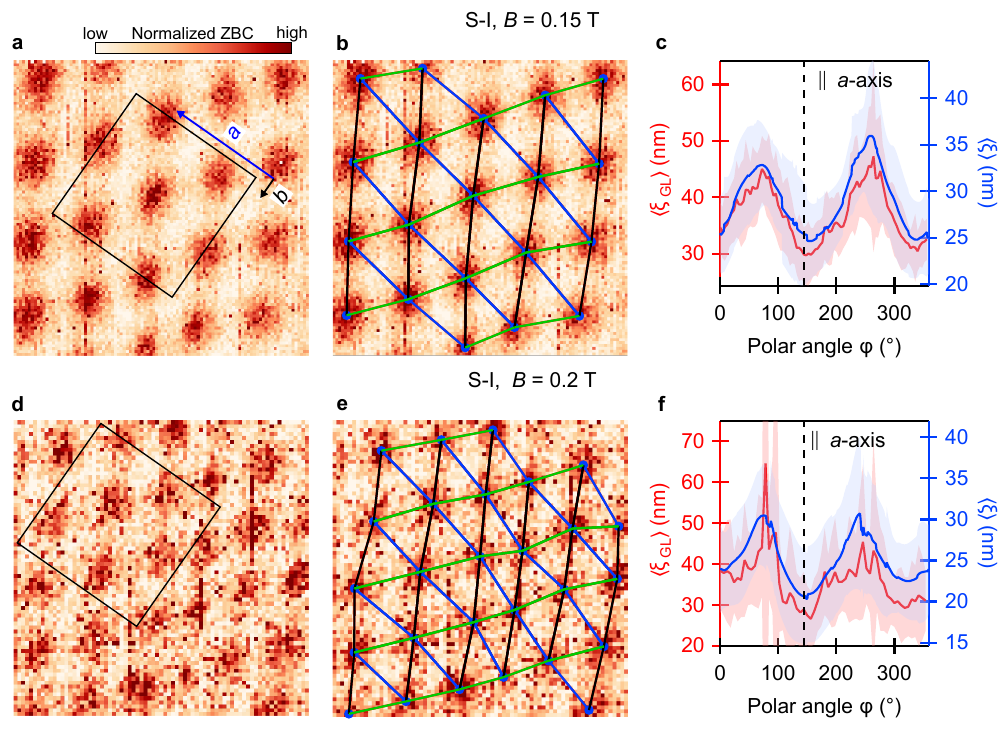}
	\caption{\JB{(a) Larger region of the vortex lattice in sample S-I in the out-of-plane magnetic field of \textit{B} = 0.15~T. The black square indicates the data show in Fig.~1(a) of the main text. The monoclinic $a$-axis and $b$-axis are given. The vortex lattice in the field of view of 500~nm $\times$ 500~nm is determined and the results are superimposed onto the raw data in (b). Blue dots mark the lattice points corresponding to vortex centres. The lattice vectors \textbf{a}1 (black), \textbf{a}2 (blue), \textbf{a}3 (green) are indicated and their average lengths are $|\textbf{a}1|=(133\pm8)$~nm, $|\textbf{a}2|=(138\pm9)$~nm and $|\textbf{a}3|=(111\pm7)$~nm. The average angle between \textbf{a}2 and the monoclinic \textit{a}-axis is $\angle_{\textbf{a}2,\textit{a}}=(12\pm4)^\circ$. Line profiles of the normalized ZBC are taken as a function of the polar angle $\varphi$ at each lattice point and the GL in-plane coherence length is determined by fitting the raw data to Eq.~(\ref{eq:vortex_prof}). (c) The average $\langle\xi_{\mathrm{GL}}\rangle$ of all lattice points is plotted in red, where the light shaded region indicates the standard deviation. For comparison, the average vortex core radius $\langle\xi \rangle$, which is determined from the half width at half maximum of the smoothed data as discussed in the main text, is shown in blue. (d,e,f) The same analysis as in (a,b,c)  for the vortex lattice in 0.2~T shown in Fig.~1(b) of the main text. The vortex lattice parameters for (e) are: $|\textbf{a}1|=(113\pm7)$~nm, $|\textbf{a}2|=(116\pm9)$~nm, $|\textbf{a}3|=(94\pm13)$~nm and $\angle_{\textbf{a}2,\textit{a}}=(15\pm6)^\circ$.}}
	\label{fig:vortex_lattice_SI}
\end{figure}

\JB{We determine the vortex lattice for each ZBC map by calculating the centre of every area of the map that exceeds an appropriate threshold value of the ZBC (typically about 0.5). From all lattice points we then determine the three average lattice vectors ($\bf{a}_1$, $\bf{a}_2$ and $\bf{a}_3$). The results of this procedure are superimposed onto the raw data in \fig{fig:vortex_lattice_SI}~(b) and (e). Interestingly, we find that the hexagonal vortex lattice observed in sample S-I is compressed along the vortex lattice vector $\bf{a}_3$ by about $15$ to $20\%$. Moreover, the vortex lattice ($\bf{a}_2$) is rotated by about $10$ to $15^\circ$ with respect to the crystallographic $a$-axis. We conservatively estimate the experimental uncertainty due to thermal drift and piezoelectric creep as $10\%$ for the compression and $10^\circ$ for the rotation.
We leave the exploration of the origin of this interesting vortex lattice deformation for future studies.}

\JB{Instead, we analyze the vortex shape at each lattice site by taking line profiles as a function of polar angle and determine $\xi$ (defined as the half width at half maximum) from smoothed data and $\xi_{\mathrm{GL}}$ by fitting the raw data to Eq.~(\ref{eq:vortex_prof}) as shown in \fig{fig:ComparisonHWHMvsGL}. We numerically average $\xi_{\mathrm{GL}}(\varphi)$ (and $\xi(\varphi)$) of all lattice sites and plot the results in \fig{fig:vortex_lattice_SI}~(c) and (f) for the vortex lattice at 0.15 and 0.2~T, respectively. On the one hand, we observe considerable variations in the vortex shape between different lattice sites, which likely reflects differences in the local scattering potential; on the other hand, a clear two-fold symmetry with a minimal average vortex extension parallel to the $a$-axis is apparent. Comparing $\langle \xi_{\mathrm{GL}} \rangle$ and $\langle \xi \rangle$ we find the same qualitative behavior.}

\begin{figure}[ht]
	\centering
	\includegraphics[scale=0.85]{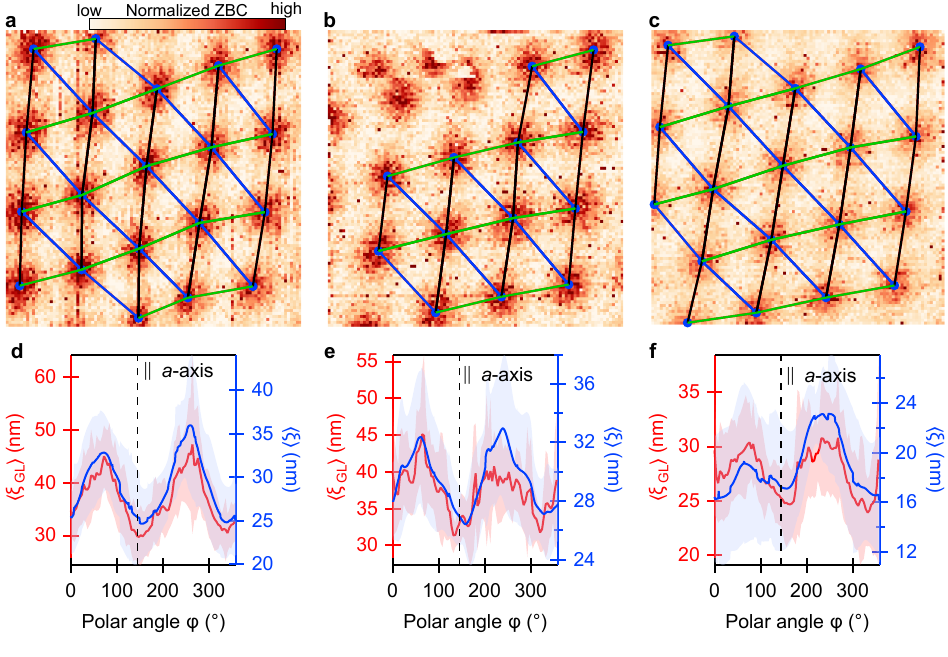}
	\vspace{-2mm}
		\caption{\YA{All the data of the vortices in sample S-I in the out-of-plane magnetic field of \textit{B} = 0.15~T used for the averaging to generate Fig.~1(f) of the main text. Panels (a,d) are the same as those shown in \fig{fig:vortex_lattice_SI}(b,c). The field of view of panels (b) and (c) is 500~nm $\times$ 500~nm, and the data are analysed and presented in the same way as in \fig{fig:vortex_lattice_SI}(b,c).} \JB{Vortex lattice parameters: $|\textbf{a}1|=(133\pm5)$~nm, $|\textbf{a}2|=(145\pm4)$~nm, $|\textbf{a}3|=(114\pm5)$~nm and $\angle_{\textbf{a}2,\textit{a}}=(13\pm3)^\circ$ for (b); $|\textbf{a}1|=(132\pm10)$~nm, $|\textbf{a}2|=(140\pm7)$~nm, $|\textbf{a}3|=(119\pm7)$~nm and $\angle_{\textbf{a}2,\textit{a}}=(12\pm3)^\circ$ for (c).}}
	\label{fig:vortex_lattice_SII150mT}
\end{figure}

\begin{figure}[ht]
	\centering
	\includegraphics[scale=0.85]{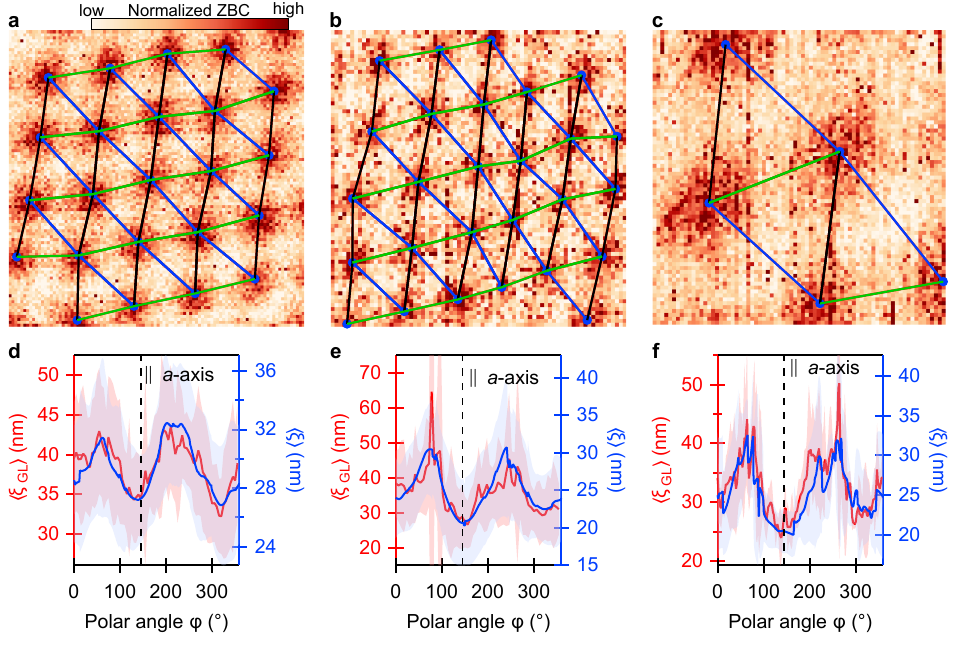}
	\vspace{-2mm}
			\caption{\JB{All the data of the vortices in sample S-I in the out-of-plane magnetic field of \textit{B} = 0.2~T used for the averaging to generate Fig.~1(f) of the main text. Panels (b,e) are the same as those shown in \fig{fig:vortex_lattice_SI}(e,f). The field of view of panels (a) and (c) is 500~nm $\times$ 500~nm and 200~nm $\times$ 200~nm, respectively. The data are analysed and presented in the same way as in \fig{fig:vortex_lattice_SI}(e,f). Vortex lattice parameters: $|\textbf{a}1|=(108\pm3)$~nm, $|\textbf{a}2|=(124\pm7)$~nm, \REV{$|\textbf{a}3|=(105\pm4)$~nm} and $\angle_{\textbf{a}2,\textit{a}}=(10\pm4)^\circ$ for (a); $|\textbf{a}1|=(107\pm3)$~nm, $|\textbf{a}2|=(109\pm6)$~nm, $|\textbf{a}3|=(94\pm6)$~nm and $\angle_{\textbf{a}2,\textit{a}}=(11\pm5)^\circ$ for (c).}}
	\label{fig:vortex_lattice_SII200mT}
\end{figure}

As mentioned above, we observed significant variations in the vortex shape anisotropy between different lattice sites in sample S-I. Therefore, to achieve a high level of confidence in the determination of the vortex elongation axis, we have analyzed a relatively large \JHB{number} of
\JB{ZBC maps at 0.15 and 0.2~T which are shown in \fig{fig:vortex_lattice_SII150mT} and \fig{fig:vortex_lattice_SII200mT}, respectively. The data shown in Fig.~1~(f) of the main manuscipt is the average of all the lattice sites indicated in \fig{fig:vortex_lattice_SII150mT} and \fig{fig:vortex_lattice_SII200mT}.}

\begin{figure}[ht]
	\centering
	\includegraphics[scale=0.85]{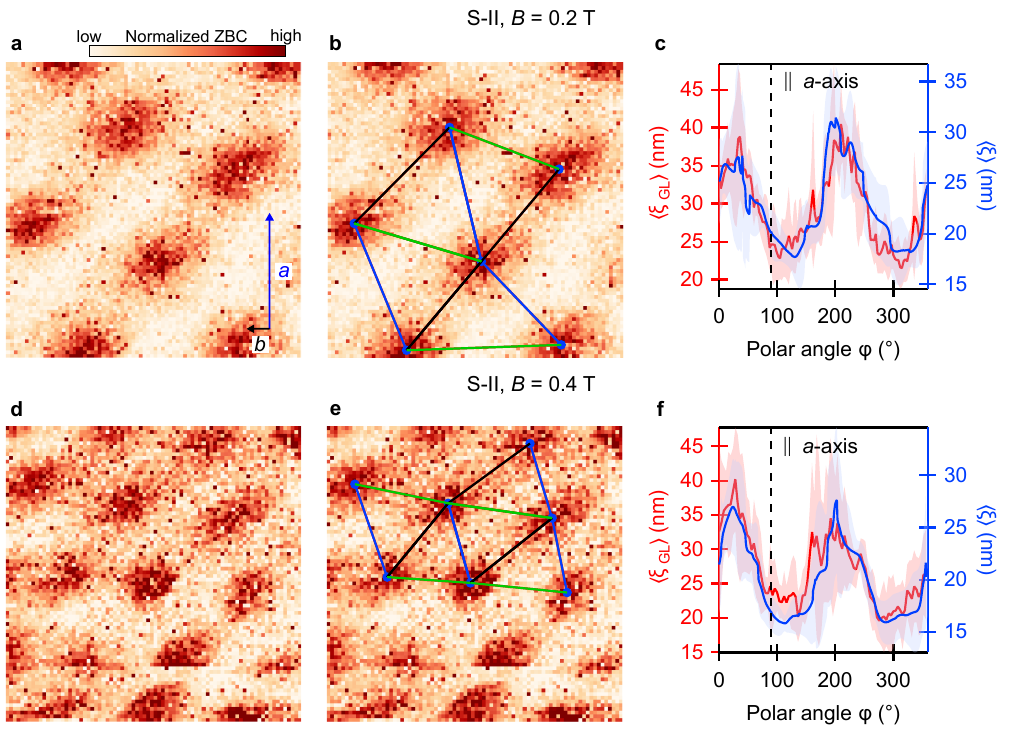}
		\caption{\JB{(a,b) The vortex lattice in sample S-II in the out-of-plane magnetic field of \textit{B} = 0.2~T show in Fig.~1(c) of the main text. The vortex lattice in the field of view of 250~nm $\times$ 250~nm is determined and the results are superimposed onto the raw data in (b). Blue dots mark the lattice points corresponding to vortex centres. The lattice vectors \textbf{a}1 (black), \textbf{a}2 (blue), \textbf{a}3 (green) are indicated and their average lengths are $|\textbf{a}1|=(105\pm8)$~nm, $|\textbf{a}2|=(113\pm10)$~nm and $|\textbf{a}3|=(115\pm13)$~nm. The average angle between \textbf{a}2 and the monoclinic \textit{a}-axis is $\angle_{\textbf{a}2,\textit{a}}=(-23\pm15)^\circ$. Line profiles of the normalized ZBC are taken as a function of the polar angle $\varphi$ at each lattice point and the GL in-plane coherence length is determined by fitting the raw data to Eq.~(\ref{eq:vortex_prof}). (c) The average $\langle\xi_{\mathrm{GL}}\rangle$ is plotted in red, where the light shaded region indicates the standard deviation. For comparison, the average vortex core radius $\langle\xi \rangle$, which is determined from the smoothed data as discussed in the main text, is shown in blue. (d-f) The same analysis as in (a-c) for the vortex lattice in 0.4~T shown in Fig.~1(d) of the main text. The vortices shown here are used for the averaging to generate Fig.~1(h) of the main text. The vortex lattice parameters for (e) are $|\textbf{a}1|=(86\pm4)$~nm, $|\textbf{a}2|=(72\pm8)$~nm, $|\textbf{a}3|=(83\pm8)$~nm and $\angle_{\textbf{a}2,\textit{a}}=(-15\pm3)^\circ$}.}
	\label{fig:vortex_lattice_SII}
\end{figure}

\JB{The same analysis as discussed for sample S-I above was also performed for the vortex lattice observed in sample S-II which is shown in Fig.~1~(c) and (e) of the main text. The results are shown in \fig{fig:vortex_lattice_SII}. On the one hand, we observe a rotation of lattice vector $\bf{a}_2$ of $-15$ to $-23^\circ$ with respect to the crystallographic $a$-axis, which is clearly different from the orientation observed for sample S-I; on the other hand, the entire vortex lattice in S-II is noticeably distorted (likely due to disorder related pinning of vortices) and does not allow for a meaningful analysis of the vortex lattice anisotropy.}

\JB{The vortex shape analysis performed at each lattice site yields the average $\langle\xi_{\mathrm{GL}}\rangle$ and $\langle\xi \rangle$ respectively plotted for 0.2 and 0.4~T in \fig{fig:vortex_lattice_SII}~(c) and (f). The noticeably smaller standard deviations compared to the data obtained for sample S-I is likely attributed to reduced vortex lattice motion due to the pinning centers and less thermal smearing due to the larger superconducting gap. Regardless of the origin, most vortices in sample S-II have similar elliptical shape with the same orientation of the major axis which is about $60^\circ$ rotated with respect to the crystallographic $a$-axis and clearly different from the orientation observed in sample S-I.}

\clearpage

\section{Influence of the in-plane magnetic field on the vortex elongation}

\begin{figure*}[ht]
	\centering
	\includegraphics[scale=0.73]{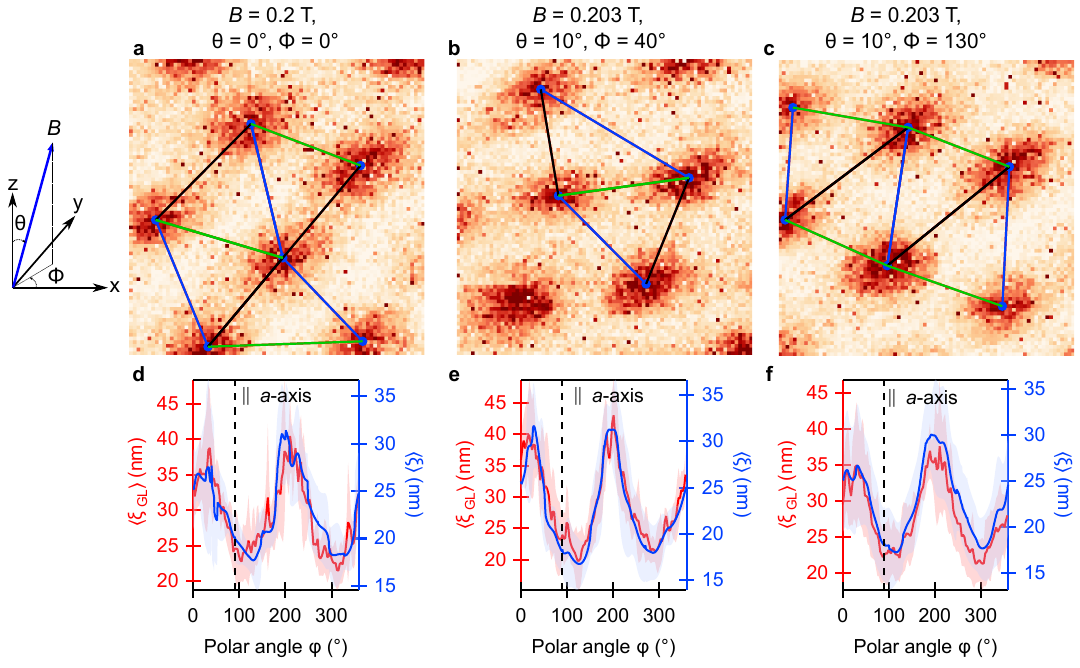}
	\caption{(a) Normalized ZBC maps in the perpendicular magnetic field of 0.2~T ($\theta$ = 0$^{\circ}$). (b,c). An additional in-plane magnetic field of 35~mT applied in the direction of $\phi$ = 40$^{\circ}$ (b) and $\phi$ = 130$^{\circ}$ (c); these in-plane magnetic fields cause $\theta$ to become 10$^{\circ}$ with the total magnetic field of 0.203 T. The definition of $\theta$ and $\phi$ is shown on the left. \stabp{} $U=1$~mV, $I=50$~pA and $U_{\text{mod}}= 100$~$\mu$V$_{\text{p}}$. \JB{(d,e,f) The average $\langle\xi_{\mathrm{GL}}\rangle$ and $\langle\xi\rangle$ calculated from the lattice sites indicated by blue dots in (a,b,c). Note that the panels (a,d) are the same as \fig{fig:vortex_lattice_SII}(b,c).}}
	\label{fig:vortex_inplane}
\end{figure*}

To dismiss the possibility that the observed vortex elongation is a mere consequence of a magnetic field that is not exactly out-of-plane but is tilted away from the \textit{z} axis, we performed the following experiment: We applied an intentional in-plane magnetic field of the magnitude of 35~mT with the azimuthal angle of 40$^{\circ}$ and 130$^{\circ}$. The resultant magnetic field is $\sim$10$^{\circ}$ off from the \textit{z} axis. As shown in \fig{fig:vortex_inplane}, we did not observe any noticeable change in the vortex anisotropy in spite of the intentional misalignment of the field. \YA{Also, the averaged orientation of the vortices did not change upon rotating the in-plane component within the experimental uncertainty of about $10^\circ$.}

\vspace{-1.1cm}

%\clearpage
\YA{\section{Absence of vortex bound states}}

\begin{figure}[h]
	\includegraphics[width=0.63\textwidth]{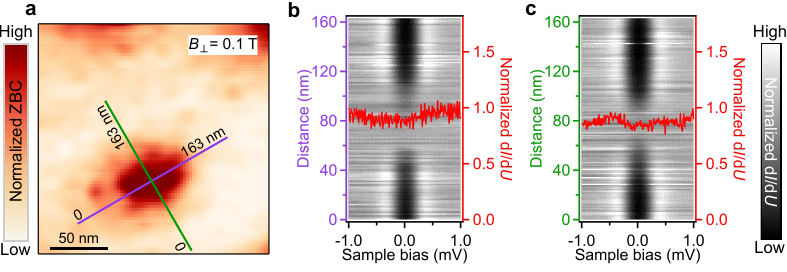}
	\caption{\YA{(a) Normalized ZBC map at \textit{B$_\perp$}=0.1~T showing a single isolated vortex. (b) Normalized \dIdU{} spectra along the purple line (long axis) in (a). Each horizontal row is a \dIdU{} spectrum vs sample bias. The color scale represents the \dIdU{} intensity and the vertical axis corresponds to the distance along the purple line in (a). The red \dIdU{} spectrum superimposed on the map is taken at the center of the vortex and highlights the absence of any vortex bound states. (c) A similar map as done in (b) for the green line (short axis) in (a), clearly illustrating the vortex anisotropy. \stabp{} $U=1$~mV, $I=50$~pA (100 sweeps) for (a); $U=3$~mV, $I=50$~pA (200 sweeps) for (b) and $U=3$~mV, $I=50$~pA (200 sweeps) for (c).}
	}
	\label{fig:single_vortex}
\end{figure}

\YA{No Caroli-de Gennes-Matricon states were observed in the core of a single isolated vortex in CPSBS (Fig.~\ref{fig:single_vortex}). The absence of bound states in the vortex core is expected for a dirty superconductor \cite{Renner1991}. Based on the values for residual resistivity ($\rho_{0}$) and carrier density ($n$) in Ref.~\cite{Andersen2020}, we estimate the mean free path ($l$) in our sample [$l=\hbar k_{\mathrm{F}}/(\rho_{0}ne^{2}$)] to be on the order of only a few nanometer, while the average in-plane coherence length ($\xi_{\mathrm{GL}}$) is measured to be about 25~nm, i.e., CPSBS is a dirty superconductor. Importantly, despite being in the dirty limit, a clear anisotropy in the vortex shape is observed.} 

%\clearpage

\clearpage
\section{Statistical overview of the SC regions observed by STM on the surface}

Tao \etal{} \cite{Tao2018} reported that the probability to find SC regions on the surface of \CuBiSe{} was less than 5\% in their STM study. Our experience was similar during the scanning on the surface of CPSBS. It is more common to find non-superconducting areas. In our experiments a single scan area corresponds to 1.5~$\mu$m by 1.5~$\mu$m, which is the maximum range for the scan piezo. A new scan area on the sample is accessed by moving the sample stage with respect to the tip holder. In Table \ref{tab:SC_NC_area}, we show the total number of scan areas along with the number of areas in which SC was observed. By multiplying the number of scan areas with the area of a single scan frame, we estimate a SC area of \JB{6.75~$\mu$m$^2$ out of 13.5~$\mu$m$^2$ in sample S-I, 6.8~$\mu$m$^2$ out of 65~$\mu$m$^2$ in sample S-II and 11.3~$\mu$m$^2$ out of 63~$\mu$m$^2$ in sample S-III}. 

We previously reported for Sr$_x$Bi$_2$Se$_3$ \cite{Bagchi2022} that spurious superconducting gap was often observed due to picking up of a  superconducting material on the tip from the sample, and it was also the case in CPSBS. Identification of a boundary between a superconducting and non-superconducting region with the same tip apex conditions was used as the criteria for ascertaining a superconducting gap to \YA{reside on the sample side, rather than on the tip side}.

\begingroup
\setlength{\tabcolsep}{8pt}
\renewcommand{\arraystretch}{1.5}
\begin{table}[h]
	\begin{tabular}{|lcc|}
		\hline
		Sample & Number of scan areas & Number of SC regions   \\
		\hline
		\JB{S-I} & 6 & 3 \\
		\JB{S-II}  & 29 & 3 \\
		\JB{S-III}  & 28 & 5  \\		
		\hline
	\end{tabular}
	\caption{Summary of total number of regions scanned and number of SC regions found on \JB{three} CPSBS samples.
		\label{tab:SC_NC_area}
	}
\end{table}
\endgroup

%\clearpage
\YA{\section{Spatial variation of the superconducting gap spectrum}\label{sec:SCvariations}}

\begin{figure*}[ht]
	\centering
	\includegraphics[width=0.95\textwidth]{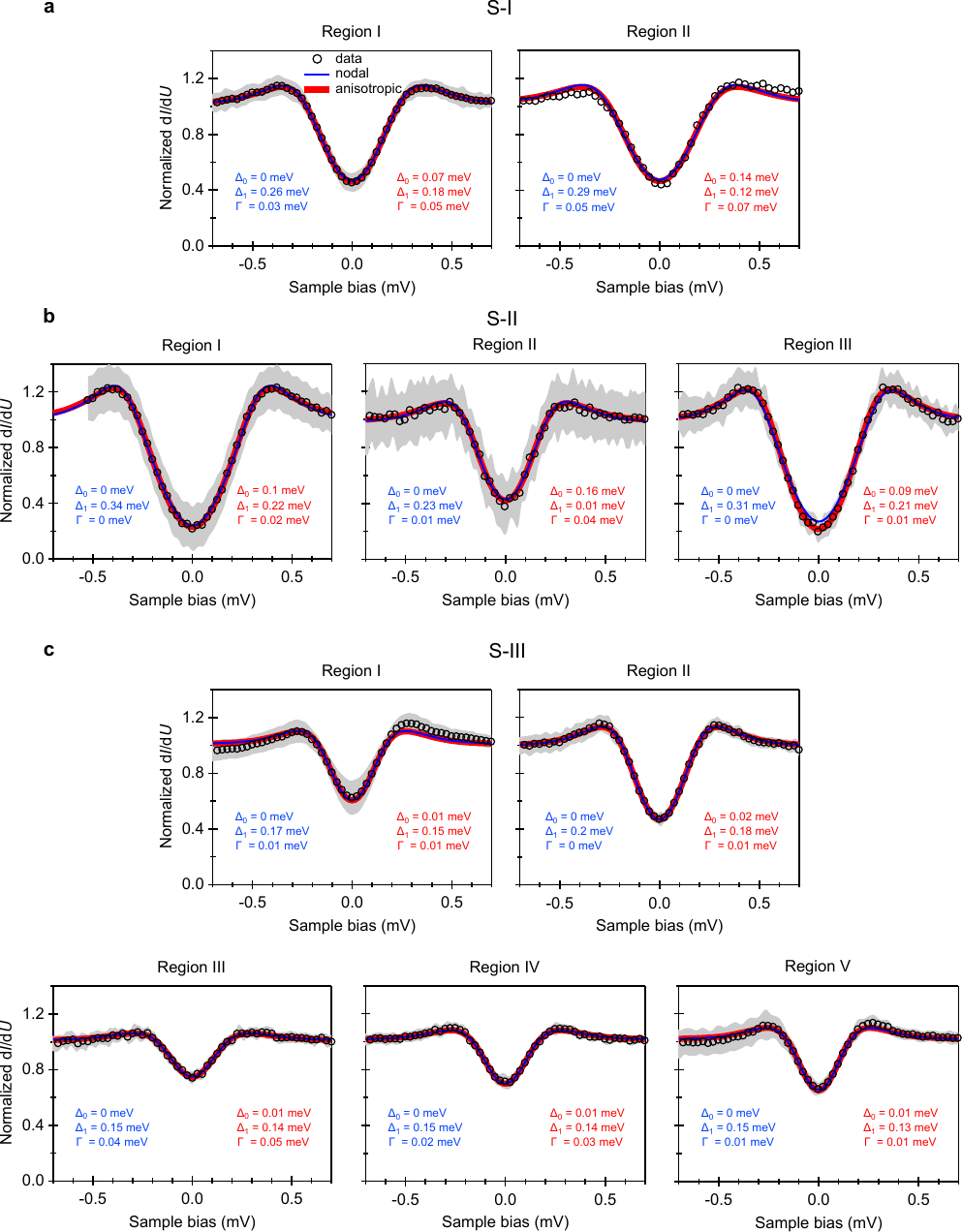}
	\caption{\YA{Variations in the superconducting gap observed in samples S-I, S-II, and S-III. For each region we show the spatial average (black open circles) and the corresponding standard deviation (light gray area). The spatial average is obtained by numerically averaging} \JB{100 STS over 500 by 500~nm$^{2}$, 900 STS over 300 by 300~nm$^{2}$, 60 STS over 25 by 500~nm$^{2}$, 28 STS over 50 by 500~nm$^{2}$, 1600 STS over 500 by 500~nm$^{2}$, 80 STS over 20 by 1500~nm$^{2}$, 200 STS over 40 by 500~nm$^{2}$, 50 STS over 25 by 1400~nm$^{2}$, 300 STS over 500 by 500~nm$^{2}$, for region I (sample S-I), region I -- III (sample S-II), and region I -- V (sample S-III), respectively. For region II of sample S-I, a point STS is shown. The blue and red traces correspond to the nodal and anisotropic fit to the data, respectively, as described in the text. Stabilization parameters: $U=1.5~\text{to}~5$~mV, $I=50~\text{or}~500$~pA.}
	}
	\label{fig:sts}
\end{figure*}

\YA{In Fig.~\ref{fig:sts} we give an overview of the superconducting gaps measured for the various superconducting regions observed in the \JB{three} different samples S-I, S-II \JB{and S-III} (see also Table~\ref{tab:SC_NC_area}). For each region, a large-area spectroscopy grid was evaluated. The open black circles and the light gray shaded area give the arithmetic mean ($\overline{x} = \sum_{i=1}^{n} X_{i}/n$) and the corresponding standard deviation ($\mathrm{SD} = \sqrt{\frac{\sum_{i=1}^{n}(x_{i} - \overline{x})^{2}}{(n-1)}}$), respectively.} \YA{One can see that when there is a large ZBC, the data can be reasonably fit with both the \JB{nodal (blue trace)} and the anisotropic \JB{(red trace)} gap functions discussed in the main text. We observe significant variations of the gap magnitude, which is presumably due to differences in local Cu-concentration and/or electric field induced weakening of superconductivity \cite{Bagchi2022}.}

%\JB{SC gaps measured in region II and III of sample S-III are not shown, since the data was acquired with a SC tip.}

%\begingroup
%\setlength{\tabcolsep}{6pt}
%\renewcommand{\arraystretch}{1.5}
%\begin{table}[ht]
	%\begin{tabular}{|cccc|}
		%\hline
		%& \multicolumn{1}{p{1.5cm}}{\centering $\Gamma$ (meV)} &  \multicolumn{1}{p{1.5cm}}{\centering  $\Delta_{0}$ (meV)} & \multicolumn{1}{p{1.5cm}|}{\centering  $\Delta_{1}$ (meV)}\\ 
		%\hline	
		%Point STS  &   0.03  & 0.20 & 0.14      \\
		%Spatially averaged STS &   0.04  & 0.18 & 0.13      \\	  
		%Spatially averaged STS + SD &   0.06  & 0.17 & 0.12       \\
		%Spatially averaged STS - SD &  0.02  & 0.19 & 0.13      \\
		%
		%\hline
		%
	%\end{tabular}
	%\caption{\YA{Fit parameters for an anisotropic gap fit for STS taken in region I, S-I. $T_{\mathrm{eff}}$ and anisotropy ratio are held constant to 0.7~K and 1.7, respectively.} 
		%\label{tab:fitting_para}. 
	%}
%\end{table}
%\endgroup

\clearpage
\YA{\section{Orientation of superconducting gap minima on sample S-III surface}}

\begin{figure*}[ht]
	\centering
	\includegraphics[width=1\textwidth]{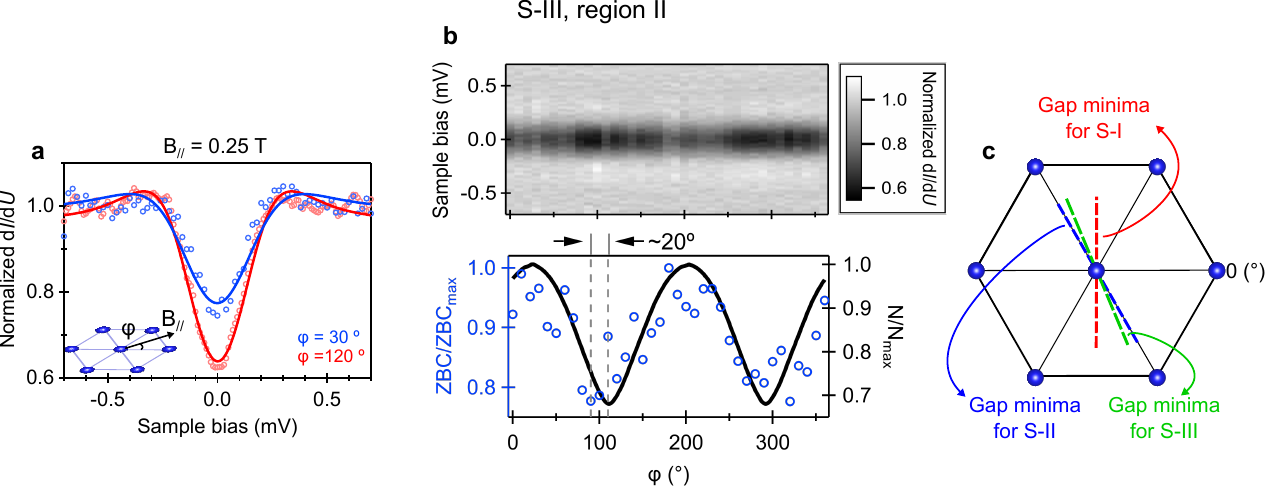}
	\caption{\YA{(a) Point STS taken in region II of sample S-III, but in the presence of an in-plane magnetic field ($B_{\parallel}$) of 0.25~T for different field orientations ($\varphi$). The definition of $\varphi$ is indicated in the inset. Solid lines are fits to the data. Here, the measured superconducting gaps are fit using $\Gamma$ as the only fitting parameter; $T_{\mathrm{eff}}=$ 0.7~K, $\Delta_{0}=0.02$~meV and $\Delta_{1}=0.18$~meV are fixed. (b) A twofold oscillation is clearly visible in the raw data of \dIdU{} at zero bias (top) and \REV{this angular dependence of the zero-bias DOS can be fit with the changes in zero energy DOS (N) calculated by Nagai~\cite{Nagai2014} (bottom)} to quantify the minima at approximately 110$^\circ$ and 290$^\circ$ . (c) \JB{The positions of the minima in the superconducting gap structure obtained for samples S-I (red dashed line), S-II (blue dashed line) and S-III (green dashed line).} Stabilization parameters: (a,b) $U=1.5$~mV, $I=100$~pA.}
	}
	\label{fig:sc_sample2}
\end{figure*}

\YA{\JB{Similar to sample S-II, no structural stripe pattern was observed in sample S-III (see Sec.~\ref{sec:topo}). Moreover,} the smaller SC magnitude paired with significant spatial inhomogeneity prohibited a reliable evaluation of the anisotropy of superconducting vortices. Due to this difficulty, we probed the anisotropy in the superconducting gap of region II of \JB{sample S-III} by measuring the response to an in-plane magnetic field ($B_{\parallel}$) of 0.25~T, a technique used for Cu$_x$Bi$_2$Se$_3$ by Tao \textit{et al.}~\cite{Tao2018}. When the magnetic field is rotated by an angle $\varphi$ in the \textit{ab}-plane of the crystal, the gap spectra show a clear variation [Fig.~\ref{fig:sc_sample2}(a,b)].} 

\REV{The influence of $B_{\parallel}$ on the zero-energy DOS was calculated within the Kramer–Pesch approximation by Nagai~\cite{Nagai2014} and the result is in good agreement with our experimental data [Fig.~\ref{fig:sc_sample2}(b)]. From this comparison, one can identify the orientation of the gap minima with respect to the crystallographic axes in Fig.~\ref{fig:sc_sample2}(c) and compare it to the orientation of the gap minima observed in sample S-I and S-II. The orientation of the gap minima in sample S-III is close to that in sample S-II and is off by only $\sim$10$^\circ$. In this regard, it is useful to note that Tao \textit{et al.}~\cite{Tao2018} reported a rotation of the gap minimia by up to $20^\circ$ away from the zero-field position when the in-plane magnetic field was applied. It is likely that the same effect is observed here.}

%for 0.05 T
%green mean(252,235,246,241) std(252,235,246,241) 244 $\pm$ 7~nm
%blue mean(232,238,247,237) std(232,238,247,237) 239 $\pm$ 6~nm
%black mean(218,237,212,235) std(218,237,212,235) 226 $\pm$ 12~nm
%
%
%for 0.1 T
%green mean(152,154,146,149) std(152,154,146,149) 150 $\pm$ 4~nm
%blue mean(172,162,166,156) std(172,162,166,156) 164 $\pm$ 7~nm
%black mean(143,153,139,148) std(143,153,139,148) 146 $\pm$ 6~nm
%
%
%for 0.2 T
%green mean(102,103,107,106) std(102,103,107,106) 105 $\pm$ 2~nm
%blue mean(116,123,120,130) std(116,123,120,130) 122 $\pm$ 6~nm
%black mean(107,107,111,110) std(107,107,111,110) 109 $\pm$ 2~nm

\clearpage
\section{Topograph and spectroscopy on PSBS and CPSBS}\label{sec:topo}

\begin{figure*}[ht]
	\centering
	\includegraphics[width=0.9\textwidth]{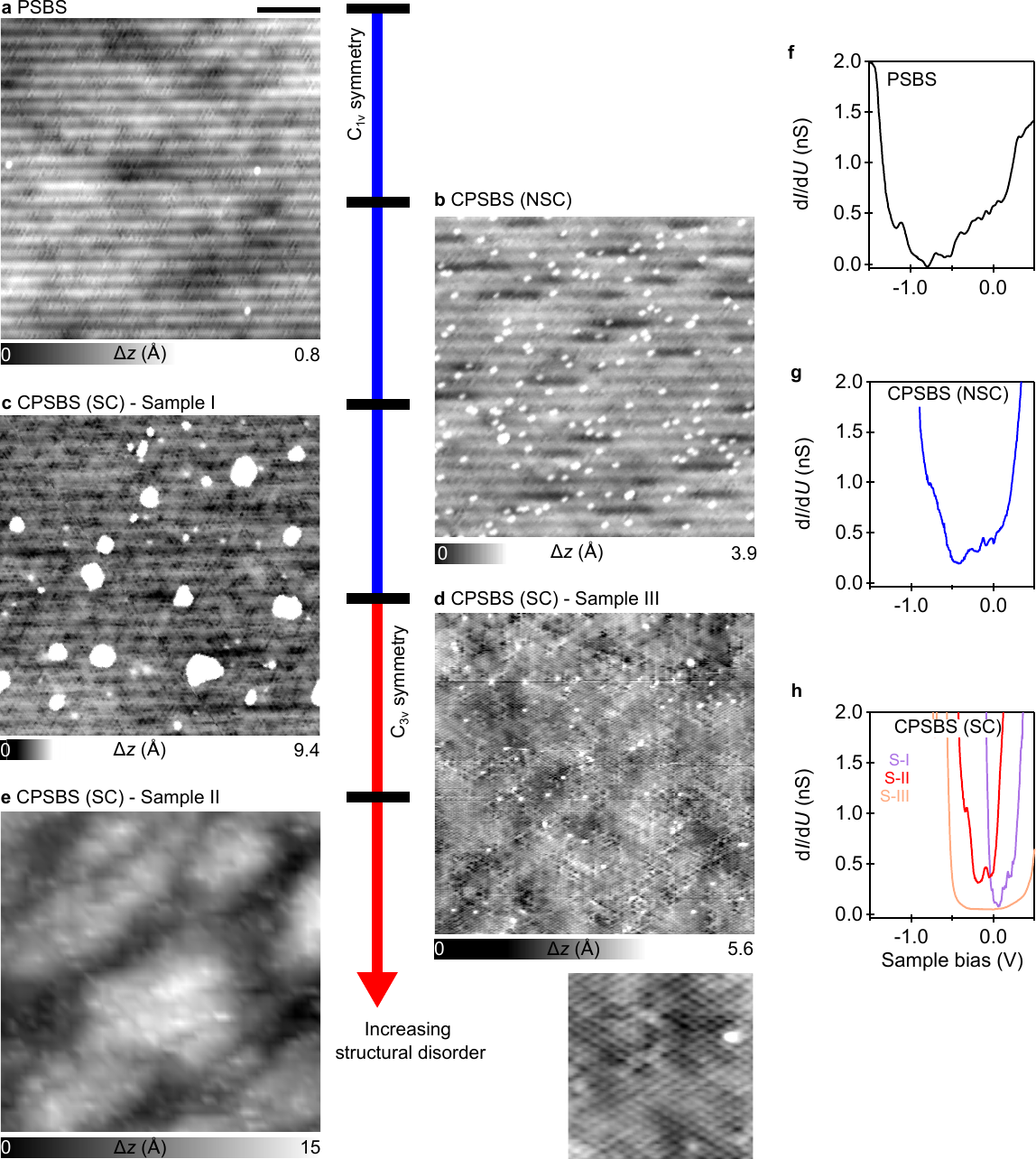}
	\caption{\JHB{(a, b, c, d, e) STM topographs (the scale bar is 10~nm) on the cleaved surface of PSBS (a), on a non-superconducting surface of CPSBS (b), and on a SC area of CPSBS in sample S-I (c), sample S-II (e) and sample S-III (d). The color scales are saturated to highlight the transition from C$_\mathrm{1v}$ (structural stripe pattern visible) to C$_\mathrm{3v}$ symmetry (no structural stripe pattern visible) in the data. Interactions between the STM-tip and Cu-clusters in panel (d) give rise to a noisy appearance. Nonetheless, the atomic lattice showing threefold symmetry is resolved.  No atomic resolution could be obtained in the SC area of sample II due to the high corrugation. 
		(f, g, h) d$I$/d$U$ spectrum on the surface of PSBS (f), CPSBS (NSC) (g) and CPSBS (SC) (S-I, S-II and S-III) (h). Scan/stabilization parameters: $U=900$~mV, $I=20$~nA for (a); $U=950$~mV, $I=50$~pA for (b); $U=900$~mV, $I=100$~pA for (c); $U=-5$~mV, $I=5$~nA for (d); $U=-1000$~mV, $I=10$~pA for (e); $U=1$~V, $I=10$~pA for (f); $U=-900$~mV, $I=500$~pA (right: $U=-5$~mV, $I=5$~nA) for (g); $U=500$~mV, $I=500$~pA (S-I), $U=-600$~mV, $I=1$~nA (S-II) and $U=-900$~mV, $I=2$~nA (S-III) for (h).}
	}
	\label{fig:topo_sts}
\end{figure*}

In \fig{fig:topo_sts}, we show large scale topography images taken on the surface of PSBS, non-superconducting (NSC) areas of CPSBS, and superconducting areas of CPSBS found on samples S-I, S-II and S-III. The clear 1D stripe on the surface of PSBS becomes more disordered on the NSC surface of CPSBS. The stripe was not observed in the SC regions of S-II and S-III but found on the SC regions of S-I. 
	
The image shown in \fig{fig:topo_sts}(e) corresponds to the area where the vortices were observed for S-II. The area was disordered, which made it impossible to obtain clear atomic-resolution images. In comparison, the SC areas on S-I and S-III \JHB{[\fig{fig:topo_sts}(c,d)]} are less disordered and atomic-resolution images could be obtained.

We note a correlation between the superconducting gap as discussed in Sec.~\ref{sec:SCvariations} and the amount of (presumably) Cu remaining on the surface. The superconducting region in sample S-III shows few and small clusters [white spots in  \JHB{\fig{fig:topo_sts}(d)]} and the smallest superconducting gap, S-I exhibits extended and larger clusters [white areas in \JHB{\fig{fig:topo_sts}(c)]} and an intermediate superconducting gap and S-II is entirely covered with Cu showing the largest superconducting gap.

We show in \JHB{\figs{fig:topo_sts}(f, g, h)} large scale \dIdU{} spectra taken on the different surfaces. As discussed in the main text, the cleaved surface of PSBS is a single QL of \BiSe{} above the PbSe layer. The ARPES data \cite{Nakayama2012} on this surface show a single, parabolic electron-like band with an onset around $\sim$-600~meV. The onset of this band is clearly seen in our \dIdU{} spectrum \JHB{[\fig{fig:topo_sts}(f)]}, since the LDOS shows a minima exactly around $\sim$-600~meV. 

On the cleaved surface of CPSBS, the ARPES data \cite{Nakayama2015} show two electron-like bands, both 2D in nature, at the onset energies of -600~meV and -280~meV. It is difficult to identify these bands in the spectroscopy curves \JHB{[\fig{fig:topo_sts}(g,h)]} due to the significant additional DOS around these energies. Neither is it possible to conclude any shift of the surface bands in comparison to the spectrum on PSBS.

\clearpage

%%%%%%%%%%%%%%%%%%%
%%%%%%%%%%%%%%%%%%%
%%%%%%%%%%%%%%%%%%%
\section{Symmetry analysis}

%%%%%%%%%%%%%%%%%%%
\subsection{Symmetry of the Bi$_2$Se$_3$ quintuple layers}

The Bi$_2$Se$_3$ quintuple-layer structure is associated with point group $D_{3d}$  \cite{Andersen2018, Zhang2009}. This group consists of twelve symmetry operations organized in six conjugacy classes: $\hat{E}$, the identity; $\hat{P}$, inversion;  2$\hat{C}_3$: two rotations along the z-axis by $\pm$120$^o$; 3$\hat{C}'_2$: three rotations by 180$^o$ along the x-axis and equivalent in-plane axes; 2$\hat{S}_6$: rotations by $\pm$60$^o$ along the z-axis followed by in-plane mirror reflection; 3$\hat{\sigma}_d$: three vertical mirrors passing through the yz-plane and equivalent planes.  The irreducible representations can be identified from the character table for $D_{3d}$, as displayed in Table \ref{Tab:D3d}.

\begin{table}[h]
\begin{center}
    \begin{tabular}{| c | c | c | c | c | c | c |}
    \hline
    irrep & $\hat{E}$ & $2\hat{C}_3$ & $3\hat{C}_2'$ & $\hat{P}$ & $2\hat{S}_6$ & $3\hat{\sigma}_d$ \\ \hline
    $A_{1g}$ & +1 & +1 & +1 & +1 & +1 & +1 \\ \hline
        $A_{2g}$ & +1 & +1 & -1 & +1 & +1 & -1 \\ \hline
                $E_{g}$ & +2 & -1 & 0 & +2 & -1 & 0 \\ \hline
    $A_{1u}$ & +1 & +1 & +1 & -1 & -1 & -1 \\ \hline
        $A_{2u}$ & +1 & +1 & -1 & -1 & -1 & +1 \\ \hline
                $E_{u}$ & +2 & -1 & 0 & -2 & +1 & 0 \\ \hline    
    \end{tabular}
        \end{center}
    \caption{  \label{Tab:D3d} Character table for the point group $D_{3d}$.}
\end{table}

%%%%%%%%%%%%%%%%%%%
\subsubsection{Description of the normal state}

The normal state Hamiltonian can be described by two effective orbitals of opposite parity, referred as $P_{1z+}$ and $P_{2z-}$ \cite{Zhang2009,Liu2010}. In the basis $\Phi^\dagger_\bk  = (c_{1\uparrow}^\dagger, c_{1\downarrow}^\dagger, c_{2\uparrow}^\dagger, c_{2\downarrow}^\dagger)_\bk$, the Hamiltonian can be parametrized as:
\begin{eqnarray}\label{Eq:H0}
\hat{H}_0(\bk)= \sum_{a,b}  h_{ab}(\bk) \hat{\tau}_a\otimes\hat{\sigma}_b,
\end{eqnarray}
where $\hat{\tau}_{a=1,2,3}$ are Pauli matrices encoding the orbital degrees of freedom (DOF), $\hat{\sigma}_{b=1,2,3}$ are Pauli matrices encoding the spin DOF, and $\hat{\tau}_0$ and $\hat{\sigma}_0$ are two-dimensional identity matrices in orbital and spin space, respectively. In the presence of time-reversal and inversion symmetries, the only allowed terms in the Hamiltonian have the subscripts $(a,b) = \{(0,0), (2,0),(3,0),(1,1),(1,2),(1,3)\}$. Time-reversal symmetry is defined as $\hat{\Theta} = K \hat{\tau}_0 \otimes (i\hat{\sigma}_2)$, where $K$ stands for complex conjugation, and the parity operator as $\hat{P}=\hat{\tau}_3\otimes \hat{\sigma}_0$. The properties of the $ \hat{\tau}_a\otimes\hat{\sigma}_b$ matrices under the point group operations allow us to associate each of these terms to a given irreducible representation of $D_{3d}$, therefore constraining the momentum dependence of the form factors $h_{ab}(\bk)$ by symmetry. Table \ref{Tab:H} provides the details on the properties of each term in the normal-state Hamiltonian and an expansion of $h_{ab}(\bk)$ for small momenta.

\begin{table}[h]
\begin{center}
    \begin{tabular}{| c | c | c | c | c | c | c | c |}
    \hline
    $(a,b)$ &   $\hat{P}$   &   $\hat{C}_3$   & $\hat{C}_2'$   & $\hat{\sigma}_d$ & Irrep & $h_{ab}(\bk)$ & Process \\ \hline
    $(0,0)$ &  +1 &  +1  & +1 & +1 & $A_{1g}$ & $C_0 + C_1 k_z^2 + C_2 (k_x^2 + k_y^2)$ & Intra-orbital hopping \\ \hline
    $(2,0)$ &  -1 &  +1  & -1 & +1 & $A_{2u}$ & $B_0 k_z $ & Inter-orbital hopping \\ \hline
    $(3,0)$ &  +1 &  +1  & +1 & +1 & $A_{1g}$ &  $M_0 + M_1 k_z^2+ M_2 (k_x^2 + k_y^2)$ & Intra-orbital hopping \\ \hline
    $(1,1)$ &  -1 &  y  & -1 & +1 & $E_{u}$ & $- A_0 k_y $ & SOC  \\ \hline
    $(1,2)$ &  -1 & x  & +1 & -1 & $E_{u}$ & $ A_0 k_x$ & SOC \\ \hline
    $(1,3)$ &  -1 &  +1  & +1 & -1 &  $A_{1u}$ & $R_1 k_x(k_x^2-3 k_y^2)$ & SOC  \\ \hline
    \end{tabular}
        \end{center}
    \caption{  \label{Tab:H} Parametrization of the normal-state Hamiltonian given in Eq. (\ref{Eq:H0}) for materials in the family of Bi$_2$Se$_3$. The columns labeled by $\hat{P}$, $\hat{C}_3$, $\hat{C}_2'$ and $\hat{\sigma}_d$ indicate how the basis matrices $\hat{\tau}_a \otimes \hat{\sigma}_b$, indicated by $(a,b)$, transform under the respective point group operations, such that one can associate these with different irreducible representations of $D_{3d}$ (Irrep). The seventh column gives the expansion of the accompanying form factors $h_{ab}(\bk)$ for small momentum, and the last column the associated physical process. The value of the coefficients for Bi$_2$Se$_3$ and other materials in this family can be found by first principles calculations (see for example Table IV in Ref. [17]).}
\end{table}

Note that the three first terms in Table \ref{Tab:H} are spin-independent: $(0,0)$ and $(3,0)$ are even and associated with intra-orbital hopping, while $(2,0)$ is odd and encodes inter-orbital hopping. The last three terms are spin-dependent and inter-orbital in character, therefore all odd. The terms $(1,1)$ and $(1,2)$ are associated with a \AR{Rashba-like spin-orbit coupling}. The term $(1,3)$ is associated with trigonal warping of the Fermi surface and is usually dropped from the effective Hamiltonians since these carry at least terms of third order in momenta. %For the chosen basis above, trigonal warping would only appear due to inter-layer hopping and would require terms of fifth order in momenta. 
Note that the coefficients $C_1$, $B_0$, and $M_1$ for CPSBS are smaller than the ones used for doped Bi$_2$Se$_3$ materials \AR{given the presence of the PbSe layers. The Fermi surface of CPSBS was experimentally shown to be cylindrical (2D Fermi surface) \cite{Nakayama2015}, while the Fermi surface in doped Bi$_2$Se$_3$ materials can be either ellipsoidal (3D) or cylindrical (2D).}

%%%%%%%%%%%%%%%%%%%
\subsubsection{Classification of order parameters with momentum-independent gap matrices in the microscopic basis}

Given these symmetries, we can classify all order parameters with $\bk$-independent gap matrices in the microscopic basis according to the irreps of $D_{3d}$. Following \cite{Andersen2020}, the order parameters can be generally written as:
\begin{eqnarray}\label{Eq:Delta}
\hat{\Delta} =  \sum_{a,b} d_{ab}  \hat{\tau}_a \otimes \hat{\sigma}_b (i\hat{\sigma}_2) .
\end{eqnarray}

The allowed momentum-independent gap matrices can be determined by first searching for matrices satisfying $\hat{\Delta }= - \hat{\Delta}^T$, following fermionic antisymmetry.  These can be classified according to the irreducible representations of $D_{3d}$, as displayed in Table \ref{Tab:SC}. These order parameters are constructed in the orbital basis, so one needs to transform them to the band basis in order to discuss the presence of nodes and their locations. Following the discussion in the SM of \cite{Andersen2020}, we find the nodes highlighted in the last two columns of Table \ref{Tab:SC}.

\begin{table}[h]
\begin{center}
    \begin{tabular}{| c | c | c | c | c | c | c | c |}
    \hline
   [a,b] & Irrep &  Spin &  Orbital   & Parity   & Matrix Form & Zero at 3D FS & Zero at 2D FS  \\ \hline
  [0,0] &   \multirow{2}{*}{$A_{1g}$}&
     \multirow{2}{*}{Singlet}&
          \multirow{2}{*}{Intra}&
               \multirow{2}{*}{Even}    & $\hat{\tau}_0\otimes \hat{\sigma}_0(i\hat{\sigma}_2)$ & - & -  \\ \cline{1-1}\cline{6-8}
     	[3,0]&				&&&& $\hat{\tau}_3\otimes \hat{\sigma}_0 (i\hat{\sigma}_2)$ & - & -   \\ \hline
	[2,3]& 	{$A_{1u}$}  & Triplet  & Inter & Odd  & $\hat{\tau}_2 \otimes\hat{\sigma}_3 (i\hat{\sigma}_2)$ & - & -  \\ \hline
	[1,0] &	{$A_{2u}$}  & Singlet  & Inter & Odd  & $\hat{\tau}_1\otimes \hat{\sigma}_0(i\hat{\sigma}_2)$  &
 \begin{tabular}{@{}c@{}}along the $k_z$ axis  \\   $k_x=k_y=0$\end{tabular}
  & - \\ \hline
	[2,1] &	 \multirow{2}{*}{$E_{u}$}&
     \multirow{2}{*}{Triplet}&
          \multirow{2}{*}{Inter}&
               \multirow{2}{*}{Odd}  
				&  $\hat{\tau}_2\otimes\hat{\sigma}_1(i\hat{\sigma}_2)$ & 
				 \begin{tabular}{@{}c@{}}along the $k_x$ axis  \\   $k_z=k_y=0$\end{tabular} &  
				 \begin{tabular}{@{}c@{}}along the $k_xk_z$ plane*  \\  $k_y=0$\end{tabular} 
				  \\ \cline{1-1} \cline{6-8}
				%?$k_z=k_y = 0$ & ?$k_x=0$   \\ \cline{1-1}\cline{6-8}
		[2,2]&	&&&& $\hat{\tau}_2\otimes \hat{\sigma}_2 (i\hat{\sigma}_2)$ & 
						 \begin{tabular}{@{}c@{}}along the $k_y$ axis  \\   $k_z=k_x=0$\end{tabular} &   \begin{tabular}{@{}c@{}}along the $k_yk_z$ plane**  \\  $k_x=0$\end{tabular}   \\ 
\hline
    \end{tabular}
               \end{center}
        \caption{\label{Tab:SC} Superconducting order parameters for materials in the family of Bi$_2$Se$_3$. Here we focus on the order parameters with momentum-independent gap matrices in the microscopic basis and highlight the associated irreducible representation (Irrep) and the spin, orbital character, and parity of the respective gap matrix. We write the order parameter in the matrix form $\hat{\tau}_a \otimes \hat{\sigma}_b (i\hat{\sigma}_2)$. Here we factor out $(i \hat{\sigma}_2)$ so that one can directly relate $b=0$ to a singlet state and $b=\{1,2,3\}$ with the $\{x,y,z\}$ components of the $d$-vector parametrization for triplet states. * Nodes are lifted if trigonal warping is introduced in the $(1,3)$ term in the normal state. ** Nodes are not lifted if trigonal warping is introduced in the $(1,3)$ term in the normal state.}
\end{table}

%%%%%%%%%%%%%%%%%%%
\subsection{Symmetry reduction for the bulk of CPSBS}

For CPSBS, the presence of the (PbSe)$_5$ layers reduces the point group symmetry to $D_{1d}$ (isomorphic to $C_{2h}$). In CPSBS $\hat{C}_3$ is not a symmetry transformation anymore, but $\hat{C}_2'$ and $\hat{\sigma}_d$ are still valid symmetry operations which allow us to generate the character table for the reduced group from the table for $D_{3d}$. The point group  $D_{1d}$ is formed by four operations organized in four conjugacy classes, therefore there are four irreducible representations, as displayed in Table \ref{Tab:D1d}. The irreducible representations from $D_{3d}$ to $D_{1d}$ are mapped as follows: $A_{1g/u} \rightarrow A_{1g/u}$, $A_{2g/u} \rightarrow A_{2g/u}$, $E_{g/u} \rightarrow \{A_{1g/u}, A_{2g/u}\}$. The last correspondence means that the two dimensional irreducible representations of $D_{3d}$ are split in $D_{1d}$.

\begin{table}[h]
\begin{center}
    \begin{tabular}{| c | c | c | c | c | c | c |}
    \hline
    irrep & $\hat{E}$ &  $\hat{C}_2'$ & $\hat{P}$ &  $\hat{\sigma}_d$ \\ \hline
    $A_{1g}$ & +1 & +1 & +1  & +1 \\ \hline
        $A_{2g}$ & +1 & -1 & +1 &  -1 \\ \hline
    $A_{1u}$ & +1 &  +1 & -1 & -1 \\ \hline
        $A_{2u}$ & +1 &  -1 & -1 & +1 \\ \hline
    \end{tabular}
        \end{center}
    \caption{  \label{Tab:D1d} Character table for the point group $D_{1d}$ obtained from a reduction of the character table for $D_{3d}$ by eliminating the columns corresponding to the conjugacy classes labelled as $2\hat{C}_3$ and $2\hat{S}_6$.}
\end{table}

%%%%%%%%%%%%%%%%%%%
\subsubsection{Description of the normal state}

Given the presence of parity and time-reversal symmetries, the only terms allowed in $\hat{H}_0(\bk)$ are the same as the ones enumerated for the case of $D_{3d}$ symmetry, but now these are mapped to different irreducible representations. The main consequence for our model is that the parameter $A_0$ does not need to be the same for the $(1,1)$ and $(1,2)$ terms (see Table \ref{Tab:HD1d}).

\begin{table}[h]
\begin{center}
    \begin{tabular}{| c  | c | c | c | c | c |}
    \hline
    $(a,b)$ &   $\hat{P}$   &    $\hat{C}_2'$   & $\hat{\sigma}_d$ & Irrep & $h_{ab}(\bk)$ \\ \hline
    $(0,0)$ &  +1 &   +1 & +1 & $A_{1g}$ & $C_0 + C_1 k_z^2 + C_2 (k_x^2 + k_y^2)$ \\ \hline
    $(2,0)$ &  -1   & -1 & +1 & $A_{2u}$ & $B_0 k_z $ \\ \hline
    $(3,0)$ &  +1   & +1 & +1 & $A_{1g}$ &  $M_0 + M_1 k_z^2+ M_2 (k_x^2 + k_y^2)$ \\ \hline
    $(1,1)$ &  -1   & -1 & +1 & $A_{2u}$ & $- A_1 k_y $  \\ \hline
    $(1,2)$ &  -1  & +1 & -1 & $A_{1u}$ & $ A_2 k_x $ \\ \hline
    $(1,3)$ &  -1  & +1 & -1 &  $A_{1u}$ & $R_1 k_x(k_x^2-3 k_y^2)$  \\ \hline
    \end{tabular}
        \end{center}
    \caption{  \label{Tab:HD1d} Parametrization of the normal-state Hamiltonian given in Eq. (\ref{Eq:H0}) for materials in the family of Bi$_2$Se$_3$ with reduced $D_{1d}$ symmetry, applicable to the bulk of CPSBS. Note the different coefficients for $(1,1)$ and $(1,2)$ terms.}
\end{table}

%%%%%%%%%%%%%%%%%%%
\subsubsection{Order parameter discussion}

Concerning the order parameters, the reduction of the point group symmetry in principle allows for new kinds of superpositions, as suggested by Table I in the main text.  Note that $[0,0]$ and $[3,0]$ belong to $A_{1g}$, $[2,3]$ and $[2,2]$ belong to $A_{1u}$, and that $[1,0]$ and $[2,1]$ belong to $A_{2u}$. According to experiments, the order parameter that is in agreement with the symmetry protected nodes in  the bulk of CPSBS is $[2,2]$. By symmetry, the most general form of the order parameter in $D_{1d}$ is a linear superposition of $[2,2]$ and $[2,3]$. A superposition of these two types of order parameters generally lifts the nodes present in the case of a pure $[2,2]$ order parameter. The mixing of [2,2] and [2,3] order parameters is only possible in the presence of \emph{both}  SOC terms $(1,2)$ and $(1,3)$ in the normal state Hamiltonian. As the trigonal warping term $(1,3)$ is negligibly small, this mixing should also be small in the bulk. The smallness of trigonal warping explains the robustness of the nodes (or their transformation to near nodes with experimentally unaccessible minima) in the bulk of CPSBS.

%%%%%%%%%%%%%%%%%%%
\subsection{Symmetry reduction at the surface of CPSBS}

At the surface of CPSBS, inversion symmetry is broken and the group is reduced to $C_{1v}$ with only two elements, therefore only two conjugacy classes and irreducible representations. The irreps are mapped according to $\{A_{1g}, A_{2u}\} \rightarrow A_1$ and $\{A_{2g}, A_{1u}\}\rightarrow A_2$, as indicated by the $C_{1v}$ character table displayed as Table \ref{Tab:C1v}. 

\begin{table}[h]
\begin{center}
    \begin{tabular}{| c | c | c | c | c | c | c |}
    \hline
    irrep & $\hat{E}$ &   $\hat{\sigma}_d$ \\ \hline
    $A_1$ & +1 &  +1 \\ \hline
        $A_2$ & +1  & -1 \\ \hline
    \end{tabular}
        \end{center}
    \caption{  \label{Tab:C1v} Character table for the point group $C_{1v}$ obtained by reducing the character table for $D_{1d}$ by eliminating the columns corresponding to $\hat{P}$ and $\hat{C}_2'$.}
\end{table}

%%%%%%%%%%%%%%%%%%%
\subsubsection{Description of the normal state}

Note that in the absence of inversion symmetry all $(a,b)$ terms in the normal state Hamiltonian are symmetry allowed. The new terms in the normal state Hamiltonian are summarized in Table \ref{Tab:HC2}.

\begin{table}[h]
\begin{center}
  \begin{tabular}{| c  | c | c | c | c | c |}
    \hline
    $(a,b)$ &  $\hat{\sigma}_d$ & TRS & Irrep & $h_{ab}(\bk)$ & Process \\ \hline
    $(0,0)$ &  +1 & + & $A_1$ & $C_0 + C_1 k_z^2 + C_2 (k_x^2 + k_y^2)$ & Intra-orbital hopping \\ \hline
     $(0,1)$ & +1   & - & $A_1$ &  $c_1 k_y + c_1' k_z$ & SOC \\ \hline
          $(0,2)$ & - 1  & -  & $A_2$ & $c_2 k_x + c_2'k_x(k_x^2-3 k_y^2)$  & SOC \\ \hline
               $(0,3)$ & -1 &  - & $A_2$ & $c_3 k_x + c_3'k_x(k_x^2-3 k_y^2)$ &   SOC \\ \hline
    $(1,0)$ & +1 & + & $A_1$ & $c_4 + c_4' k_z^2 + c_4'' (k_x^2 + k_y^2)$  &  Inter-orbital hopping \\ \hline
    $(1,1)$ &  +1 & - & $A_1$ & $- A_1 k_y + a_1 k_z  $ & SOC \\ \hline
    $(1,2)$ &   -1 & -& $A_2$ & $ A_2 k_x  $ & SOC \\ \hline
    $(1,3)$ &   -1 & - &  $A_2$ & $R_1 k_x(k_x^2-3 k_y^2) + r_1 k_x$ & SOC \\ \hline
    $(2,0)$ &   +1 & - & $A_1$ & $B_0 k_z + b_0 k_y $ & Inter-orbital hopping \\ \hline
     $(2,1)$ & +1   & + & $A_1$ & $c_5 + c_5' k_z^2 + c_5'' (k_x^2 + k_y^2)$  & SOC \\ \hline
          $(2,2)$ & -1 & + & $A_2$ & $c_6 k_xk_z + c_6'k_yk_z$ & SOC \\ \hline
               $(2,3)$ & -1  & + &  $A_2$ &  $c_7 k_xk_z + c_7'k_yk_z$& SOC \\ \hline
    $(3,0)$ &   +1 & + & $A_1$ &  $M_0 + M_1 k_z^2+ M_2 (k_x^2 + k_y^2)$ & Intra-orbital hopping \\ \hline
     $(3,1)$ & +1  & - & $A_1$  & $c_8 k_y + c_8' k_z$  & SOC \\ \hline
          $(3,2)$ & -1  & - & $A_2$ & $c_9 k_x + c_9'k_x(k_x^2-3 k_y^2)$  & SOC \\ \hline
               $(3,3)$ & -1 &- & $A_2$  &  $c_{10} k_x + c_{10}'k_x(k_x^2-3 k_y^2)$ & SOC \\ \hline
    \end{tabular}
        \end{center}
    \caption{  \label{Tab:HC2} Parametrization of the normal-state Hamiltonian given in Eq. (\ref{Eq:H0}) for materials in the family of Bi$_2$Se$_3$ with reduced $C_{1v}$ symmetry, applicable to the surface of CPSBS. Here we entered new symmetry allowed coefficients $b_0$, $a_1$, $r_1$, and $c_i$, $c_i'$, and $c_i''$, for $i=\{1,...,10\}$.}
\end{table}

%%%%%%%%%%%%%%%%%%%
\subsubsection{Order parameter discussion}

The order parameters are now mapped such that $[0,0]$, $[3,0]$, $[1,0]$, and $[2,1]$ belong to the $A_1$ irrep, while $[2,3]$ and $[2,2]$ belong to the $A_2$ irrep. Note that the latter is the same type of mixing found for $D_{1d}$ symmetry.  Again, according to experiments, the order parameter that is in agreement with the symmetry protected nodes in CPSPS is $[2,2]$. By symmetry, the most general form of the order parameter in $C_{1v}$ is a linear superposition of $[2,2]$ and $[2,3]$. As discussed above, the mixing of these two types of order parameters would generally lift the nodes. Note that, in presence of only $C_{1v}$ symmetry, these two order parameter components can be coupled by multiple pairs of terms in the normal state Hamiltonian.

%%%%%%%%%%%%%%%%%%%
\subsection{Symmetry reduction at the surface of CPSBS with disorder}

At the surface of CPSBS, superconductivity was only observed in areas with no stripe pattern. This suggests that the influence of the PbSe layers in the superconducting areas is weakened and the three-fold rotational symmetry of the Bi$_2$Se$_3$ layers in the bulk is restored in these regions. We consider then the point group symmetry of the surface of Bi$_2$Se$_3$, identified as $C_{3v}$. The irreps are mapped according to $\{A_{1g}, A_{2u}\} \rightarrow A_1$, $\{A_{2g}, A_{1u}\}\rightarrow A_2$, and $\{E_u, E_g\}\rightarrow E$, as indicated by the $C_{3v}$ character table displayed as Table \ref{Tab:C3v}. 

\begin{table}[h]
\begin{center}
    \begin{tabular}{| c | c | c | c | c | c | c |}
    \hline
    irrep & $\hat{E}$ &   $2\hat{C}_3$ & $3\hat{\sigma}_d$ \\ \hline
    $A_1$ & +1 &  +1 & +1 \\ \hline
        $A_2$ & +1  & +1 & -1 \\ \hline
        $E$ & 2 & -1 & 0 \\ \hline
    \end{tabular}
        \end{center}
    \caption{  \label{Tab:C3v} Character table for the point group $C_{3v}$ obtained by reducing the character table for $D_{3d}$ by eliminating the columns corresponding to $\hat{P}$, $3\hat{C}_2'$, and $2\hat{S}_6$.}
\end{table}

%%%%%%%%%%%%%%%%%%%
\subsubsection{Description of the normal state}

Note that in the absence of inversion symmetry all $(a,b)$ terms in the normal state Hamiltonian are symmetry allowed. The new terms in the normal state Hamiltonian are summarized in Table \ref{Tab:HC3v}.

\begin{table}[h]
\begin{center}
    \begin{tabular}{| c  | c | c | c | c | c |}
    \hline
    $(a,b)$ &  $\hat{\sigma}_d$ & TRS & Irrep & $h_{ab}(\bk)$ & Process \\ \hline
    $(0,0)$ &  +1 & + & $A_1$ & $C_0 + C_1 k_z^2 + C_2 (k_x^2 + k_y^2)$ & Intra-orbital hopping \\ \hline
     $(0,1)$ & +1   & - & $E$ &  $m_1 k_y $ & SOC \\ \hline
          $(0,2)$ & - 1  & -  & $E$ & $-m_1 k_x $  & SOC \\ \hline
               $(0,3)$ & -1 &  - & $A_2$ & $m_2 k_x(k_x^2-3 k_y^2)$ &   SOC \\ \hline
    $(1,0)$ & +1 & + & $A_1$ & $m_3 + m_3' k_z^2 + m_3'' (k_x^2 + k_y^2)$  &  Inter-orbital hopping \\ \hline
    $(1,1)$ &  +1 & - & $E$ & $- A_0 k_y  $ & SOC \\ \hline
    $(1,2)$ &   -1 & -& $E$ & $ A_0 k_x  $ & SOC \\ \hline
    $(1,3)$ &   -1 & - &  $A_2$ & $R_1 k_x(k_x^2-3 k_y^2)$ & SOC \\ \hline
    $(2,0)$ &   +1 & - & $A_1$ & $B_0 k_z $ & Inter-orbital hopping \\ \hline
     $(2,1)$ & +1   & + & $E$ & $m_4 (k_x^2 - k_y^2) $  & SOC \\ \hline
          $(2,2)$ & -1 & + & $E$ & $-m_4 k_xk_y $ & SOC \\ \hline
               $(2,3)$ & -1  & + &  $A_2$ &  $m_5  k_x k_z(3k_x^2- k_y^2)$& SOC \\ \hline
    $(3,0)$ &   +1 & + & $A_1$ &  $M_0 + M_1 k_z^2+ M_2 (k_x^2 + k_y^2)$ & Intra-orbital hopping \\ \hline
     $(3,1)$ & +1  & - & $E$  & $m_6 k_y $  & SOC \\ \hline
          $(3,2)$ & -1  & - & $E$ & $-m_6 k_x $  & SOC \\ \hline
               $(3,3)$ & -1 &- & $A_2$  &  $m_7 k_x(k_x^2-3 k_y^2)$ & SOC \\ \hline
    \end{tabular}
        \end{center}
    \caption{  \label{Tab:HC3v} Parametrization of the normal-state Hamiltonian given in Eq. (\ref{Eq:H0}) for materials in the family of Bi$_2$Se$_3$ with reduced $C_{3v}$ symmetry, applicable to the surface of doped Bi$_2$Se$_3$ and the disordered surfaces of CPSBS. Here we entered new symmetry allowed coefficients $m_i$, $m_i'$, and $m_i''$, for $i=\{1,...,7\}$.}
\end{table}

%%%%%%%%%%%%%%%%%%%
\subsection{Parametrization of normal state Hamiltonian and gaps for \YA{Figure 5} in the main text}

For the generation of \YA{Figure 5}, we used the following first principle parameters extracted from Ref. \cite{Liu2010} for the starting Hamiltonian with $D_{3d}$ symmetry:
\begin{eqnarray}
A_0&=&3.33 \text{ eV$\AA^{-1}$}, \\ \nonumber
C_0&=&-0.0083 \text{ eV}, \\ \nonumber
C_2&=&30.4\text{ eV$\AA^{-2}$}, \\ \nonumber
M_0&=&-0.28 \text{ eV}, \\ \nonumber
M_2&=&44.5 \text{ eV$\AA^{-2}$}, \\ \nonumber
R_1 &=& 50.6  \text{ eV$\AA^{-3}$}.
%r_2&=& 50.6,
\end{eqnarray}

We set $B_0=C_1=M_1=0$ (in the respective units) to eliminate the $k_z$ dependence.

In the presence of $D_{1d}$ symmetry, $A_1\neq A_2$. As we do not have access to first principles calculations, we choose $A_{1,2} = A_0 \pm A_0/10$. The gap anisotropy does not seem to strongly depend on the ratio $A_1/A_2$.

For the new terms at the surface, we also do not have access to a first principles calculation, so we set these to be a fixed percentage ($p$) of the value of terms with similar momentum dependence in the original normal state Hamiltonian with inversion symmetry. 

For the case with $D_{1d}$ symmetry we set
\begin{eqnarray}
%A1 couples to ky and A2 couples to kx
&&c_1= b_0 = c_8= -p A_ 1\\ \nonumber
&&c_2= c_3= c_9 = c_{10}= p A_2 \\ \nonumber
&&c_2' = c_3'  = c_9' = c_{10}' p R_1\\ \nonumber
&&c_4= c_5 = p C_0 \\ \nonumber
&&c_4'=c_5' = p C_1,\\ \nonumber
&&c_4'' = c_5'' = p C_2\\  \nonumber
&&r_1=p A_2 
\end{eqnarray}
and take $a_1 = c_6=c_6' = c_7 = c_7' = c_8' = 0$ to eliminate the $k_z$ dependence.

For the case with $C_{3v}$ symmetry we set
\begin{eqnarray}
&&m_1 = m_6= p A_0 \\ \nonumber
&&m_2=m_7 = p R_1 \\ \nonumber
&&m_3 = p C_0  \\ \nonumber
&& m_3' = p C_1  \\ \nonumber
&& m_3'' = p C_2 \nonumber
\end{eqnarray}
and take $m_4=0$ as there is no such terms in our parametrization of the normal state Hamiltonian with $D_{3d}$ symmetry, and we take $m_5=0$ to eliminate the $k_z$ dependence.

For \YA{Figure 5} in the main text, we used a chemical potential shift of $\mu=0.4$eV, the order parameter magnitude of $d=0.01$eV, and the surface factor $p=0.01$.  For the gap with $C_{1v}$ symmetry we choose $d_{22}/d_{23} = \sqrt{2}$ and for the gap with $C_{3v}$ symmetry we choose \YA{$d_{22}/d_{21} = 1$}. The choice of $p$ and ratio of order parameters were such that we could find order parameter superpositions that have gap anisotropy and gap minima rotation comparable to the experimental observations. A more microscopic derivation of such parameters would be desirable, but goes beyond the scope of this work.